\documentclass[11pt]{article}
\usepackage{epsfig, amsmath, amssymb, amsthm, times}
\usepackage{graphicx}
\newtheorem {thm}{Theorem}[section]
\newtheorem {lem}[thm]{Lemma}

\newtheorem {cor}[thm]{Corollary}

\theoremstyle{defintion}
\newtheorem {df}[thm]{Definition}

\theoremstyle{remark}
\newtheorem{rem}[thm]{Remark}
\theoremstyle{example}

\theoremstyle{assumption}

\def\pf{{\it Proof.\;}}

\def\P{\mathbb P}
\def\R{{\mathbb R}}
\def\N{{\mathbb N}}
\def\C{{\mathbb C}}
\def\Z{{\mathbb Z}}

\def\T{{\mathbb T}}
\def\F{\mathcal{F}}

\DeclareMathOperator{\E}{\mathbb E}

\DeclareMathOperator{\re}{Re}
\DeclareMathOperator{\im}{Im}
\DeclareMathOperator{\Cay}{Cay}
\renewcommand\ln{\operatorname{ln}}
\newcommand{\iu}{{i\mkern1mu}}

\renewcommand\Im{\operatorname{Im}}

\providecommand{\keywords}[1]{\textbf{\textit{Key words:}} #1}
\providecommand{\subjclass}[1]{\textbf{\textit{AMS subject classifications:}} #1}

\def\lbl{\label}
\def\be{\begin{equation}}
\def\ee{\end{equation}}
\def\p{\partial}
\def\qed{\square}
\def\t{\intercal}
\def\1{\mathbf{1}}

\title{Stability of twisted states in the Kuramoto model on 
Cayley and random graphs
}\author{Georgi S. Medvedev  and Xuezhi Tang
\thanks{
Department of Mathematics, Drexel University, 3141 Chestnut Street,
Philadelphia, PA 19104, {\tt medvedev@drexel.edu}, {\tt  xt32@drexel.edu} 
}
}

\begin{document}
\maketitle
\begin{abstract}
The Kuramoto model of coupled phase oscillators on complete,
Paley, and 
Erd\H{o}s-R\'{e}nyi (ER) graphs is analyzed
in this work. 
As quasirandom graphs, the complete, Paley, and ER graphs share many
structural properties. For instance, they exhibit  the same asymptotics
of the edge distributions, homomorphism densities, graph spectra, and
have constant graph limits. Nonetheless, we show that the asymptotic
behavior of solutions in the Kuramoto model on these
graphs can be qualitatively different. Specifically, we identify twisted states, steady state solutions
of the Kuramoto model on complete and Paley graphs, which are stable for one
family of graphs but not for the other. 
On the other hand, we show that the solutions of the initial value
problems for the Kuramoto model
on complete and random graphs remain close on finite time intervals,
provided they start from close initial conditions and the graphs are sufficiently large.
Therefore, the results of this paper elucidate the relation between
the network structure and dynamics in coupled nonlinear dynamical
systems. Furthermore, we present new results on synchronization
and stability of twisted states for the Kuramoto model on Cayley and random graphs.
\end{abstract}

\section{Introduction}
Collective dynamics of large systems of coupled oscillators feature
prominently in the mathematical modeling of many physical, biological,
social, and technological networks.
Examples include regulatory and neuronal
networks \cite{Bre12, MZ12},
Josephson junctions and coupled lasers \cite{WatStr94}, 
power networks  and consensus protocols \cite{DorBul12, Med12}, to
name a few. Mathematical models of individual oscillators comprising
real-world  networks can be quite complex, which can obstruct the view
of global mechanisms controlling collective dynamics. For weakly coupled
networks, the problem can be substantially simplified by recasting the 
coupled model as a system of equations for the phase variables only
\cite{Malkin-Poincare, Kur84, HopIzh-book}. This is the basis of the 
following model of coupled phase oscillators due to Kuramoto.

Let $\Gamma=\langle V, E \rangle$ be an undirected
graph. 
Denote the neighborhood of $x\in V$ by 
$$
N(x)=\{y\in V:\; xy\in E\}.
$$ 
The Kuramoto model of coupled phase oscillators on $\Gamma$ has
the following form
\be\lbl{KM}
{\p \over \p t} u(x,t)=\omega(x) +{(-1)^\alpha\over |N(x)|} \sum_{y\in N(x)}
\sin \left(2\pi (u(y,t)-u(x,t))\right), 
\ee
where $u(x,t)$ denotes the phase of the oscillator at $x\in V$ at time $t\in\R$ and $\omega(x)\in\R$
is its intrinsic frequency. The coupling in (\ref{KM}) is called
attractive if $\alpha=0$ and repulsive if $\alpha=1$.
The derivation of (\ref{KM}) assumes that the frequencies $\omega(x), x\in V,$
are close to each other (cf.~\cite{Kur84}). In this paper, we study 
(\ref{KM}) for $\omega(x)=\mbox{const}$.
In this case,  after recasting (\ref{KM}) into a moving frame of coordinates, we have
\be\lbl{KM-before}
{\p \over \p t} u(x,t)={(-1)^\alpha\over |N(x)|} \sum_{y\in N(x)}
\sin \left(2\pi (u(y,t)-u(x,t))\right).
\ee
\begin{figure}
\begin{center}
\textbf{a} \includegraphics[height=1.8in,width=2.2in]{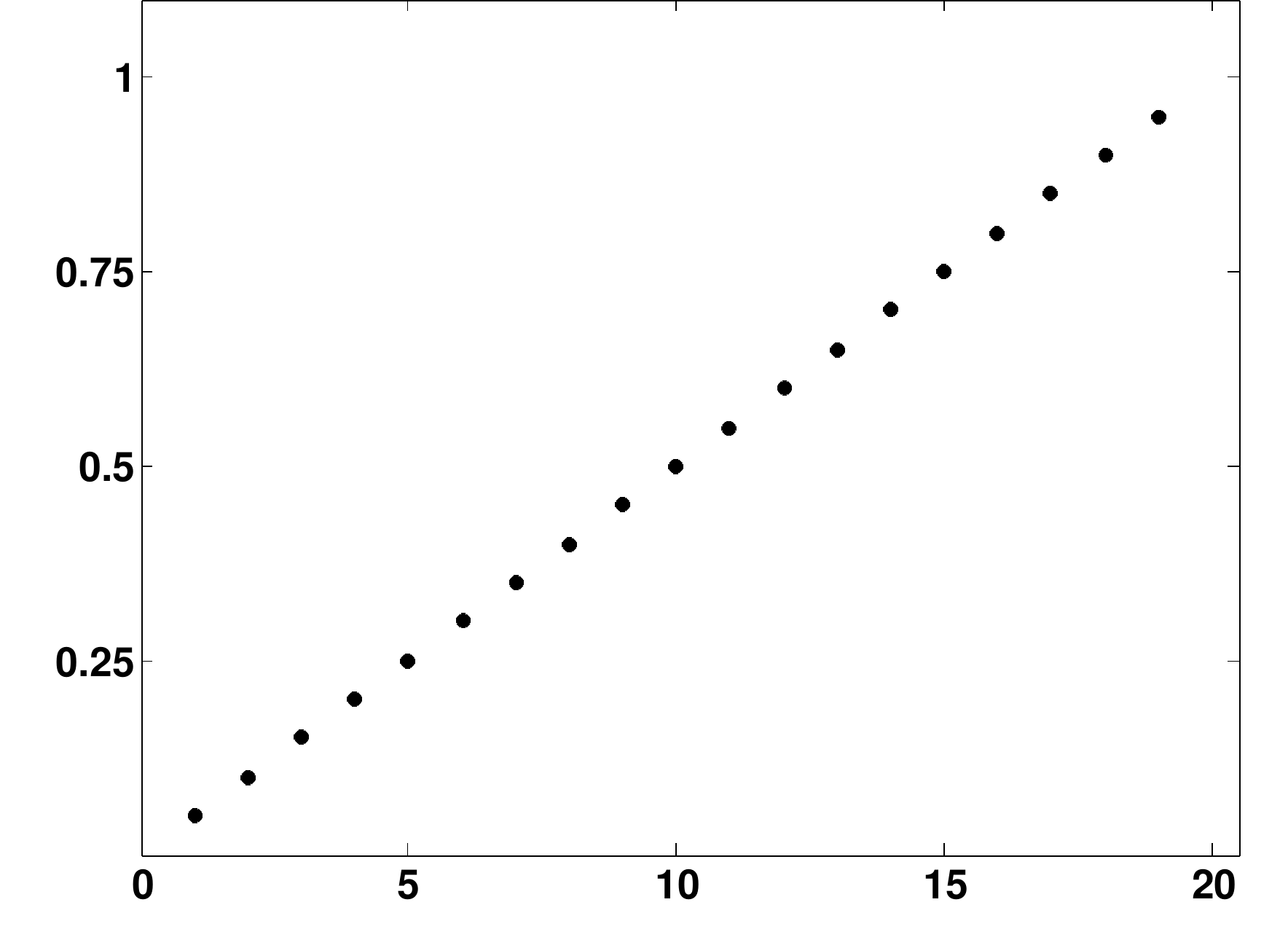} \qquad
\textbf{b} \includegraphics[height=1.8in,width=2.2in]{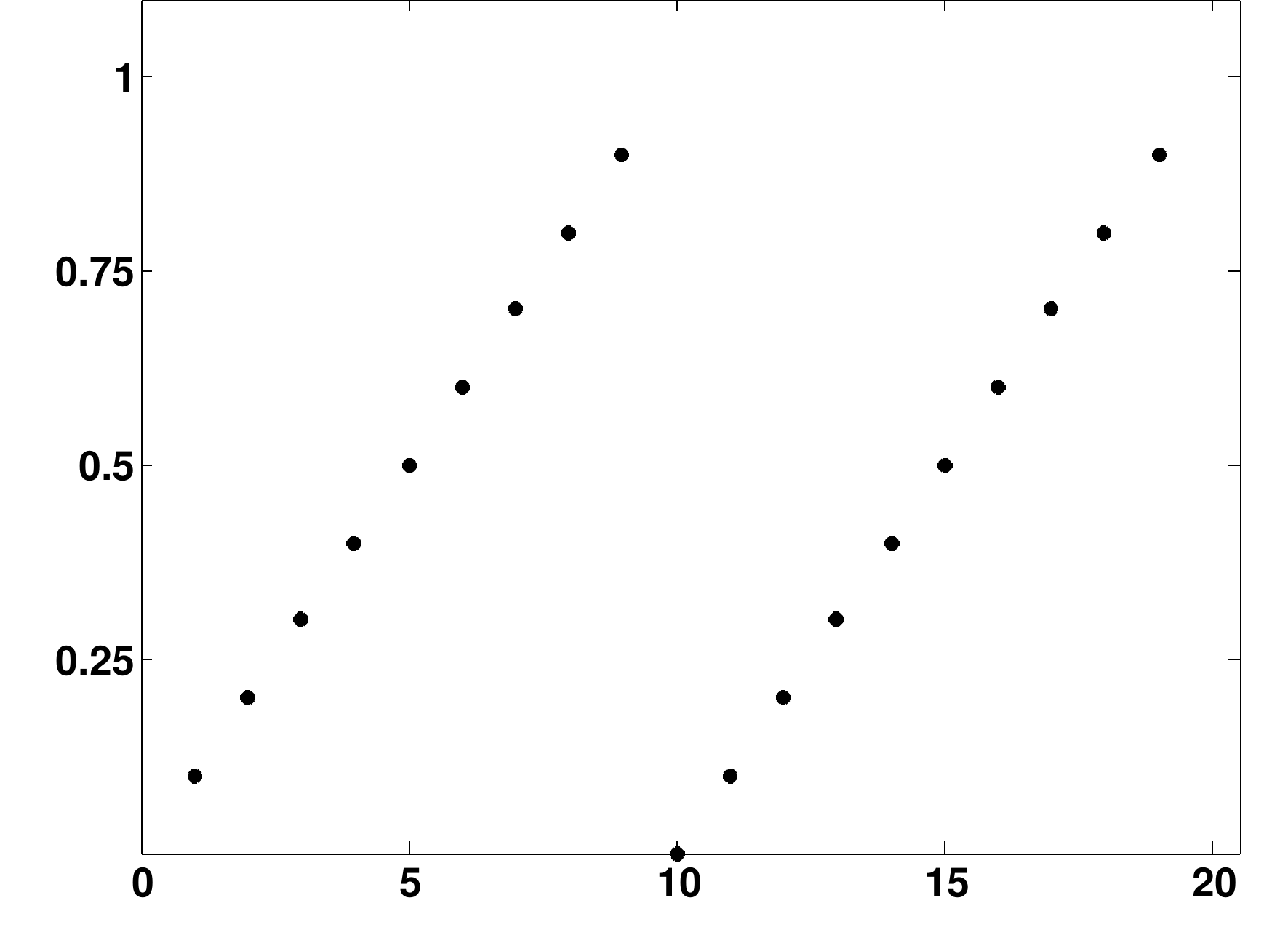}
\end{center}
\caption{Twisted states shown in \textbf{a}) and \textbf{b}) are steady state solutions of 
the Kuramoto model (\ref{KM}) on Cayley graphs.
}
\lbl{f.3}
\end{figure}

The Kuramoto model (\ref{KM}) and its numerous variations have formed an influential
framework for studying synchronization \cite{Kur84, Kur84-book, HopIzh-book}. 
More recently, 
new spatio-temporal patterns in the Kuramoto model with nonlocal coupling have been 
discovered and received a great deal of attention.   These are first of all chimera states, 
spectacular patterns combining regions of coherent and incoherent oscillations 
\cite{KurBat02, AbrStr06, WolfOme11, Ome13}, and so-called twisted or
splay states \cite{WilStr06}.
The latter are defined as follows.
\begin{df}\lbl{df.q-twist}
For $q\in \{0,1,\dots, n-1\}$ consider $u_q: [n] \to \T:=\R/\Z$ defined by
\be\lbl{q-twist}
u_n^{(q)}(x)= {qx\over n}+c \pmod{1} , c\in \R.
\ee
Function $u_n^{(q)}$ is called a $q$-twisted state (see Fig.~\ref{f.3}). 
\end{df}

Twisted states were identified as an important family of steady-state solutions in the 
Kuramoto model on $k$-nearest-neighbor graphs by Wiley, Strogatz, and Girvan in \cite{WilStr06}. 
They have been studied in the Kuramoto model with repulsive coupling in \cite{GirHas12},
in that on small-world graphs in \cite{Med13c}, and in the Kuramoto model with randomly distributed
frequencies \cite{OmeLai14}. The stability of twisted states, which is intimately related
to the network connectivity \cite{WilStr06}, provides  valuable insights into the 
structure of the phase space of the Kuramoto model and is important for understanding more complex
spatial patterns such as chimera states \cite{XieKno14}. 
In this paper, we study stability of twisted states in the Kuramoto model on certain important  regular and 
random graphs. Our goal is to elucidate the link between the network structure 
and emergent spatio-temporal patterns in coupled nonlinear dynamical systems such as the Kuramoto model. 
To this end, the problem of stability of twisted states provides a convenient setting for highlighting
a subtle relation between the network topology and dynamics. 

Dynamics in coupled networks is shaped by the interplay of the properties of the local dynamical systems 
at the individual nodes of the graph and by the  structure of connections between them.
The connectivity patterns that are of interest in such applications as power grids, neuronal networks, 
or the Internet, can be extremely complex. Therefore, one has to understand what
structural features of large networks are important for various aspects
of its dynamics. As a first step in this direction, we ask the following question. Do  networks with 
similar structural properties exhibit similar dynamics?
Large graphs that have similar combinatorial properties do not have to look alike.
This is perhaps best illustrated by quasirandom graphs 
\cite{Tho87a, ChuGra88, KriSud06, AloSpe-PMethod}. Roughly speaking, quasirandom graphs 
are the graphs that behave like truly random 
Erd\H{o}s-R\`{e}nyi (ER) graphs \cite{AloSpe-PMethod}. Surprisingly, some
very regular graphs like  complete and Paley graphs turn out
to be quasirandom (see Fig.~\ref{f.1}).
To quantitatively compare connectivity of disparate graphs such as complete,
Paley and ER graphs, one can use 
the edge distributions, or employ the densities of the homomorphisms
from simpler  graphs (e.g., triangles, $n$-cycles, etc) into these (large) graphs, 
or compare the eigenvalues  of the corresponding adjacency matrices. 
Surprisingly, all these tests turn out to be equivalent
for quasirandom graphs, i.e., it is sufficient to use any of them to determine whether
a given graph is quasirandom \cite{ChuGra88}. In particular, the infinite sequences 
of the complete, Paley, and ER graphs exhibit the same (up to rescaling by degree) asymptotics
of all of the above quantities. Thus, all three graph sequences have a great deal of 
similarity from a combinatorial viewpoint. But what does this say about the dynamics
on these graphs? To tackle this question, in this paper we undertake the study of stability
of twisted states in the Kuramoto model on these three families of graphs. 
\begin{figure}
\begin{center}
{\bf a}\hspace{0.1 cm}\includegraphics[height=1.8in,width=2.0in]{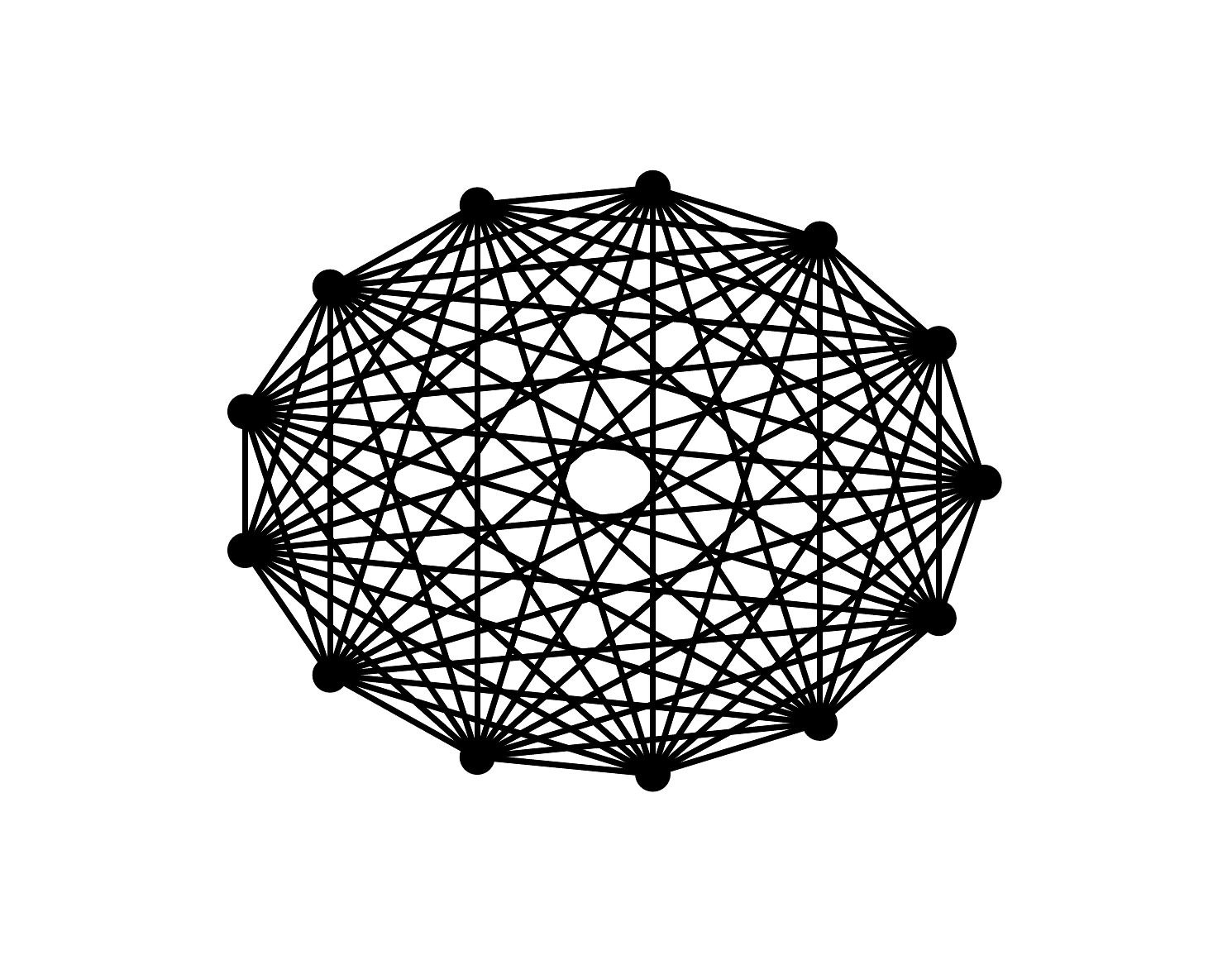}
{\bf b}\hspace{0.1 cm}\includegraphics[height=1.8in,width=2.0in]{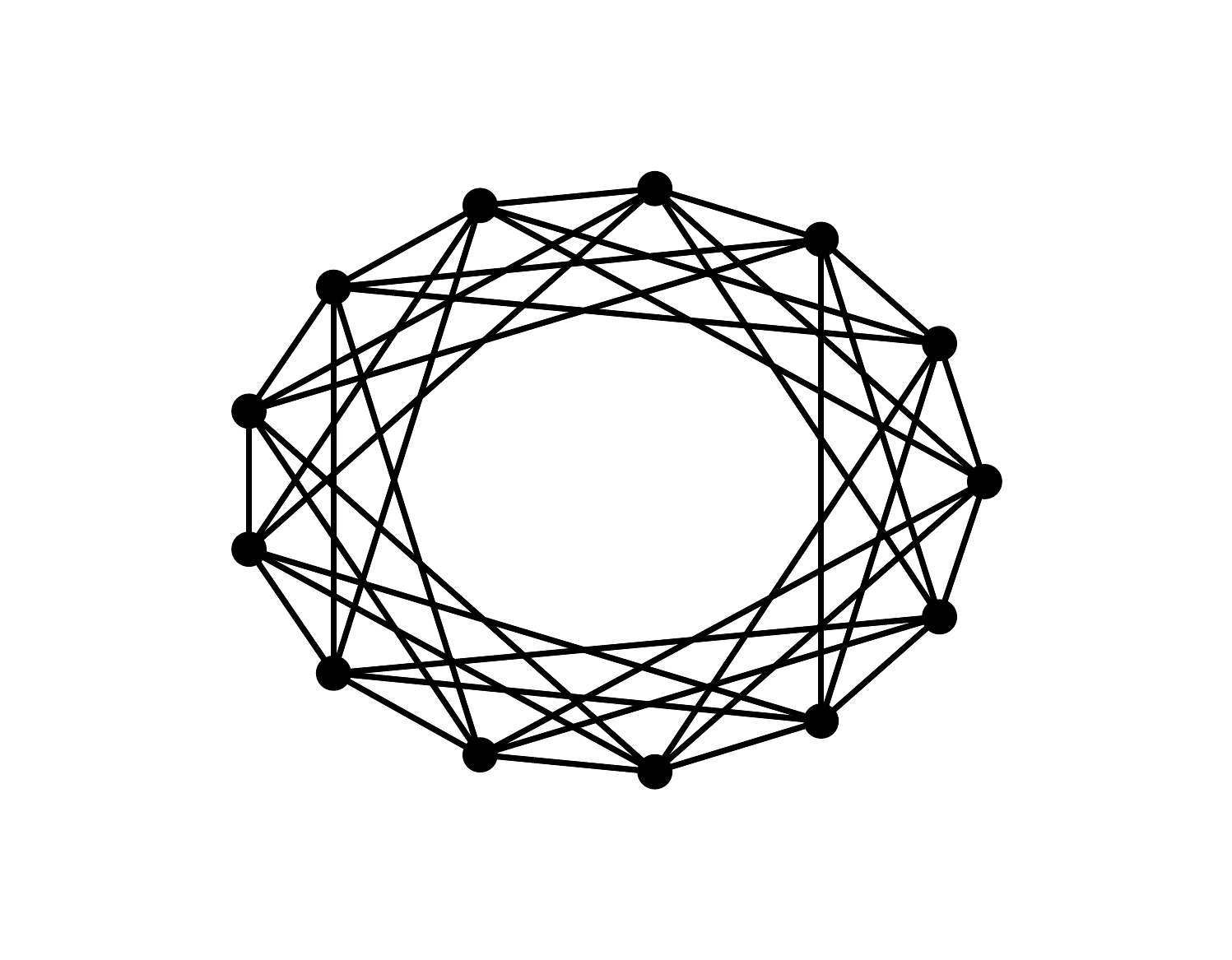}
{\bf c}\hspace{0.1 cm}\includegraphics[height=1.8in,width=2.0in]{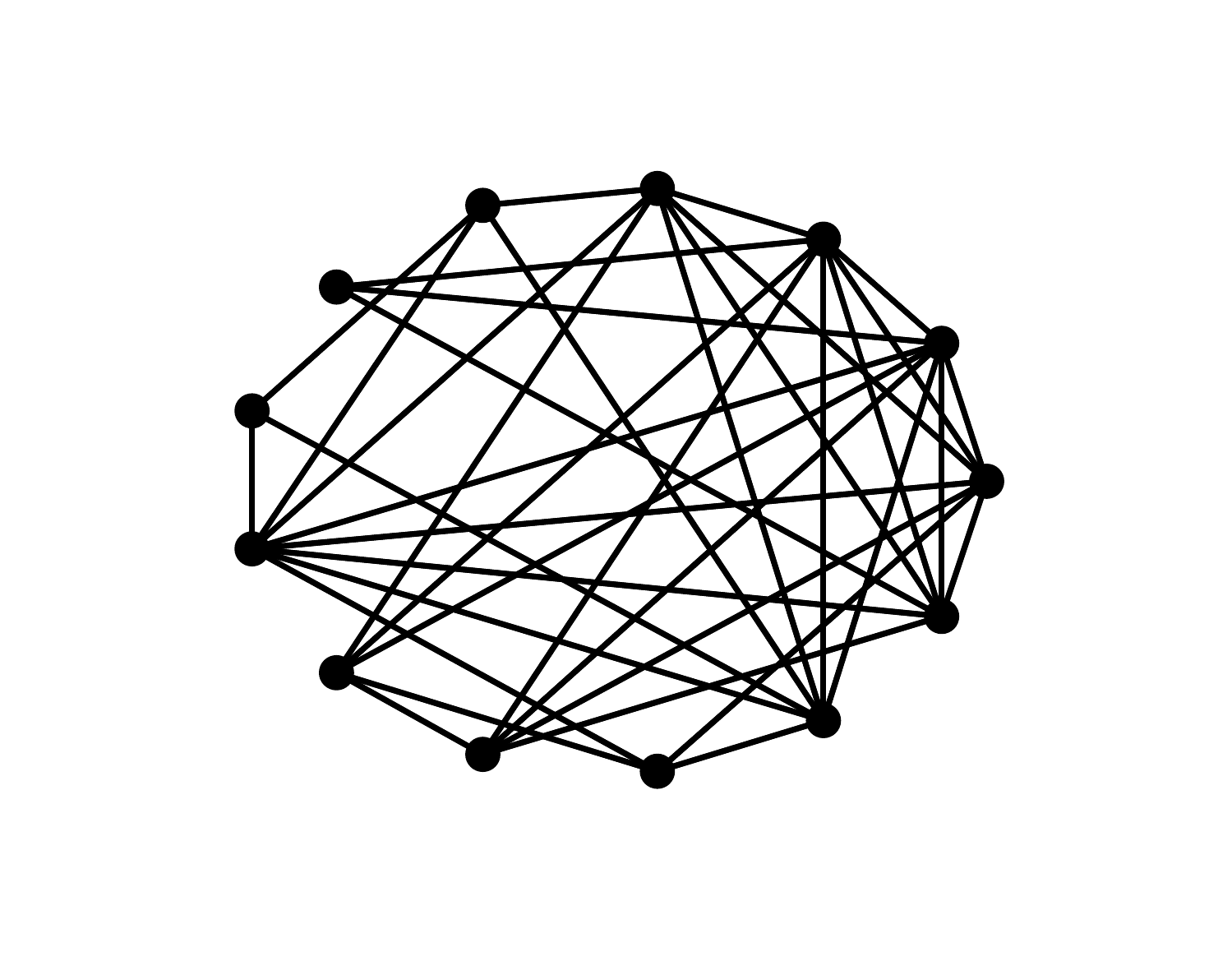}
\end{center}
\caption{  {\bf a}) Complete, {\bf b}) Paley, and  {\bf c}) an ER graphs. 
}
\lbl{f.1}
\end{figure}

In Section~\ref{sec.twist}, we show that twisted states are steady state solutions of the 
Kuramoto model on any Cayley graph generated by a cyclic group. This, in particular, includes the
models on Paley and complete graphs\footnote{For convenience, we consider complete
graphs with odd number of vertices, so that they can be interpreted as Cayley graphs.}.
Furthermore,  one can show that the Kuramoto model on ER graphs supports
twisted states almost surely, albeit in the limit as the number of oscillators tends to infinity
(cf.~Theorem~\ref{thm.steady}, see also \cite{Med13c}).
Thus, by comparing stability of twisted states on the complete, Paley, and ER graphs
we can distill the effects of the finer structural features of these graphs on the dynamics
of the Kuramoto model. 
The analysis of this paper shows that while there is a significant overlap in the sets
of stable twisted states in the Kuramoto models on the complete and Paley graphs, surprisingly there are 
twisted states with opposite stability properties.  This indicates  that the stability 
of steady state solutions may differ even in networks with such close
asymptotic properties as the families of Paley and complete graphs. On the other hand,
we show that for the repulsively coupled Kuramoto model on finite (albeit large) ER graphs, 
all nontrivial twisted states are metastable in the following sense. The solutions of the
IVP with the initial data near nontrivial twisted states remain near
them on for a long time. 
The same conclusion obviously holds for models
on Paley and complete graphs by the continuous dependence on initial values. 
Thus, the solutions of the IVPs for all three models 
that started sufficiently close to a given twisted state remain close on finite time intervals,
but not necessarily asymptotically.

The organization of the paper is as follows.
Before turning to stability of twisted states, in the preliminary Section~\ref{sec.graphs}
we review necessary background information from graph theory. Our goal here is
to explain the similarity between the infinite families of complete, Paley, and ER graphs. 
In addition, to the results about edge distributions and the eigenvalues of quasirandom graphs,
in this section we  include several basic facts about graph limits, which provide another way
to highlight the similarities between different quasirandom graphs. In Section~\ref{sec.twist},
we develop two methods for studying stability of twisted states on Cayley graphs.
The first approach is based on linearization. To study the stability of the linearized system
we use the discrete Fourier transform \cite{Ter99}. This leads to a sufficient condition
for stability of twisted states generalizing the condition for stability of twisted 
states in the $k$-nearest-neighbor coupled Kuramoto model in \cite{WilStr06}. In the same section,
we present an alternative variational method for studying stability of twisted states.
Specifically, we show that stable steady states in the Kuramoto model with attractive (repulsive) coupling
are local minima (maxima) of the quadratic form for the graph Laplacian $L\in\R^n$ on
the $n$-torus
$$
Q(z)=z^\ast Lz,\; z=(z_1,z_2,\dots,z_n)^\t, z_i\in\C, |z_i|=1, i\in [n].
$$
We use this observation to show that synchrony is always stable in an attractively coupled model,
whereas twisted states corresponding to the largest eigenvalue of $L$ are stable in the model with 
repulsive coupling. This provides an interesting relation between the spectral properties of
the graph $\Gamma$  and stability of twisted states. Furthermore, the variational interpretation
helps to determine stability of twisted states when the spectrum 
of the linearized problem has multiple zero eigenvalue\footnote{One zero eigenvalue is always present due to 
the translational invariance of twisted states.} and the linear stability analysis is inconclusive. 
In Section~\ref{sec.sync}, we take a closer look at 
synchronization. We prove that the synchronization subspace is asymptotically stable in the 
attractively coupled model (cf.~Theorem~\ref{thm.synchrony}). Here, we use the gradient structure 
of the Kuramoto model to construct a Lyapunov function. Likewise, we show that twisted states corresponding to
the largest eigenvalue of the graph Laplacian  are stable in the model
with repulsive coupling. In Section~\ref{sec.Kn}, we study twisted states in the Kuramoto model on the complete
graph. For this system the variational approach of Section~\ref{sec.twist} implies that all nontrivial twisted states
are stable (unstable) if the coupling is repulsive (attractive). To get a more detailed picture of the flow
near twisted states we use linearization. The linearized problem for the model on the complete
graph on $n$ nodes, $K_n$, is highly degenerate:
the spectrum contains two negative eigenvalues and $n-2$ zero eigenvalues. We show that each nontrivial
$q$-twisted state ($q\neq 0$) is stable,  because it lies in an $(n-2)$-dimensional smooth manifold 
formed by the equilibria
of the Kuramoto model. In Section~\ref{sec.Pn}, we turn to the Kuramoto model on Paley graphs. Here, we use the combination of the 
linear stability analysis and the variational analysis to determine stability for most of the twisted states
in the Kuramoto model with attractive and repulsive coupling. Interestingly, we find many unstable twisted states in
the  repulsively coupled model. Recall that in the same model on $K_n$, all nontrivial twisted states are stable.
In Section~\ref{sec.gnp}, we study the Kuramoto model on the ER random graphs. We first establish 
that the solutions of the IVPs for the Kuramoto model on ER and complete graphs on $n$ nodes
that started from the same initial condition remain $O(n^{-1/2})$ close in the appropriate metric
on finite time intervals with probability
tending to $1$ as the graphs' size tends to infinity. This result is an analog of homogenization in the 
discrete setting. Next, we show that the Kuramoto model on $G(n,p)$ supports twisted states almost surely as $n\to\infty$.
Finally, we prove that in finite random networks solutions starting near twisted states remain 
near them for a long time provided the network is sufficiently large. In other words, 
in the Kuramoto model on ER graphs twisted states are metastable.  Section~\ref{sec.discuss} offers concluding remarks.
  
\section{The graphs}\lbl{sec.graphs}
\setcounter{equation}{0}
In this section, we present the background material from graph theory,
which is meant to explain why we study the Kuramoto model on the complete, Paley, and ER graphs
in the remainder of this paper. The facts collected below show that these three families of graphs
are very similar from the combinatorial, probabilistic, and algebraic viewpoints. In particular, all three 
graph sequences would yield the same (up to rescaling) formal mean field limit for the Kuramoto
model (see \textbf{Fact 3} \S\ref{sec.limit}). Nonetheless, the dynamics of the Kuramoto model
on these graphs may be qualitatively different, as follows from the stability analysis of the 
twisted states in the second half of this paper.


\subsection{Preliminaries}

Let $\Gamma=\langle V, E\rangle$ be an undirected graph. Here, $V$
stands for the set of nodes and 
$E$ denotes the set of edges, i.e., unordered pairs from $V$. An edge joining $x\in V$ and
$y\in V$ is denoted by $xy$. Note that $xy$ and $yx$ mean the same edge. We assume
that $\Gamma$ does not have multiple edges and loops, i.e., it is a simple graph. For
dynamical systems defined on graphs,
edges represent connections between the local dynamical systems located at nodes of $\Gamma$.
Below we consider dynamical systems on several different
types of graphs. We start with symmetric  Cayley graphs defined on the
additive cyclic group $\Z_n=\Z/n\Z, n\in\N,$ with respect to a symmetric subset $S\subset \Z_n$,
i.e., $s\in S\,\implies\, -s\in S$.

\begin{df}\lbl{df.Cayley} 
$\Gamma=\langle V, E\rangle$ is called a Cayley graph of $\Z_n$ with respect to $S$ and
denoted $\Cay(\Z_n,S)$ if 
 $a,b\in \Z_n,$ $ab\in E$ if $b-a\in S$.
\end{df}  

We restrict to Cayley graphs of finite cyclic groups (i.e., circulant graphs), because they provide 
a natural setting for studying twisted states in the Kuramoto model. 
This is a special class of Cayley graphs\footnote{Note, however, that any finite 
abelian group is isomorphic to a
direct product of cyclic groups. This can be used to extend the linear stability analysis of this 
section to the Kuramoto model on Caley graphs of abelian groups (cf. \cite[Chapter 10]{Ter99}).} 
(see \cite{Magnus-Karrass} for a 
general definition).
The following examples of graphs are used throughout this paper.

\begin{df}\lbl{df.ball}
\begin{description}
\item[A)]
Let $n\ge 3, r\le\lfloor n/2\rfloor$ and $B(r)=\{\pm 1, \pm 2, \dots, \pm r \}.$
$\Gamma =\Cay(\Z_n, B(r))$ is called a Cayley graph based on the ball $B(r)$ \cite{Ter99}.
\item[B)] 
Let  $n= 1\pmod 4$ be a prime and denote 
$$
\Z^\times_n=\Z_n/\{0\} \quad\mbox{and}\quad Q_n=\{ x^2 \pmod{n}:\; x\in \Z_n^\times\}.
$$  
$Q_n$ is viewed as a set (not multiset), i.e., each element has multiplicity $1$. Then $Q_n$ is 
a symmetric subset of $Z_n^\times$ and $|Q_n|=2^{-1}(n-1)$ (cf. \cite[Lemma~7.22]{KreSha11}).
$P_n=\Cay(Z_n, Q_n)$ is called a Paley graph. For the details of construction of Paley graphs and
the discussion of their properties, we refer an interested reader to  \cite[Chapter 7, \S 6]{KreSha11}.
\item[C)] A simple graph $K_n=\langle V,E\rangle$ on $n$  with $V=[n]$
and $E=\{ ij:~ i,j\in [n]\; \&\; i\neq j\}$ is called the complete graph.
\end{description}
\end{df}

\begin{rem}\lbl{rem.complete} For analytical convenience, in this paper we consider  complete
graph $K_n$ on the odd number of nodes.
In this case, $K_{2r+1}$ can 
be viewed as a Cayley graph $\Cay(\Z_{2r+1}, B(r))$.
\end{rem}

Along with highly regular Cayley graphs, we consider random graphs. The ER graphs 
defined below do not look like Cayley graphs, but it turns out that they have much in 
common with $K_n$ and $P_n$, $n\gg 1$.

\begin{df} \lbl{df.ER}\cite{KriSud06} Let $n\in\N$ and $p\in (0,1]$. The ER graph 
$G(n,p)$ is the probability space of all labeled graphs on vertex set $V=[n]$ such that 
every pair $(i,j)$ ($1\le i<j\le n$) forms an edge  with probability $p$ independently
from any other pair.
\end{df}

\begin{figure}
\begin{center}
{\bf a}\hspace{0.1 cm}\includegraphics[height=1.8in,width=2.2in]{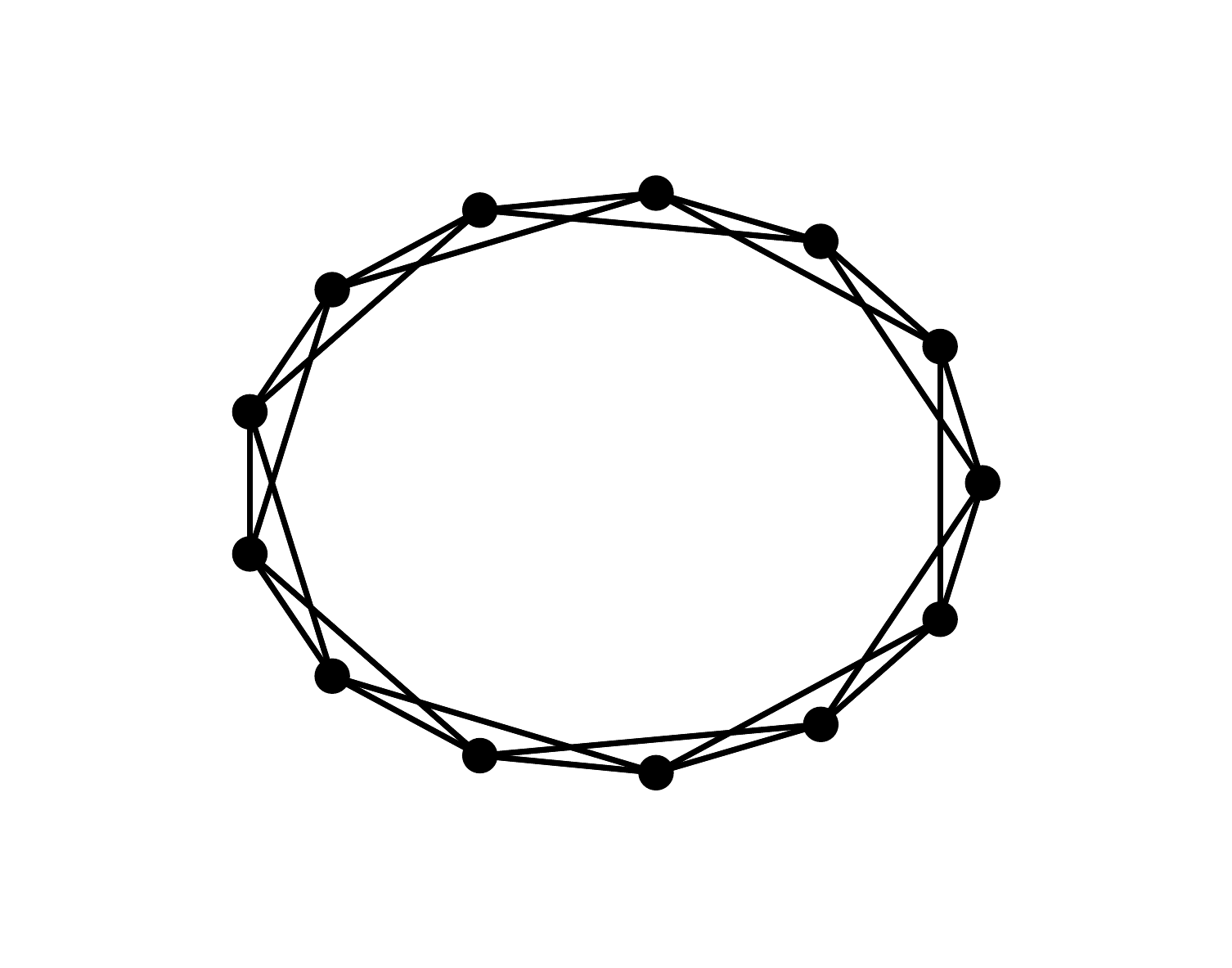} \qquad
{\bf b}\hspace{0.1 cm}\includegraphics[height=1.8in,width=2.2in]{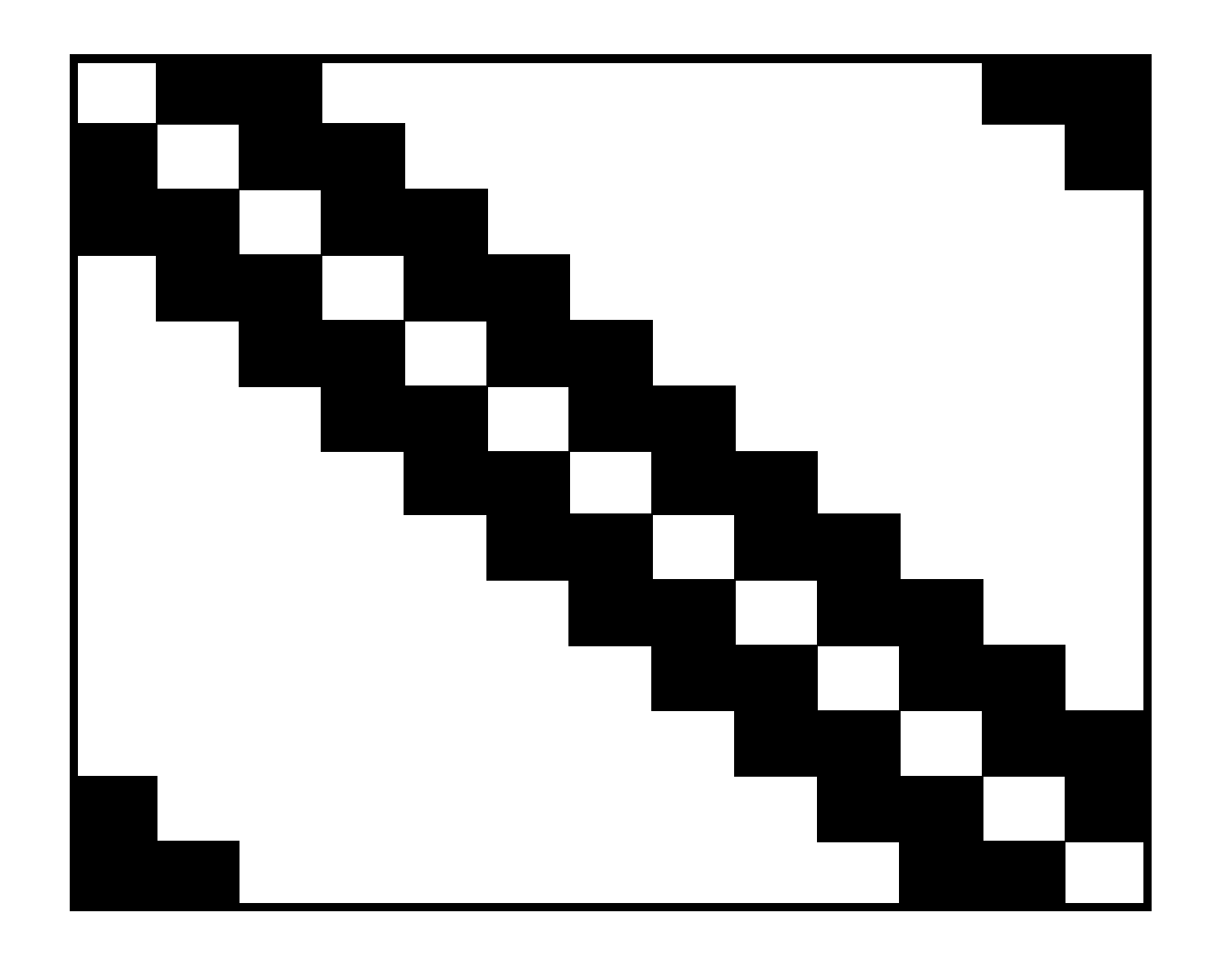}
\end{center}
\caption{  \textbf{a)}  A Cayley graph based on a ball. \textbf{b}) The pixel picture of the 
graph in \textbf{a}).
}
\lbl{f.pixel}
\end{figure}

Next, we review some tools for describing connectivity of graphs. The structure of
of $\Gamma=\langle [n],E\rangle$  is encoded in  its adjacency matrix $A=(a_{ij})\in\R^{n\times n}$
defined as follows 
\be\lbl{adjacency}
a_{ij}=\left\{ \begin{array}{ll} 
1,& ij \in E, \\
0,& \mbox{otherwise}.
\end{array}
\right.,\; (i,j)\in [n]^2.
\ee
Throughout this paper, we will often refer to the following geometric realization of the 
adjacency matrix $A$. Consider a $\{0,1\}$-valued function 
$W_\Gamma$ defined on the unit square $[0,1]^2$ as follows
\be\lbl{pixel}
W_{\Gamma}(x,y)=\left\{ \begin{array}{ll} 1, & \mbox{if}\; ij \in E\;\mbox{and}\;
(x,y)\in \left[{i-1\over n}, {i\over n}\right)\times \left[{j-1\over n}, {j\over n}\right),\\
0,& \mbox{otherwise}.
\end{array}\right.
\ee
The plot of the support of $W_\Gamma$ is called the pixel picture of $\Gamma$ \cite{LovGraphLim12}. 
It provides a convenient visualization of the structure of the
graph. Fig.~\ref{f.pixel} presents a schematic representation of the
Cayley graph based on a ball (\textbf{a}) and its pixel picture (\textbf{b}).
In this and other pixel pictures of graphs, we place the origin at the top left corner of the 
plot to emphasize the relation between $W_\Gamma$ and the adjacency matrix $A$.

The adjacency matrix and the pixel plot yield algebraic and geometric means for comparing
connectivity of distinct graphs. A complementary analytic way is provided by the edge distribution 
of $\Gamma=\langle V, E\rangle$,  $e_\Gamma:2^V\to\Z$. This function 
for every  $U\subset V$ yields the number of edges in $U$ as an induced subgraph
of $\Gamma$. With these tools at hand, we are now in a position to discuss the similarities 
the complete, Paley, and ER random graphs. 

\begin{figure}
\begin{center}
{\bf a}\hspace{0.1 cm}\includegraphics[height=1.8in,width=2.2in]{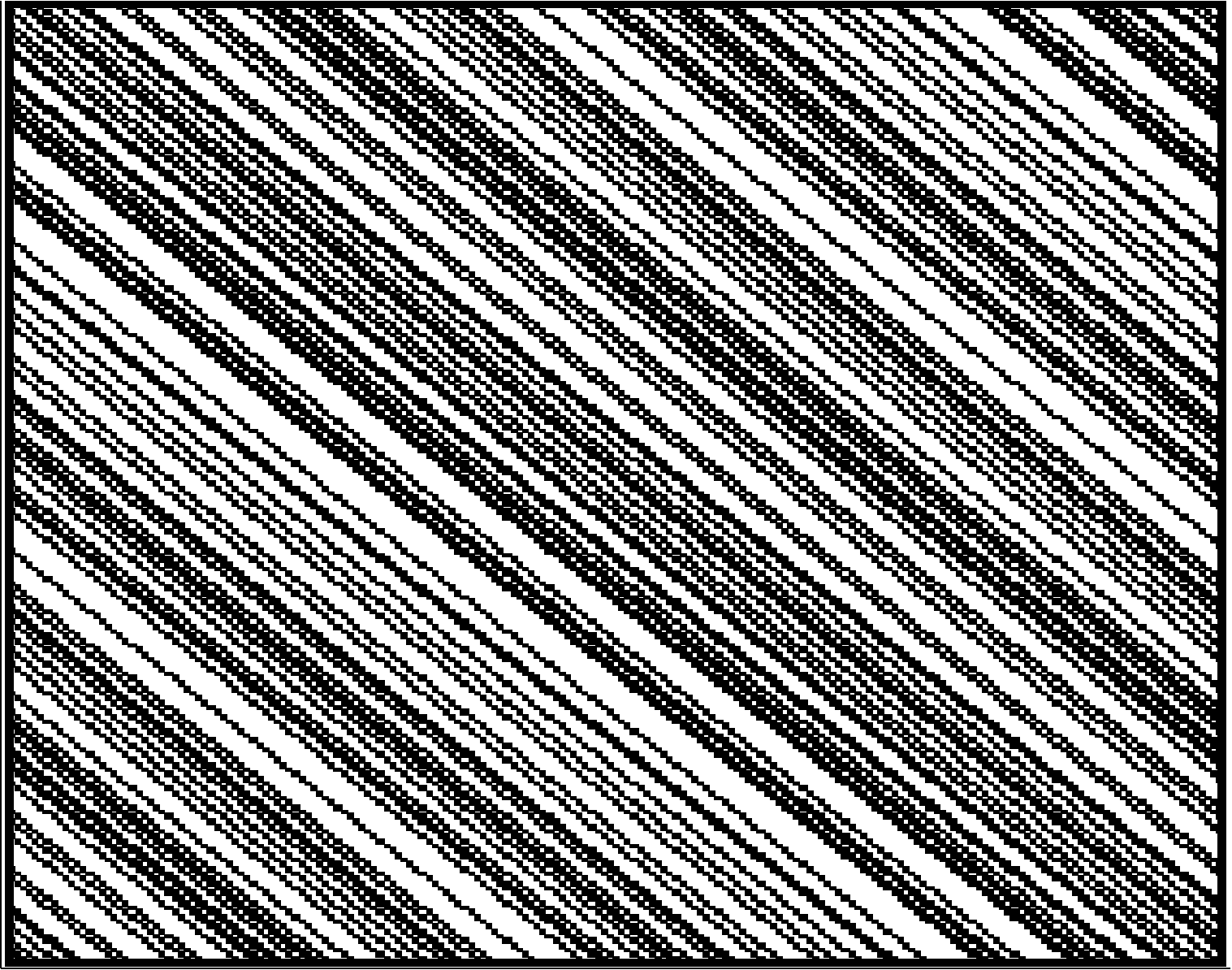} \qquad
{\bf b}\hspace{0.1 cm}\includegraphics[height=1.8in,width=2.2in]{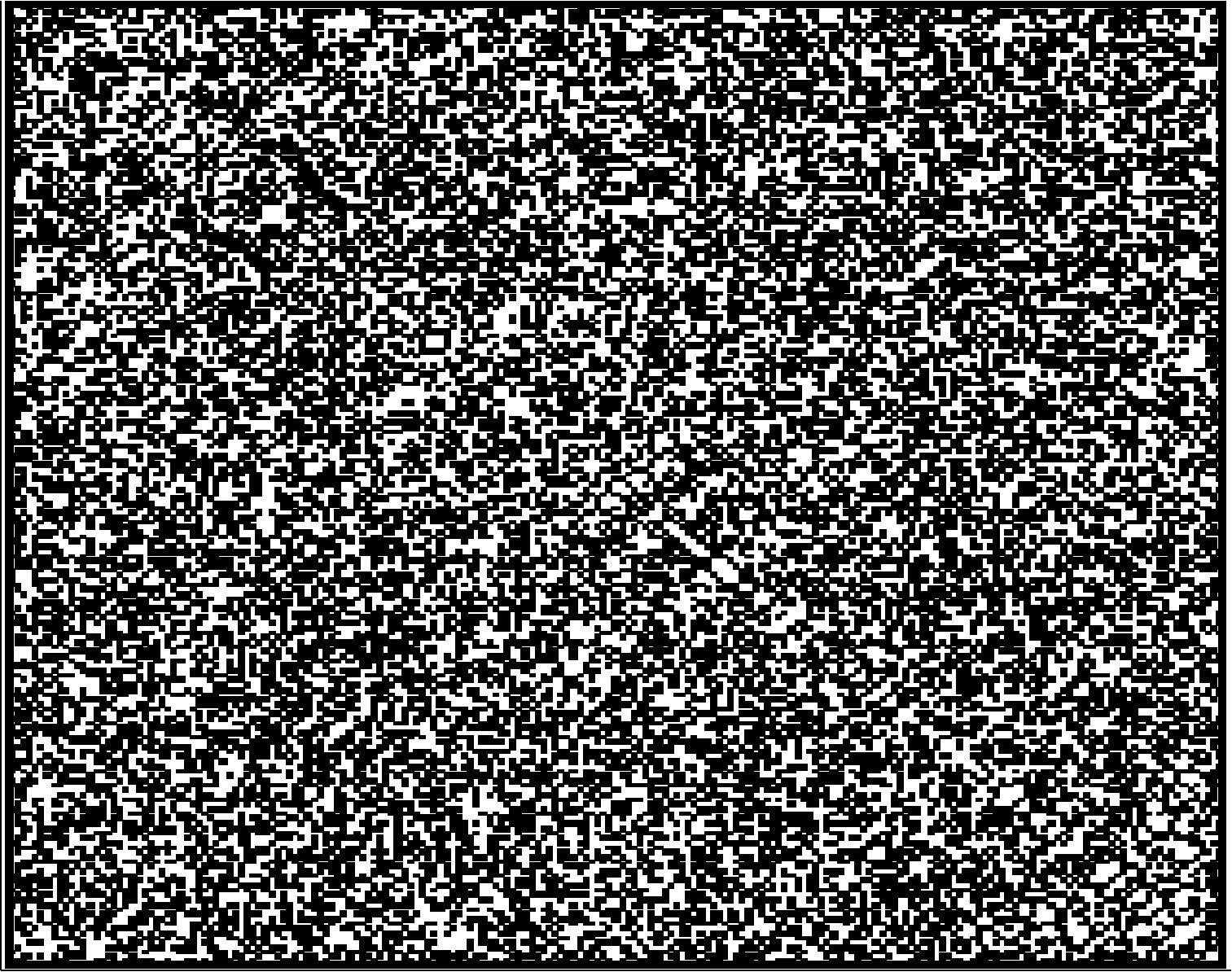}
\end{center}
\caption{  Pixel pictures of the adjacency matrices of a
Paley (\textbf{a}) and an ER   (\textbf{b}) graphs. 
}
\lbl{f.2}
\end{figure}

\subsection{The edge distributions of $K_n$, $P_n$, and $G(n,p)$}
At first glance, the pixel pictures of the large Paley and ER graphs shown in Fig.~\ref{f.2}
look quite different. The former shows distinct band structure, while the latter appears
scrambled. Nonetheless, what the two plots have in common is that the pixels are distributed
approximately uniformly. This becomes more evident for pixel plots of graphs of larger size.
The widths of the black and white bands in the pixel plots of $P_n$ decrease with increasing 
$n$. Thus, for large $n$ the pixel plots for $P_n$ and $G(n,1/2)$ look increasingly more alike
at least if viewed from a distance. The uniform pixel distribution obviously holds for the pixel
plots of the complete graphs. These observations suggest that all three families of graphs
may in fact have qualitatively similar edge distributions.  This 
brings us to the first result that highlights the similarity
of $K_n$, $P_n$ and $G(n,p)$.

\noindent \textbf{Fact 1.}~{\it Let $p\in (0,1), n\gg 1,$ and $\Gamma\in\{K_n,P_n, G(n,p)\}$.
Then 
\be\lbl{pseudo}
e_\Gamma(U)={\gamma_\Gamma\over 2} |U|^2 +o(n^2)\quad \forall U\subset V(\Gamma), 
\ee
where $\gamma_\Gamma=|E(\Gamma)|/ {|V(\Gamma)| \choose 2}$  stands for the edge
density of $\Gamma.$  In the case $\Gamma=G(n,p)$, (\ref{pseudo}) holds with probability 
$1$.
}
\begin{rem}\lbl{rem.pseudo}
Graphs satisfying (\ref{pseudo}) are called quasirandom  or pseudorandom
graphs \cite{ChuGra88,KriSud06}.
\end{rem}

Asymptotic relation (\ref{pseudo}) obviously holds for  complete graphs $K_n$. Clearly, $\gamma_{K_n}=1$ and
$$
e_{K_n}(U)={|U|\choose 2},
$$ 
in this case. For the random graph, $e_{G(n,p)}(U)$ is a binomial random variable with parameters
$p$ and ${|U| \choose 2}$. Thus,
$$
 \E e_{G(n,p)}(U)= p {|U|\choose 2} \quad\mbox{and}\quad \E\gamma_{ G(n,p)} =p.
$$
Using the estimates for the tail of the binomial distribution, one can further show that (\ref{pseudo})
with even smaller error term holds for $G(n,p)$ almost surely. As to  Paley graphs, we do not 
know any simple way of verifying (\ref{pseudo}) directly. However,
for $P_n$, (\ref{pseudo}) can be shown using an equivalent to (\ref{pseudo}) spectral characterization 
of quasirandom graphs \cite{ChuGra88, AloSpe-PMethod}, which
we discuss in the next paragraph.

\subsection{The graph eigenvalues}

The graph Laplacian of $\Gamma$ is defined as follows
$$
L=D-A, 
$$
where $A$ is the adjacency matrix and 
$$
D=\operatorname{diag}(d_1,d_2,\dots,d_n)
$$
is the degree matrix. Here, $d_i=|N(i)|$ is the degree of node $i\in [n]$. 
The normalized graph Laplacian of $\Gamma$ is defined by
$$
\tilde L=I-D^{-1/2} A D^{-1/2}.
$$
If the degree of every node of $\Gamma$ is equal to $d$, i.e., $\Gamma$ is a $d$-regular
graph, then the normalized graph Laplacian of $\Gamma$ is equal to $d^{-1}L$.
The eigenvalues of $L$ and $\tilde L$ carry important information about  $\Gamma$ \cite{Chung-Spectral}. 
In this subsection, we review 
certain facts about the eigenvalues of Cayley and quasirandom graphs that will be 
needed below.

Let $\Gamma=\Cay(\Z_n, S)$ for some symmetric subset $S\subset \Z_n$.
Then the  eigenvalues  of $L$ can be computed using characters 
of $\Z_n$
\be\lbl{character}
\mathbf{e}_x(y)=\exp\left\{ {2\pi\iu xy\over n}\right\}, \; y\in\Z_n.
\ee
Let $\mathbf{e}_x:\Z_n\to\C,\; x\in \Z_n,$ be a complex-valued
function on $\Z_n$:
\be\lbl{ex}
\mathbf{e}_x=(\mathbf{e}_x(0), \mathbf{e}_x(1), \dots,
\mathbf{e}_x(n-1))^\t, \; x\in\Z_n.
\ee
 The space of all complex-valued  functions  on $\Z_n$ is denoted by
$L_2(\Z_n,\C)$.
For the eigenvalues of Cayley graphs on a cyclic group, we have the following lemma.

\begin{lem}\lbl{lem.EV-cyc}
The eigenvalues of the graph Laplacian of $\Gamma=\Cay(\Z_n, S)$ are given by
\be\lbl{EV-C}
\lambda_x= |S|- \sum_{y\in S} e_x(y)=|S|- \sum_{y\in S}\cos\left({2\pi xy\over n}\right), \; x\in\Z_n.
\ee
The corresponding eigenvectors are  $\mathbf{e}_x, \; x\in\Z_n.$
\end{lem}
\pf
For any $x,y\in\Z_n,$ we have
$$
(A\mathbf{e}_x)(y)=\sum_{s\in S}\mathbf{e}_x(y+s)=\left(\sum_{s\in S}\mathbf{e}_x(s)\right) \mathbf{e}_x(y).
$$
Thus,
\be\lbl{EV-AC}
A\mathbf{e}_x=\mu_x \mathbf{e}_x, \quad \mu_x=\sum_{s\in S}\mathbf{e}_x(s).
\ee
Since characters $\mathbf{e}_x, x\in\Z_n,$ are mutually orthogonal, (\ref{EV-AC}) 
gives the full spectrum of $A$. The statement of the lemma follows from  (\ref{EV-AC}) and $L=|S|I-A$.\\
$\qed$

\begin{cor}\lbl{cr.zero} The spectrum of $L$ contains a simple zero eigenvalue
$
\lambda_0=0
$
corresponding to the constant eigenvector $\mathbf{1_n}=(1,1, \dots,1)^\t$.
\end{cor}

At the level of eigenvalues, the similarity between the sequences of complete, Paley, and 
ER graphs
can be expressed as follows.

\noindent\textbf{Fact 2.}~{\it The $n-1$ nonzero eigenvalues of the normalized Laplacian
of $\Gamma\in \{K_n, P_n, G(n,p)\}$ are equal to $1+o(1)$ for $n\gg 1$.
}

From Lemma~\ref{lem.EV-cyc}, for complete graphs,  we immediately have
\be\lbl{EV-Kn}
\lambda_x(K_n)=n-1, \; x\in \Z^\times_n.
\ee
 By (\ref{EV-C}), the eigenvalues of $P_n$\footnote{Whenever we refer to $P_n$, 
we assume implicitly that $n$ is a prime and $n=1\pmod 4$.} 
are
\be\lbl{EV-Pn-1}
\lambda_x(P_n)= {n-1\over 2}-2^{-1}\sum_{k=1}^{n-1} \mathbf{e}_x(k^2)=
2^{-1}(n-G_n(x)), \; x\in\Z_n^\times,
\ee
where 
\be\lbl{gauss-def}
G_n(x)=\sum_{k=0}^{n-1} \mathbf{e}_x(k^2)
\ee
stands for the Gauss sum \cite[\S 7.6]{KreSha11}. By the well-known
properties of the Gauss sum (see, e.g., \cite{KreSha11}), 
\be\lbl{Gsum}
G_n(x)=
\left\{\begin{array}{ll}
0 & x\pmod{n}=0,\\
\sqrt{n}, & y^2=x\pmod n \; \mbox{has a solution}, \\
-\sqrt{n},& y^2=x\pmod n \; \mbox{does not have a solution}.
\end{array}
 \right.
\ee
If for a given $x$ equation $y^2= x\pmod n$ is solvable,
we say that $x$ is a quadratic residue  (QR) modulo $n$. Thus,
for $x\in\Z_n^\times$, we have
 \be\lbl{EV-Pn}
\lambda_x(P_n)=
\left\{\begin{array}{ll}
2^{-1} (n - \sqrt{n}), & x \mbox{ is a QR}~\pmod{n}, \\
2^{-1} (n + \sqrt{n}), & x \mbox{ is not a QR}~\pmod{n}.
\end{array}
 \right.
\ee
From (\ref{EV-Kn}) and (\ref{EV-Pn}), it follows that the nonzero eigenvalues of the normalized
Laplacians of the complete and Paley graphs are all $1+o(1)$. 
Furthermore, Theorem~4 in \cite{ChuRad11} implies that $n-1$ eigenvalues of the normalized Laplacian 
of $G(n,p)$ converge to $1$ in probability. Thus, the spectra of all three families of graphs
coincide to leading order for $n\gg 1$. The localization of the nonzero eigenvalues of the normalized
graph Laplacian is one of several equivalent characterization of quasirandom graphs.
In particular, the formulas for the nonzero eigenvalues (\ref{EV-Pn}) imply (\ref{pseudo}) for the 
sequence of Paley graphs (cf.~\cite{ChuGra88}; see also \cite[\S 9.3]{AloSpe-PMethod}).

\subsection{Graph limits}\lbl{sec.limit}
In  the previous subsections, we explored the similarities between the sequences of
complete, Paley, and ER graphs through the prisms of the edge distributions and 
graph eigenvalues. Here, we will use graph limits as another way for comparing these graph
sequences. Below,  we review some facts about graph limits that will be needed in
the remainder of this paper. For more details, we refer the interested reader to 
\cite{LovSze06,BorChay08, LovGraphLim12}.

Let 
$\Gamma_n=\langle V(\Gamma_n), E(\Gamma_n)\rangle, n\in\mathbb{N},$ be a sequence of dense 
(simple) graphs, i.e., $|E(\Gamma_n)|=O(|V(\Gamma_n)|^2)$. The convergence of the graph sequence  
$\{\Gamma_n\}$ is defined in terms of the homomorphism densities
\be\lbl{hdense}
t(F,\Gamma_n)={\mbox{hom}(F,\Gamma_n)\over \left|V(\Gamma_n)\right|^{|V(F)|}}.
 \ee
Here, $F=\langle V(F), E(F)\rangle$ is a simple graph and $\mbox{hom}(F,\Gamma_n)$ stands for
the number of homomorphisms (i.e., adjacency preserving maps $V(F)\to V(\Gamma_n)$).
In other words,  (\ref{hdense}) is the probability that a random
map $h:~V(F)\to V(\Gamma_n)$ is a homomorphism. 

\begin{df}\lbl{df.convergent}\cite{LovSze06, BorChay08}
The sequence of graphs $\{\Gamma_n\}$
is called convergent if $t(F,\Gamma_n)$ is convergent for every simple graph 
$F$.
\end{df}
\begin{rem}
Sometimes it is more convenient  instead of the homomorphism density (\ref{hdense})
 to use the injective or induced homomorphism densities defined 
respectively by
\be\lbl{hinj}
t_0(F,\Gamma_n)={\mbox{inj}(F,\Gamma_n)\over (n)_k } \quad\mbox{and}\quad
t_1(F,\Gamma_n)={\mbox{ind}(F,\Gamma_n)\over (n)_k},
\ee
where $(n)_k=n(n-1)\dots (n-k+1)$. $t_1(F,\Gamma_n)$ is the density of the imbeddings of 
$F$ into $\Gamma_n$ as an induced subgraph (i.e., it preserves adjacency as well as nonadjacency
of nodes). Asymptotically for $n\gg 1$, all three densities are equivalent in the sense
that $|t(F,\Gamma_n)-t_{0,1}(F,\Gamma_n)|=o(1)$ \cite{LovSze06, LovGraphLim12}.
\end{rem}

For a convergent graph sequence, the limiting object is represented by a measurable 
symmetric function $W: I^2\to I$. Here and below, $I$ stands for $[0,1]$. 
Such functions are called graphons. The set of all graphons is 
denoted by $\mathcal{W}_0$.
\begin{thm}\lbl{thm.graphon}\cite{LovSze06}
For every convergent sequence of simple graphs, there is  $W\in\mathcal{W}_0$
such that
\be\lbl{t-to-t}
t(F,G_n)\to t(F,W):=\int_{I^|V(F)|} \prod_{(i,j)\in E(F)} W(x_i,x_j)
dx_1dx_2\dots dx_{|V(F)|} 
\ee
for every simple graph $F$.  
Moreover, for every $W\in\mathcal{W}_0$ there is a sequence of graphs 
$\{G_n\}$ satisfying 
(\ref{t-to-t}).
\end{thm}

Graphon $W$ in (\ref{t-to-t}) is the limit of the convergent sequence $\{\Gamma_n\}$.
It is unique up to measure preserving transformations, meaning that for any other
limit $W^\prime\in\mathcal{W}_0$ there are measure preserving maps $\phi,\psi:I\to I$
such that $W(\phi(x),\phi(y))=W^\prime(\psi(x),\psi(y))$.
Geometrically, graphon $W$ representing the limit of a convergent sequence of
simple graphs $\{\Gamma_n\}$ can be interpreted as the limit of pixel pictures
of the adjacency matrices of $\Gamma_n$, $A(\Gamma_n)$  (cf.~(\ref{pixel})).
After relabeling the nodes of $\{\Gamma_n\}$ if necessary, 
$\{W_{\Gamma_n}\}$ converges to $W$
in the cut norm\footnote{The cut-norm of graphon $W\in\mathcal{W}_0$ is defined by  
$$
\|W\|_\qed =\sup_{S,T\in \mathcal{L}_I} \left| \int_{S\times T} W(x,y) dxdy\right|
$$
where $\mathcal{L}_I$ stands for the set of all Lebesgue measurable subsets of  $I$.}. 
Note that the pixel picture of $A(\Gamma_n)$ coincides with the plot of
 the support of $W_{\Gamma_n}$ (see Fig.~\ref{f.2} for the pixel pictures of $A (P_n)$ and
$A(G(n,1/2))$). Thus, the graph limit $W$ can be thought  as
the limit of the pixel pictures of adjacency matrices  $A(\Gamma_n)$ in the appropriate metric.
The similarity of the pixel pictures of the adjacency matrices of $K_n$, $P_n,$ and $G(n,p)$
reflect the following fact about their limits.

\noindent\textbf{Fact 3.}~{\it Graph sequences $\{K_n\}$, $\{P_n\}$, and $\{G(n,p)\}$ 
converge to the constant graphons $\operatorname{Const}(1),$ $\operatorname{Const}(1/2)$,
and $\operatorname{Const}(p)$, respectively.
}  

Clearly, $\{K_n\}$ is a convergent sequence as $t_0(F, K_n)=1$ for any simple graph $F$ 
and sufficiently large $n$.
It is not hard to see that the expected value $\E t_0(F,G(n,p))=p^{|E(F)|}$ for 
$n\ge |V(F)|$. From this using
concentration inequalities, one can further show that $t(F,G(n,p))\to p^{|E(F)|}$ 
almost surely \cite[Corollary~2.6]{LovSze06}.  The sequence of Paley graphs $P_n$, 
as quasirandom graphs, satisfies
\be\lbl{P1}
t_1(F,\Gamma_n)=(1+o(1))n^{|V(F)|}2^{-{|V(F)| \choose 2}}, \quad n\gg 1
\ee
\cite[Property $P_1(s)$, \S 9.3]{AloSpe-PMethod}. This property,  follows
the concentration of eigenvalues (cf.~(\ref{EV-Pn})  
\cite[Property $P_3$, \S 9.3]{AloSpe-PMethod}. Thus, all three sequences converge 
to constant graphons.
Furthermore, any quasirandom sequence with edge density $p$ converges to 
$\operatorname{Const}(p)$ \cite{BorChay08}. In particular, $\{P_n\}$ converges
to $\operatorname{Const}(1/2)$.

\section{Twisted states in the Kuramoto model on Cayley graphs} \lbl{sec.twist}
\setcounter{equation}{0}
In this section, we formulate the Kuramoto model and show that on Cayley graphs it has a 
family of steady state solutions called twisted states. We present two methods for
studying  stability of twisted states. The first method uses linearization and the Fourier
transform on $\Z_n$, while the second relies on a Lyapunov function. Both methods
elucidate the link between the structure of a Cayley graph and stability of the twisted states.

Suppose  $\Gamma=\Cay(\Z_n,S)$. Then (\ref{KM-before}) can be rewritten as follows
\be\lbl{Cayley}
{\p \over \p t} u(x,t)={(-1)^\alpha \over |S|} \sum_{y\in S}
\sin \left(2\pi (u(x+y,t)-u(x,t))\right), \; x\in V(\Gamma),\; t\in \R.
\ee

\subsection{Stability of twisted states via linearization}

\begin{lem}\lbl{lem.twist}
Twisted states  (\ref{q-twist}) are steady state solutions of (\ref{Cayley}).
\end{lem}
\pf 
Using the symmetry of $S$, we have
\begin{eqnarray}
\nonumber
&&\sum_{y\in S}\sin \left(2\pi (u_n^{(q)}(x+y,t)-u_n^{(q)}(x,t))\right)=\\
\nonumber
&& 2^{-1}\sum_{y\in S}\sin \left(2\pi (u_n^{(q)}(x+y,t)-u_n^{(q)}(x,t))\right)
+2^{-1}\sum_{y\in -S}\sin \left( 2\pi (u_n^{(q)}(x+y,t)-u_n^{(q)}(x,t))\right)
=\\
\lbl{use-symmetry} 
&& 2^{-1}\sum_{y\in S}\left\{ \sin \left(2\pi (u(x+y,t)-u(x,t))\right) 
\sin \left(2\pi (u(x-y,t)-u(x,t))\right)\right\}.
\end{eqnarray}
By plugging (\ref{q-twist}) into  (\ref{use-symmetry}), we obtain
$$
\sum_{y\in S}\sin \left(2\pi (u_n^{(q)}(x+y,t)-u_n^{(q)}(x,t))\right)=
2^{-1}\sum_{y\in S}\left\{\sin \left({2\pi qy\over n}\right) -\sin\left( {2\pi qy\over n}\right) \right\}=0.
$$
$\qed$

Next, we turn to the stability of the twisted states. By plugging
$u=u_n^{(q)}+\eta(x,t)$ into (\ref{Cayley}), we have
\be\lbl{linearize}
{\p \over \p t}\eta(x,t)={(-1)^\alpha 2\pi\over |S|} \sum_{y\in S} \cos \left({2\pi q y\over n}\right)
\left[\eta(y+x,t)-\eta(y,t)\right]+O(\|\eta(\cdot,t)\|^2).
\ee
The linearized equation can be rewritten as
\be\lbl{linear}
{\p \over \p t} \eta= \delta_S\ast \eta - m_S\eta,
\ee
where $f\ast g$ stands for the convolution of two functions from $L_2(\Z_n,\C)$
$$
f\ast g= \sum_{y\in \Z_n} f(y) g(x-y),
$$
\be\lbl{delta-S}
\delta_S(x)=\left\{ \begin{array}{ll} 
 {(-1)^\alpha 2\pi\over |S|} \re \exp \left({2\pi\iu q x\over n}\right), & x\in S,\\
0,& \mbox{otherwise},
\end{array}
\right.
\quad\mbox{and}\quad
 m_S=\sum_{y\in S} \delta_S (y).
\ee

\begin{lem}\lbl{lem.B-EVs}
Let $B:L_2(\Z_n,\C)\to L_2(\Z_n,\C)$ be the linear operator defined by the 
right-hand side of (\ref{linear}). The eigenvalues of $B$ are
\be\lbl{EVs-B}
\lambda_x(B)=
{(-1)^\alpha 2\pi\over |S|} \sum_{y\in S} \left\{\cos\left({2\pi (q+x)y\over n}\right) 
-2 \cos\left({2\pi qy\over n}\right) 
+\cos\left({2\pi (q-x)y\over n}\right) \right\}, \; x\in\Z_n.
\ee
\end{lem}
\begin{rem}\lbl{zero-EV} It follows from (\ref{EVs-B}) that the
spectrum of $B$ always has a zero eigenvalue $\lambda_0=0$. The 
corresponding eigenspace
\be\lbl{diagonal-subspace}
\mathcal{D}=\operatorname{span}\{\1_n\}. 
\ee
The presence of
the zero eigenvalue reflects the fact that the set of $q$-twisted states
is invariant under spatial translations.
\end{rem}
\pf\;
We compute the eigenvalues of $B$ using the discrete Fourier transform.
For $f\in L_2(\Z_n,\C)$ the latter is defined by
$$
\F f(x)=\hat f(x):= \sum_{y\in\Z_n} f(y) e_x (-y),
$$
where $e_x, x\in \Z_n,$ stand for the characters of $\Z_n$ 
(cf. \cite[Equation (\ref{character})]{Ter99}). By applying the
Fourier transform to the right hand side of (\ref{linear}), we have
\be\lbl{apply-Fourier}
\F B\eta(x)= \F\left( \delta_S\ast \eta -
  m_S\eta\right)=\left(\hat\delta_S(x)-m_S\right)\F\eta(x).
\ee
Setting $h=\F\eta$, we rewrite (\ref{apply-Fourier}) as 
\be\lbl{diagonal}
\left[ \F B\F^{-1}(h)\right](x)=\left(\hat\delta_S(x)-m_S\right) h(x).
\ee
Thus, the Fourier transform diagonalizes $B$. The eigenvalues of $B$ are
\be\lbl{lambda-B}
\lambda_x(B)=\hat\delta_S(x)-m_S, \; x\in\Z_n,
\ee
and $\mathbf{e}_x, x\in \Z_n,$ are the corresponding eigenvectors.
Finally, from
\be\lbl{transform-delta}
\hat\delta_S(x)= {(-1)^\alpha 2\pi\over |S|}\sum_{y\in S}  \cos \left( {2\pi qy\over n}\right) 
\exp\left( {-2\pi\iu xy\over n}\right)
\ee
we  obtain
\begin{eqnarray}\nonumber
\re\hat\delta_S(x)&=& {(-1)^\alpha 2\pi\over |S|}\sum_{y\in S} \cos\left({2\pi qy\over n}\right) \cos\left({2\pi xy\over n}\right)\\
\lbl{real-delta}
&=&{(-1)^\alpha\pi\over |S|} \sum_{y\in S} \left\{\cos\left({2\pi (q+x)y\over n}\right) +\cos\left({2\pi (q-x)y\over n}\right) \right\},\\
\lbl{imaginary-delta}
\im\hat\delta_S(x)&=&0,
\end{eqnarray}
where the symmetry of $S$ was used to obtain (\ref{imaginary-delta}).
The expressions for the eigenvalues in (\ref{EVs-B}) follow from (\ref{lambda-B}),
(\ref{transform-delta}), (\ref{real-delta}), and (\ref{imaginary-delta}).\\
$\qed$

\begin{thm}\lbl{thm.stability}
A $q$-twisted state $u_n^{(q)}, q\in \Z_n,$ is a stable equilibrium of (\ref{Cayley}) provided
\be\lbl{stability-condition}
 (-1)^\alpha \sum_{y\in S} \left\{\cos\left({2\pi (q+x)y\over n}\right) 
-2 \cos\left({2\pi qy\over n}\right) 
+\cos\left({2\pi (q-x)y\over n}\right) \right\}<0 \; \forall x\in\Z^\times_n.
\ee
\end{thm}
\pf
The spectrum of the linearized problem about $q$-twisted state $u_n^{(q)}$
contains the zero eigenvalue (see Remark~\ref{zero-EV}). Below, we show that
$u_n^{(q)}$ is a stable twisted state provided the remaining eigenvalues are
negative (cf.~(\ref{stability-condition})).

Recall that $\mathcal{D}=\operatorname{span} \{\1_n\}$.
 Suppose $u(x,t)$ is a solution of (\ref{Cayley}). Then
\be\lbl{sum-zero}
{d\over dt} \langle u(\cdot,t), \1_n\rangle = {(-1)^\alpha \over |S|}
\sum_{ xy \in E} \left\{ \sin\left(2\pi(u(y,t)-u(x,t))\right)+
\sin\left(2\pi(u(x,t)-u(y,t))\right)\right\} =0,
\ee
where $\langle\cdot,\cdot\rangle$ is the inner product in $L_2(\Z_n,\C)$.
Thus, $\mathcal{D}^\perp$ is an invariant subspace of the nonlinear system (\ref{Cayley}).

Further, recall the linearization Ansatz $u=u_n^{(q)}+\eta(x,t)$ (cf.~(\ref{linearize})).
Decompose 
$$
\eta=c_1\1_n+\tilde\eta, \; \tilde\eta\in\mathcal{D}^\perp.
$$
Then
$$
u(x,t)=u_n^{(q)}(x)+\eta(x,t)=\tilde u_n^{(q)}(x) +\tilde\eta(x,t),
$$
where $\tilde u_n^{(q)}$ is a shifted twisted state
$$
\tilde u_n^{(q)}:=u_n^{(q)}+c_1\1_n \pmod 1
$$
and $\tilde\eta(\cdot,0)\in\mathcal{D}^\perp$. Note that $\tilde u_n^{(q)}$ depends continuously on  
$\|\eta(\cdot,0)\|$.  Further, $\tilde\eta(\cdot, t)\in\mathcal{D}^\perp$ for all $t> 0$, because
by (\ref{sum-zero}),
$$
\langle u(\cdot, t), \1_n\rangle =\langle u_n^{(q)}, \1_n\rangle + 
\langle\tilde\eta(\cdot, t), \1_n\rangle=\operatorname{const},
$$
and, therefore,
$$
\langle \tilde\eta(\cdot, t), \1_n \rangle = \langle \tilde\eta(\cdot,0), \1_n\rangle =0.
$$
This shows that for studying stability of $q$-twisted state $u_n^{(q)}$, we can restrict to perturbations 
from $\mathcal{D}^\perp$. 

The linearization about $\tilde u_n^{(q)}$ yields
\be\lbl{restrict}
{\partial \over \partial t} \tilde\eta(\cdot,t)=B\tilde\eta(\cdot,t)+ O(\|\tilde\eta(\cdot,t)\|^2).
\ee
since $\tilde\eta(\cdot,t)\in\mathcal{D}^\perp$ for all $t>0$, we restrict (\ref{restrict}) to 
$\mathcal{D}^\perp$. The spectrum of $B$
restricted to $\mathcal{D}^\perp$ consists of the eigenvalues $\lambda_x(B),$ $x\in\Z_n^\times$,
because the eigenvector corresponding to the zero eigenvalue $\mathbf{e}_0\in\mathcal{D}$.
By the assumption (\ref{stability-condition}), the eigenvalues of the linear operator $B$ restricted to 
$\mathcal{D}^\perp$ are negative. Therefore, $\tilde u_n^{(q)}$ is asymptotically stable to small perturbations
from $\mathcal{D}^\perp$, and, $u_n^{(q)}$ is a stable steady state of (\ref{Cayley}).\\
$\qed$

\subsection{The variational approach to stability of twisted states}\lbl{sec.var}
In this section, we present a variational interpretation of stability and instability of
certain twisted states in the Kuramoto model. In particular, we establish the correspondence between
twisted states of the Kuramoto model and the eigenvalues of the graph Laplacian $L$. We show that twisted states
corresponding to the smallest (largest) eigenvalue are stable (unstable) if the coupling is attractive.
The converse relations hold for systems with repulsive coupling. In particular, synchrony is stable for
the Kuramoto models with attractive coupling and is unstable for those with repulsive coupling.
  
First, we rewrite the Kuramoto model on $\Gamma=\langle E,V\rangle$ in the form
convenient for the analysis of this section:
\be\lbl{theta}
\dot\theta_i ={(-1)^\alpha\over 4\pi} \sum_{j: ij\in E}
\sin\left(2\pi(\theta_j-\theta_i)\right), \; i\in [n].
\ee
\begin{rem} 
For convenience, in (\ref{theta}) we use a different scaling on the right hand side
(compare (\ref{theta}) with (\ref{Cayley})). Clearly, this does not affect existence and stability
of twisted state. After rescaling time (\ref{theta}) covers Kuramoto models on Cayley
graphs (\ref{Cayley}), as well as on other undirected graphs.
\end{rem}

The coboundary matrix  of $\Gamma=\langle V, E\rangle$ ($m=|E|$, $n=|V|$), $H\in \R^{m\times n}$,
is defined as follows  \cite{Biggs}. For each edge  $e:=ij\in E$, we 
choose the starting node $s(e)\in \{i,j\}$  and  the terminal node $t(e)=\{i,j\}/s(e)$.
Then the entries of $H$ are given by
\be\lbl{cobound}
(H)_{ev}=\left\{ \begin{array}{ll} 1, & \mbox{if} \;t(e)=v,\\
-1,& \mbox{if} \;s(e)=v,\\
0,&\mbox{otherwise},
\end{array}\right. \;\; (e,v)\in [m]\times [n].
\ee
Consider
\be\lbl{def-Phi}
\Phi(\theta)= \langle H e^{2\pi\iu\theta}, H e^{2\pi\iu\theta}\rangle,
\ee
where   $\theta=(\theta_1,\theta_2,\dots,\theta_n)^\t\in \R^n$
and
$
e^{2\pi\iu\theta} :=(e^{2\pi\iu\theta_1}, e^{2\pi\iu\theta_2},\dots,e^{2\pi\iu\theta_n})^\t.
$
Since $\Phi(\theta)$ is a $1$-periodic function, we consider $\theta$ as an element of $\T^n.$
On the other hand, $e^{2\pi\iu\theta}$ is convenient to interpret as an element of 
$\T_\C^n=\T_\C\times \T_\C\times\dots\times \T_\C,$ where $\T_\C=\{z \in\C: |z|=1\}$. 
The exponential map 
\be\lbl{bijection}
\phi \mapsto R(\phi)=e^{2\pi\iu\phi}
\ee
provides a bijection between $\T$ and $\T_\C$.

\begin{lem}\lbl{lem.Lyapunov} 
$\Phi$ is a nonnegative bounded function 
\be\lbl{2-bdd}
0\le \Phi(\theta)\le 4m \quad \forall \theta\in \T^n.
\ee
Moreover, (\ref{theta}) can be rewritten as a gradient system
\be\lbl{grad}
\dot\theta=-\nabla U_\alpha (\theta), \; U_\alpha(\theta):=(-1)^{\alpha} \Phi(\theta),
\ee
and, thus, for any solution $\theta(t)$ of (\ref{theta}), we have
\be\lbl{Phidot}
\dot \Phi(\theta):={d\over dt}\Phi(\theta(t))=(-1)^{\alpha+1} \|\nabla \Phi(\theta)\|^2.
\ee
\end{lem}
\pf
\begin{eqnarray}
\nonumber
0\le \Phi(\theta)&=&\|He^{\iu 2\pi\theta}\|^2= \sum_{ kj \in E} 
\left\{ \left(\cos(2\pi \theta_j) -\cos(2\pi \theta_k)\right)^2+
\left(\sin(2\pi\theta_j)-\sin(2\pi\theta_k)\right)^2 \right\}\\
\lbl{rewrite-Phi}
&=&  2 \sum_{ kj\in E} \left(1-\cos\left(2\pi(\theta_j-\theta_k)\right)\right) \le 4m.
\end{eqnarray}
This shows (\ref{2-bdd}). By differentiating (\ref{rewrite-Phi}), we have
\be\lbl{dPhi}
{\p\over \p \theta_j} \Phi(\theta)=-4\pi \sum_{k: kj\in E} \sin(2\pi(\theta_k-\theta_j)), \; j\in [n].
\ee
Thus, (\ref{grad}) follows.\\
$\qed$ 

Since the coupled system (\ref{theta}) is  a gradient
system with the real analytic potential function $U_\alpha$
(\ref{grad}), the set of stable equilibria of (\ref{theta}) coincides with the 
set of local minima of $U_\alpha$ \cite[Theorem~3]{AbsKur06}. The latter
are the local minima of $\Phi$ if the coupling is attractive, and are the 
local maxima of $\Phi$ if the coupling is repulsive. Thus, the problem
of locating the attractors for (\ref{theta}) for both attractive and repulsive
coupling is reduced to the variational problem
\be\lbl{variational}
\Phi(\theta)\rightarrow \min/\max, \; \theta\in\T^n.
\ee
We consider the minimization problem first
\be\lbl{Phimin}
\Phi(\theta)=\langle He^{2\pi\iu\theta}, He^{2\pi\iu\theta}\rangle\; \rightarrow \; \min, \, \theta\in \T^n.
\ee
This problem can be rephrased as the minimization problem for the quadratic form
\be\lbl{Qmim}  
Q(z)=z^*Lz\; \rightarrow \; \min, \, z\in\T_\C^n,
\ee
where $L$ is the Laplacian of $\Gamma$.  By the variational characterization of the 
eigenvalues of Hermitian matrices \cite{HornJohn-Matrix}, the minimum of $Q$ on
$\T_\C^n$ is achieved on the intersection of $\T_\C^n$ and the one-dimensional eigenspace
corresponding to the smallest eigenvalue of $L$, $\lambda_{min}=0$:
$$
\mathbf{z}_{min}=e^{2\pi\iu c} \1_n, \; c\in\R.
$$
Therefore, the minimum of $\Phi$ on $\T^n$ is achieved on
$$
u_{min}={1\over 2\pi \iu} \ln \mathbf{z}_{min}= c\1_n  \pmod{1},
$$
where $\ln z$ stands for the univalent branch of the complex logarithm
with values in the strip $0\le \Im\ln z< 2\pi.$ Thus, we have proved the 
following theorem.

\begin{thm}\lbl{thm.sync} A synchronous solution $c \1_n$ is a stable (unstable)
equilibrium of the Kuramoto model (\ref{theta}) with attractive (repulsive) coupling
 on any undirected graph $\Gamma$ and for any $c\in \T$.
\end{thm}

Next, we turn to the repulsive coupling case, which leads us to consider the
maximization problem 
\be\lbl{Phimax}
\Phi(\theta)=\langle He^{2\pi\iu\theta}, He^{2\pi\iu\theta}\rangle\; \rightarrow \; \max, \, \theta\in \T^n,
\ee
which, in turn, is equivalent to the following problem
\be\lbl{Qmax}  
Q(z)=z^*Lz\; \rightarrow \; \max, \, z\in\T_\C^n.
\ee
For the remainder of this section, we assume that $\Gamma$ is a Cayley
graph on a cyclic group
so that its eigenvalues and the corresponding eigenvectors are given in Lemma~\ref{lem.EV-cyc}.
Denote the largest EV of L
\be\lbl{bar-lambda}
\bar\lambda=\max \{\lambda_x:~x\in\Z_n\}.
\ee
Since the corresponding eigenvectors 
\be\lbl{bar-em}
\mathbf{e}_m, \; m\in M=\{ j\in Z_n:~\lambda_j=\bar\lambda\},
\ee
belong to $\T_\C^n$, $e^{2\pi\iu c}\mathbf{e}_m$ maximize $Q$ on
$\T_\C^n$ for any $c\in\R$ and $m\in M$. By arguing as in the
attractive coupling case, we conclude that the corresponding twisted states
\be\lbl{bar-um}
\mathbf{u}_m =(u_m(0), u_m(1), \dots, u_m(n-1)), \;\;
u_m(x)={mx\over n}+c \pmod 1,\; c\in\R, m\in M
\ee
are stable equilibria of (\ref{Cayley}). Thus, we arrive at the following result.

\begin{thm}\lbl{thm.lambda-max}
Suppose $\Gamma$ is a Cayley graph on a cyclic group and $L$ is the graph
Laplacian of $\Gamma$. Then twisted states (\ref{bar-um}) corresponding to the
largest eigenvalue of $L$ (\ref{bar-lambda}) are stable (unstable) equilibria of the Kuramoto model 
model on $\Gamma$ is the coupling is repulsive (attractive).
\end{thm}

\section{Synchronization}\lbl{sec.sync}
\setcounter{equation}{0}
In this section, we study the stability of the family of synchronous solutions
of (\ref{theta}) in more detail. Theorem~\ref{thm.sync} shows that synchronous 
solutions are stable equilibria
of (\ref{theta}) with attractive coupling. Below, we show that the invariant subspace formed by such 
solutions is, in fact, asymptotically stable. To this end, we use the properties 
of the system at hand rather than relying on the general stability result for gradient
systems in \cite{AbsKur06}. Specifically, the 
analysis below uses on the Lyapunov function $U_\alpha$.

 Let
$$
0=\lambda_1<\lambda_2\le \dots\le \lambda_n
$$
denote the eigenvalues of the Laplacian $L=H^\t H$ and let $v_1,v_2,\dots,
v_n$ be an orthonormal set of 
the corresponding eigenvectors. We choose $v_1=n^{-1/2}\1_n,$ where $\1_n=(1,1,\dots,1)^\t  \in \R^n$.
Recall 
$$
\mathcal{D}=\operatorname{span} \{c\1_n\pmod 1,~c\in\R\},
$$
the invariant subspace of (\ref{grad}) corresponding to the synchronous dynamics.
Next, we construct a system of tubular neighborhoods of $\mathcal{D}$.

Let $0<\delta<1/2$ and denote
\be\lbl{ellipse}
E_\delta=\left\{ \theta=c_1v_1+V\bar c\pmod 1:\; c_1\in \R, \,
  \sum_{i=2}^n\lambda_i c_i^2< {\delta^2\over (2\pi)^2} \right\},
\ee
and its boundary
\be\lbl{dE}
\partial E_\delta=\left\{ \theta=c_1v_1+V\bar c\pmod 1:\; c_1\in \R, \,
  \sum_{i=2}^n\lambda_i c_i^2= {\delta^2\over (2\pi)^2} \right\}.
\ee
Here, $\bar c=(c_2,c_3,\dots,c_n)$ and $V=\operatorname{col} (v_2,v_3,\dots, v_n)$.
The following lemma describes the level sets of $\Phi$.

\begin{lem}\lbl{lem.level}
There exists $\delta_0>0$ such that
\begin{eqnarray}\lbl{Phi<}
\Phi\left|_{\partial E_\delta}\right. &\le& \delta^2,\\
\lbl{Phi>}
\Phi\left|_{\partial E_{2\delta}}\right. &\ge & 2\delta^2,
\end{eqnarray}
provided $0<\delta\le\delta_0$.
\end{lem}
\pf
Recall (\ref{rewrite-Phi})
\be\lbl{cos}
\Phi(\theta)=2 \sum_{ kj\in E}
\left(1-\cos(2\pi(\theta_k-\theta_j))\right).
\ee
After expanding the right hand side of (\ref{cos}) into Taylor series
and using 
the remainder estimates for the resultant 
alternating series, we have
\be\lbl{pre-double}
\sum_{k: kj\in E} (2\pi(\theta_k-\theta_j))^2-{1\over 12}
\sum_{k: kj\in E} 
(2\pi(\theta_k-\theta_j))^4
\le \Phi(\theta)\le  \sum_{k: kj\in E} (2\pi(\theta_k-\theta_j))^2.
\ee

Using the definitions of $H$ and $L$, we have
\be\lbl{quadratic}
\sum_{ kj\in E} (\theta_k-\theta_j)^2=\langle H\theta, H\theta\rangle 
=\theta^\t L\theta.
\ee
By plugging  $\theta=c_1\1_n+V\bar c$ into  (\ref{pre-double}) and using (\ref{quadratic}),
we have
\be\lbl{double}
(2\pi)^2\sum_{i=2}^n \lambda_i c_i^2 -\psi(\bar c)
\le \Phi(\theta)\le (2\pi)^2 \sum_{i=2}^n \lambda_i c_i^2,
\ee
where
\be\lbl{psi}
0\le \psi(\bar c)={(2\pi)^4\over 12} \sum_{kj\in E} \left((V\bar c )_k-(V\bar c)_j\right)^4\le C_1\|\bar c\|^4.
\ee
for some $C_1>0$.

The right-hand side inequality in  (\ref{double}) and the definition of $\partial E_\delta$ (cf.~(\ref{dE}))
imply (\ref{Phi<}). Further, suppose 
\be\lbl{delta0}
0<\delta<\delta_0=\min\left\{ {\lambda_2\over 2\sqrt{2} C_1}, {1\over 2}\right\}.
\ee
Then from the left-hand side inequality in (\ref{double}), we have
\be\lbl{Phi2delta}
\Phi\left|_{\partial E_{2\delta}}\right. \ge 4\delta^2 - \psi\left|_{\partial E_{2\delta}}\right..
\ee
For $\theta\in \partial E_{2\delta}$,
$$
\lambda_2\|\bar c\|^2\le \sum_{i=2}^n \lambda_i c_i^2=4\delta^2,
$$
and, thus,
\be\lbl{c4}
\|\bar c\|^4\le {16\delta^4\over \lambda_2^2}.
\ee
The combination of (\ref{Phi2delta}), (\ref{c4}), and (\ref{psi}) yields 
$$
\Phi\left|_{\partial E_{2\delta}}\right. \ge 4\delta^2- {16 C_1\delta^2\over \lambda_2}\ge 2\delta^2,
$$
 where we used $0<\delta\le \delta_0$ to derive the last inequality (cf.~(\ref{delta0})).\\
$\qed$

\begin{thm}\lbl{thm.synchrony}
Invariant subspace $\mathcal{D}$ of (\ref{grad}) is asymptotically stable if 
$\alpha=0$ and is unstable if $\alpha=1$.
\end{thm}
\pf We consider the case of $\alpha=0$ first. Let $\theta_0\in E_{\delta_1}$ for some
$0<\delta_1\le\delta_0$.
We want to show that $\omega$-limit set of $\theta_0$, $\omega(\theta_0)\in \mathcal{D}$.
This is clearly true for $\theta_0\in\mathcal{D}$. We, therefore, assume $\theta_0\in E_{\delta_1}/\mathcal{D}$.
Then $\theta_0\in \partial E_{\delta}$ for some $0<\delta<\delta_1$.

By (\ref{Phi<}) and (\ref{Phidot}),
$$
\Phi(\theta(t))\le \delta^2, \; t\ge 0.
$$
Thus, $\theta(t)\in E_{2\delta}$ for all $t\ge 0$. By \cite[Lemma~11.1]{Hartman-ODEs},
$\omega(\theta_0) \in E_0=\{\theta\in \T^n:\, \dot\Phi(\theta)=0\}\cap E_\delta$. 

It remains to show that $E_0=\mathcal{D}$. To this end, let $\theta=c_1\1_n+V\bar c$ and note
that 
\be\lbl{Phi_local}
\Phi(\theta)= (2\pi)^2\sum_{i=2}^n \lambda_i c_i^2 + O(\|\bar c\|^4),
\ee
by (\ref{double}) and (\ref{psi}). It follows from (\ref{Phi_local}) that there are no
critical points of $\Phi$  outside $\mathcal{D}$ in $E_{\delta_1}$ for sufficiently small $\delta_1>0$.   
Thus, 
\be\lbl{isolate}
\nabla\Phi(\theta)\neq 0, \quad \theta\in E_{\delta}/\mathcal{D}\; \forall 0<\delta<\delta_1.
\ee
 The combination of (\ref{isolate}) and  (\ref{Phidot}), we have
$E_0=\mathcal{D}$. Thus, $\omega(\theta_0)\in \mathcal{D}$.

Next, we consider the repulsive coupling case. Set $\alpha=1$ in (\ref{grad}) and fix 
$0<\delta_2<\delta_1$
such that (\ref{isolate}) holds. Let $\theta_0\in E_{\delta_2}/\mathcal{D}$. Then $\theta_0\in \partial E_\delta$
for some $0<\delta<\delta_2$.  Function $\|\nabla \Phi(\theta)\|^2$ is bounded away from zero
in $\overline{ E_{\delta_2}/E_\delta}$. By (\ref{Phidot}), the trajectory starting from $\theta_0$
leaves $E_{\delta_2}$ in finite time. Thus, $\mathcal{D}$ is unstable.\\
$\qed$

We now turn to the repulsive coupling case. Here, we will need an additional assumption
that $\Gamma$ is a Cayley graph on the cyclic group $\Z_n$. Thus, in the remainder 
of this section, instead of (\ref{theta}) we consider (\ref{Cayley}).
Recall that $\mathbf{u}_m, m\in M,$ denote  twisted states corresponding to the largest
eigenvalue of $L$ (see (\ref{bar-um}). Below, we show that $\mathbf{u}_m, m\in M$ are  
stable twisted steady states
of the Kuramoto model with repulsive coupling under an additional nondegeneracy condition
on the Hessian matrix
\be\lbl{Hess}
h_{ij}:=(H(\Phi)(\mathbf{u}_m))_{ij}=
{\partial^2\Phi\over \partial \theta_i\partial\theta_j}(\mathbf{u}_m), \; m\in M.
\ee

By differentiating (\ref{dPhi}) with respect to $\theta_i$ and using (\ref{bar-um}), we have 
\be\lbl{hij}
h_{ij}=\left\{ \begin{array}{cc} -4\pi\sin\left({2\pi m(j-i)\over n}\right), & ij\in E,\\
0,& \mbox{otherwise}
\end{array}
\right.
\ee
Since $\sin$ is an odd function and $\Gamma$ is a Cayley graph, we have $h_{ij}=-h_{ji}$.
Thus, the Hessian matrix 
$H(\Phi)(\mathbf{u}_m)$ has zero row sums. Therefore, $-H(\Phi)(\mathbf{u}_m)$ has a zero eigenvalue 
$\tilde\lambda_1=0$ 
with the corresponding eigenvector 
$\tilde v_1=n^{-1/2}\1_n$. Denote the remaining eigenvalues of $-H(\Phi)(u_m)$ arranged in the 
increasing order
by $\tilde\lambda_k, \; k=2,3,\dots,n.$ 
Choose an orthonormal basis in $\R^n$ from the eigenvectors of $H(\Phi)(\mathbf{u}_m)$, $\{v_k, k\in [n]\}$.
Because $\mathbf{u}_m$ is a point of maximum of $\Phi$, all eigenvalues $\tilde\lambda_k, k\in [n],$ are nonnegative:
\be\lbl{order-mu}
0=\tilde\lambda_1\le\tilde\lambda_2\le\dots\le\tilde\lambda_n.
\ee

\begin{thm}\lbl{thm.max}
Let $\mathbf{u}_m$ be a twisted state corresponding to the largest eigenvalue of $L$, i.e., $m\in M$. Suppose that
$\tilde\lambda_0=0$ is a simple eigenvalue of $-H(\Phi)(\mathbf{u}_m)$. Then 
\be\lbl{Dm}
\mathcal{D}_m=\{\theta=u_m+c\1_n \pmod 1: \; c\in\R\}
\ee
is asymptotically stable.
\end{thm}
\pf
Let 
\be\lbl{tildePhi}
\tilde\Phi(\theta)=\Phi(u_m)-\Phi(\theta)\; \mbox{and}\;
\tilde E_\delta=\left\{ \theta=u_m +c_1\1_n+V\bar c\pmod 1:\; c_1\in \R, \,
  \sum_{i=2}^n\tilde\lambda_i c_i^2< {\delta^2\over (2\pi)^2} \right\}.
\ee
Then $\tilde\Phi$ is nonnegative and is monotonically decreasing along the trajectories of (\ref{grad})
($\alpha=1$). Following the lines of the proof of Lemma~\ref{thm.synchrony}, one can show that
for initial condition $\theta_0\in\tilde E_\delta$, the trajectory $\theta(t)$ remains in $\tilde E_{2\delta}$,
provided  $\delta>0$ is sufficiently small. Furthermore, for small $\delta>0$, there are no zeros of 
$\tilde\Phi$ in $\tilde E_{2\delta}$ outside $\mathcal{D}_m$. From this, we conclude that $\omega(\theta_0)$
is contained in maximal invariant subspace contained in the set of zeros of $\dot{\tilde\Phi}$ in
$\tilde E_\delta$, i.e., in $\mathcal{D}_m$.\\
$\qed$

\section{The Kuramoto model on $K_n$}
\lbl{sec.Kn}
\setcounter{equation}{0}
In this and in the following sections, we apply the results of 
Sections~\ref{sec.twist} and \ref{sec.sync} to study stability of twisted states 
on complete and Paley graphs. We first focus on the Kuramoto model on the family
of complete graphs $K_n$. To simplify the 
presentation, we consider $K_n$ for odd $n$, so that they can be viewed as 
Cayley graphs. 

By (\ref{EV-Kn}) the graph Laplacian of $K_n$ has a simple zero
eigenvalue and $n-1$ multiple eigenvalues equal to $n-1$. This immediately implies that
all nontrivial twisted states ($q=0$) are unstable if the coupling is attractive and
are stable for repulsive coupling (cf.~Theorem~\ref{thm.lambda-max}).
Thus, the stability of all twisted states is determined completely by the spectrum   
of the graph Laplacian. In the remainder of this section, we develop
a complementary approach to studying stability of twisted states using linearization.
The main difficulty in implementing this approach is dealing with the highly
degenerate matrix of the linearized system about the twisted states. 
This problem does not come up in the stability analysis of
phase-locked solutions in the Kuramoto model with distributed
intrinsic frequencies $\omega(x)\neq \mbox{const}$ \cite{MirStr05, BDP12}.
Our analysis
shows that this degeneracy is due to the fact that nontrivial twisted states lie
in an $(n-2)$-dimensional manifold formed by the equilibria of the Kuramoto model.
Therefore, in addition to recovering the stability results, known from the 
variational properties of the Kuramoto model, the second approach yields a detailed picture
of the phase flow near the nontrivial twisted states.

We start by computing the eigenvalues of the linearized problem (\ref{linear}).
\begin{lem}\lbl{lem.eigenvalue-K}
The eigenvalues of the linearization of the Kuramoto model on $K_n$ about the $q$-twiste state
$u_n^{(q)}$ are 
\begin{description}\item[$q=0:$]
\be\lbl{EV-K=0}
\lambda_0(0)=0 \quad\mbox{and}\quad \lambda_x(0)=(-1)^{\alpha+1}4\pi\left( 1+(n-1)^{-1}\right), \;
x\in\Z_n^\times,
\ee 
\item[$q\in\Z_n^\times:$]
\be\lbl{EV-Kneq0}
\lambda_x(q)=\left\{\begin{array}{ll}
(-1)^\alpha 2 \pi [1+(n-1)^{-1}], & x\in\{q, n-q\},\\
0,& \mbox{otherwise},
\end{array}
\right.
\ee
\end{description}
where $n\in\N$ is odd.
\end{lem} 
\pf
We compute the eigenvalues for the Kuramoto model on Cayley graphs generated
by the ball $B(r)$, $\Cay(Z_n, B(r)),$ for any $r\in [n-1/2]$
first. We then use this result to  compute the eigenvalues of 
$K_{2r+1}=\Cay(\Z_{2r+1},B(r)).$ 

By Lemma~\ref{lem.B-EVs},
\be\lbl{EV-ball}
\lambda_x(q)={(-1)^\alpha \pi\over r} \sum_{y=-r}^r
\left\{ 
\cos\left({2\pi (q+x)y\over n}\right) -2 \cos\left({2\pi qy\over n}\right) 
+\cos\left({2\pi (q-x)y\over n}\right) 
\right\}, \; q\in\Z_n.
\ee
Using the formula for the partial sum of the geometric series, 
we obtain
\begin{eqnarray*}
\sum_{k=-r}^r w^{qk}&=&w^{-rq} {w^{q(2r+1)}-1\over w^q -1}=
 w^{-q(r+1/2)} { w^{q(2r+1)}-1\over w^{q/2}- w^{-q/2}}\\
&=& { w^{q(r+1/2)}- w^{-q(r+1/2)}\over w^{q/2}- w^{-q/2}},
\end{eqnarray*}
which for $w=e^{2\pi\iu\over n}$ becomes
\be\lbl{useful-identity}
\sum_{k=-r}^r \cos\left({ 2\pi q k\over n}\right)=
\left\{ \begin{array}{cl} 
{\sin (\pi q (2r+1)/n)\over \sin(\pi q/n)},& q\in \Z^\times_n,\\
2r+1, & q=0,
\end{array}
\right.
\ee 
The combination of (\ref{EV-ball}) and (\ref{useful-identity})
yields
\be\lbl{EV-B}
\lambda_x(q)={(-1)^\alpha \pi\over r} \left[ S(q+x, 2r+1)-2S(q,2r+1)+S(q-x, 2r+1)\right],\;
x,q\in\Z_n,
\ee
where 
$$
S(q,m)= \left\{ \begin{array}{cl}  \sin(\pi qm/n)/\sin(\pi q/n), &  q\in \Z^\times_n,\\
m, & q=0,
\end{array}
\right.
$$ 
The formulas for the eigenvalues of the linearized 
problem in (\ref{EV-K=0}) and (\ref{EV-Kneq0}) follow from (\ref{EV-B}) with $n=2r+1$.
\; $\qed$

By Theorem~\ref{thm.synchrony}, the synchronous state ($q=0$)
is stable when the coupling is attractive. By (\ref{EV-Kneq0}), there are no other 
stable twisted states in the attractive coupling case. 
Thus, in the remainder of this section we focus on the repulsive coupling case.
Theorem~\ref{thm.synchrony} implies that the synchronous solutions are unstable
if the coupling is repulsive.
If $q\neq 0$ and $\alpha=1$,
(\ref{EV-Kneq0}) shows that the spectrum of the linearized problem contains two negative
eigenvalues and $n-2$ zero eigenvalues. The linearization alone, therefore, does not resolve stability of the   
the twisted states. The argument below deals with the presence of the zero eigenvalues.

\begin{lem}\lbl{lem.preF}
Consider $F:\T^n\to\C$ defined by
\be\lbl{def-F}
F(u)=\sum_{k=1}^n e^{2\pi \iu u_k},\; u=(u_1,u_2,\dots,u_n).
\ee
Then every element in $F^{-1}(0)=\{ u\in\T^n:\; F(u)=0\}$ is a steady state solution
of (\ref{KM-before}). Furthermore, $q$-twisted states $u_n^{(q)}\in F^{-1}(0)$ for all $q\in\Z_n$.
\end{lem}
\pf
From $F(u)=0$, it follows 
\be\lbl{ImF}
\im F(u) e^{-2\pi\iu u_j}=0 \; \forall j\in [n],
\ee
which, in turn, implies
\be\lbl{implies-steady}
\sum_{k=1}^n \sin( 2\pi (u_k-u_j))=0\; \forall j\in [n].
\ee
Thus, $F^{-1}(0)$ is formed by steady states of (\ref{KM-before}).

Next, we show that $u_n^{(q)} \in F^{-1}(0)$ for any $q\in \Z_n$. Let $n=2r+1$, $q\in [2r]$, and
$j\in [n-1]\cup \{0\}$. Then 
$$
\sum_{k=0}^{2r} \cos \left( {2\pi q(k-j)\over n}\right)=e^{-2\pi jq \iu\over n}
\sum_{k=0}^{n-1} e^{2\pi qk \iu\over n}= 
e^{-2\pi jq \iu\over n} {e^{2\pi q \iu}-1\over e^{2\pi q \iu\over n}-1}=0.
$$ 
$\qed$

\begin{thm}\lbl{thm.stable-twist}
The twisted state $u_n^{(q)}$ is a stable steady state of the Kuramoto model on $K_n$ with
repulsive coupling for any odd $n\in\N$ and $q\in [n-1].$ 
\end{thm}
\pf
By (\ref{EV-Kneq0}), the spectrum of the linearized problem in the repulsive coupling 
case contains two negative eigenvalues and $n-2$ zero eigenvalues. Therefore, the stability of the 
equilibrium is decided by the dynamics on the center manifold. Below, we show that
$u_n^{(q)}$ lies in the $(n-2)$-dimensional smooth manifold of equilibria. From this we will derive stability.

Let  $^{\R}F(u)$ stand for the realification of $F(u)$:
$$
^{\R}F(u)=\begin{pmatrix} \sum_{k=0}^{n-1} \cos(2\pi u_k) \\
\sum_{k=0}^{n-1} \sin(2\pi u_k) \end{pmatrix}.
$$
From the definition of $u_n^{(q)}$ and (\ref{useful-identity}), we have
\be\lbl{value}
^{\R}F(u_n^{(q)})=0.
\ee
Further,
\be\lbl{DF}
{\partial \over \partial u} ^{\R}F (u_n^{(q)})=2\pi \begin{pmatrix}
-\sin(2\pi q/n)& -\sin(4pq/n) &\dots & -\sin(2(n-1)\pi q/n)\\
\cos(2\pi q/n)& \cos(4pq/n) &\dots & \cos(2(n-1)\pi q/n)
\end{pmatrix}
\ee
and 
\be\lbl{check-rank}
\begin{vmatrix}
-\sin(2\pi q/n) & -\sin(4\pi q /n)  \\
\cos(2\pi q/n) & \cos(4\pi q/n)
\end{vmatrix}
=-\sin(2\pi q/n) \neq 0,
\ee
because $n$ is odd.
This implies $\operatorname{rank} {\partial \over \partial u} ^{\R}F
(u_n^{(q)})=2$. Thus, $0$ is a regular value of $^{\R}F$. By the
Regular Value Theorem  \cite[Theorem~3.2]{Hirsch-DiffTop}, near 
$u_n^{(q)}$, $^{\R}F^{-1}(0)$ is an $(n-2)$-dimensional
differentiable manifold, which we denote by $\mathcal{M}$. Let $\psi$ be a local parametrization 
of $\mathcal{M}$ near $u_n^{(q)}$. $\psi$ maps diffeomorphically $U$, a neighborhood of $0\in\R^{n-2}$,
onto a neighborhood of $u_n^{(q)}$ in $\mathcal{M}$ such that $\psi(0)=u_n^{(q)}$.

Consider the change of variables $v=\psi^{-1}(u)$. Then
\be\lbl{center}
\dot v= {\partial \over\partial u} \psi^{-1}(u) f(\psi(u))=0,
\ee
because $\psi(U)\subset \mathcal{M}$ and $f\left|_\mathcal{M}\right.=0$.
Equation (\ref{center}) represents the flow on the $(n-2)$-dimensional 
center manifold of $u_n^{(q)}$. Since the spectrum of the linearized problem 
has two negative eigenvalues, by the Reduction Theorem (cf. \cite{ArnAfrShi99}), in a small
neighborhood of $u_n^{(q)}$, the flow is topologically equivalent to the standard
saddle suspension over its restriction to the center manifold
\begin{eqnarray*}
\dot x &=& -x, \\
\dot y&=& 0, \; (x,y)\in\R^2\times\R^{n-2}.
\end{eqnarray*}
This shows that $u_n^{(q)}$ is stable. $\qed$

\section{The Kuramoto model on $P_n$} \lbl{sec.Pn}
\setcounter{equation}{0}
In this section, we focus on stability of twisted states in the Kuramoto model on
Paley graphs $P_n$. Thus, we consider (\ref{Cayley}) with $\Gamma=P_n$.
It is instructive to discuss the linearization
about the twisted state $u_n^{(q)}, q\in \Z_n,$ first.
By Lemma~\ref{lem.B-EVs},  the eigenvalues of the linearized
problem are
\be\lbl{EV-P-1}
\lambda_x(q)= {(-1)^\alpha 4\pi\over n-1} \sum_{k=1}^{n-1}
\cos(2\pi (q+x) k^2/n) -2 \cos(2\pi qk^2/n) + \cos(2\pi (q-x) k^2/n), \; x\in\Z_n.
\ee
We rewrite (\ref{EV-P-1}) in terms of the Gauss sums (\ref{gauss-def})
as follows
\be\lbl{EV-P-2}
\lambda_x(q)= {(-1)^\alpha 4\pi\over n-1} \left\{ G_n(q+x)-
  2G_n(q)+G_n(q-x)\right\}, \; x\in\Z_n.
\ee
The value of $G_n(q)$ in (\ref{EV-P-2}) depends on whether or not $q$ is a QR modulo $n$.
This leads to several cases for stability of twisted states on Paley graphs. We summarize the 
information about the spectrum of the linearized problem in the following lemma.

\begin{lem}\lbl{lem.linearize} The eigenvalues of the linearization of the Kuramoto model
  (\ref{Cayley}) with $\Gamma=P_n$ about 
$q$-twisted state $u_n^{(q)}$, $q\in\Z_n,$ are as follows. For all $q\in\Z_n$, $\lambda_0(q)=0$.
The remaining eigenvalues $\lambda_x(q), x\in\Z^\times_n, q\in\Z_n,$ are given in the following table.
\begin{center} Table~1
\end{center}
\begin{tabular}{l c|c| c}
\hline 
& & \textbf{1}) $\alpha=0\;$  (attractive coupling)  & \textbf{2}) $\alpha=1$ \; (repulsive coupling)\\
\hline
\textbf{A})&  $q=0$ &  $\lambda_x(q)\le -4\pi (1+O(n^{-1/2}))$ & $\lambda_x(q)> 0$ \\
\hline
\textbf{B})&  $q$  is a QR $\pmod{n}$ & $\lambda_x\le 0$ & $\lambda_x(q)\ge 0$ \\
\hline
\textbf{C})& $q$ is a not  QR $\pmod{n}$ & $\lambda_x\ge 0$ & $\lambda_x(q)\le 0$ \\
\hline
\end{tabular}

Moreover, in each of the above cases, the spectrum contains a nonzero eigenvalue.
\end{lem}
\begin{cor}\lbl{cor.instability}
If $q$  is not a QR modulo $n$ then $q$-twisted states are unstable in the Kuramoto model 
(\ref{Cayley}) with $\Gamma=P_n$ with repulsive coupling.
\end{cor}
\pf (Lemma~\ref{lem.linearize}) Consider (\textbf{C1}) first: $\alpha=0$ and $q$ is not a QR.
Then $G_n(q)=-\sqrt{n}$ by (\ref{Gsum}).  Further, pick $y\in \Z_n^\times$ such that $y$ is a QR.
Using $x=|q-y|$ in (\ref{EV-P-2}), we have $\lambda_x(q)\ge \pi n^{-1/2}>0$.
This shows (\textbf{C1}) in Table~1. The results for all other case follow similarly from (\ref{EV-P-2})
and (\ref{Gsum}).\\
$\qed$

Aside from the stability of the synchronous solutions in (\textbf{A}) and instability result in 
(\textbf{C1}), the linear stability analysis of $q$-twisted states on Paley graphs is inconclusive,
because we can not exclude the presence of zero eigenvalues. Thus, we can not decide on stability 
in (\textbf{B1}) and in (\textbf{C2}) based on linearization.
To clarify stability of twisted states in these cases, we employ the variational principle 
of \S\ref{sec.var}. To this end, we compute the eigenvalues of the graph Laplacian of $P_n$.
Using (\ref{EV-Pn}), we compute 
$\lambda_0(P_n)=0$ and  
\be\lbl{EV-Pn-recall}
\lambda_x(P_n)=
\left\{\begin{array}{ll}
2^{-1} (n - \sqrt{n}), & x~ \mbox{ is a QR}~\pmod{n}, \\
2^{-1} (n + \sqrt{n}), & x~ \mbox{ is not a QR}~\pmod{n},
\end{array}
 \right.
\ee
for $x\in\Z_n^\times$ (cf.~\ref{EV-Pn}). Theorem~\ref{thm.lambda-max}
implies that 
if $q$ is not a QR $\pmod{n}$, the 
$q$-twisted states are stable (unstable) if the coupling is repulsive (attractive).
Likewise, Theorem~\ref{thm.sync} yields stability (instability) of synchronous 
solutions ($q=0$) for attractive (repulsive) coupling. Thus, we have resolved stability
of the $q$-twisted states in (\textbf{A}), (\textbf{C}), and (\textbf{B2}). For part 
(\textbf{B1}), we have to resort to numerics, which suggests that twisted states
are unstable in this case. We summarize information about stability of twisted
states on Paley graphs in the following table.

\begin{center} Table~2
\end{center}
\begin{tabular}{l c|c| c}
\hline
 & & \textbf{1}) $\alpha=0\;$  (attractive coupling)  & \textbf{2}) $\alpha=1$ \; (repulsive coupling)\\
\hline
\textbf{A})&  $q=0$ &  stable (V,L) & unstable (VL) \\
\hline
\textbf{B})&  $q$  is a QR & unknown & unstable (L) \\
\hline
\textbf{C})& $q$ is a not  QR $\pmod{n}$ &  unstable (V) & stable (V)  \\
\hline
\end{tabular}

In the table above, we used L and V to indicate that the conclusion about the stability
of a twisted state was based on the linearization or the variational argument respectively. 

\section{The Kuramoto model on $G(n,p)$} \lbl{sec.gnp}
\setcounter{equation}{0}
In this section, we study the Kuramoto model on the ER graph $G(n,p)$. The random graph 
$G(n,p)$ is not a Cayley graph. Therefore, neither of the techniques of Section~\ref{sec.twist}
applies to the analysis of the Kuramoto model on $G(n,p)$. Moreover, in contrast to  the 
graphs that we considered above, $G(n,p)$ in general does not support twisted
states as steady state solutions of the Kuramoto model for finite $n$. Nonetheless, in this section, we show
that the dynamics of the Kuramoto model on random and complete graphs are closely related. 
Specifically, for large $n$ the solutions of the IVP for the Kuramoto model on $G(n,p)$ can be approximated by those
for the Kuramoto model on $K_n$. This result is a discrete version of the homogenization, because the Kuramoto model
model on the complete graph can be viewed as an averaged model on the ER graph.
Having established the link between the Kuramoto models on 
ER and complete graphs, we use it to elucidate the dynamics of the former model.
To this effect, we show that twisted states become steady state solutions of the Kuramoto model on 
$G(n,p)$ almost surely in the limit as $n\to\infty$. Further, we show that 
all nontrivial twisted states in the Kuramoto model on $G(n,p)$ with repulsive 
coupling are metastable for finite albeit large $n$.

\subsection{Discrete homogenization}
Our next goal is to show that the solutions of the IVPs for coupled dynamical
systems on the large  complete and ER graphs remain close on finite
time intervals with high probability, provided they start from close initial
data. To achieve this, we employ the method developed in \cite{Med13b}. 

Let $A_n=(a_{nij})$ be the adjacency matrix of the ER random graph $G(n,p)$. 
$A_n$ is a symmetric matrix with zero diagonal. Entries $a_{nij},$ $1\le i<j\le n,$ 
are independent identically distributed  random variables (RVs) 
from the binomial distribution with parameter $p\in (0,1)$.
The homogenization result, that
we prove below, holds of the following class of models which covers the Kuramoto model.
Specifically, for a fixed $n\in\N$ assume that $u_{ni}, i\in [n]$ satisfy the following 
system of differential equations
\be\lbl{gnp}
{d\over dt} u_{ni}(t) = (np)^{-1} \sum_{i=1}^n a_{nij} D(u_{nj}-u_{ni}), \; i\in [n],
\ee
where $D$ is a Lipschitz continuous function with constant $L>0$
\be\lbl{Lip}
|D(u)-D(v)|\le L|u-v|, \; u,v\in\R
\ee
Along with (\ref{gnp}), we consider the averaged model on the  
complete graph $K_n$
\be\lbl{kn}
{d\over dt} v_{ni}(t) = n^{-1} \sum_{i=1}^n  D(v_{nj}-v_{ni}), \; i\in [n].
\ee
\begin{rem}\lbl{rem.scaling} Note that we are using a different scaling 
compared with that in (\ref{KM-before}). The right hand side of (\ref{gnp}) 
is scaled by the expected value of
the degree of a node of $G(n,p)$.  The analysis of the model scaled by 
the actual degree can be done similarly. We choose the former scaling
for analytical convenience. For the same reason, we scale the right hand side
of the Kuramoto model on $K_n$ (\ref{kn}) by $n$ rather than by $n-1$, the degree of $K_n$.
\end{rem}

The following weighted Euclidean inner product
\be\lbl{in-pro}
(u,v)_n={1\over n} \sum_{i=1}^n u_iv_i, \; u=(u_1,u_2,\dots, u_n)^\t, \; 
v=(v_1, v_2, \dots, v_n)^\t,
\ee
and the corresponding norm $\|u\|_{2,n}=\sqrt{(u,u)_n}$
will be used to measure the distance between solutions 
of the dynamical equations 
on random and complete graphs. We are ready to state
our first result.

\begin{thm}\lbl{thm.approx}
Let
$$
u_n(t)=(u_{n1}(t), u_{n2}(t),\dots,u_{nn}(t)) \;\mbox{and}\;
v_n(t)=(v_{n1}(t), v_{n2}(t),\dots,v_{nn}(t))
$$
denote the solutions of (\ref{gnp})  and (\ref{kn})  subject to the
same initial condition $u_n(0)=v_n(0)$.  Assume  that for a given $T>0$
there are constants $0<C_1\le C_2$ such that
\be\lbl{nondeg}
C_1\le\liminf_{n\to\infty} \min_{t\in [0,T]}  n^{-2}\sum_{i,j=1}^n D(v_{nj}-v_{ni})^2 \le
\limsup_{n\to\infty} \max_{t\in [0,T]}  n^{-2}\sum_{i,j=1}^n D(v_{nj}-v_{ni})^2\le C_2.
\ee
Then 
\be\lbl{lln}
\lim_{n\to\infty} \P\left\{ \max_{t\in[0,T]} \| u_n(t)-v_n(t)\|_{n,2}\ge Cn^{-1/ 2}\right\}=0,
\ee
where a positive constant $C=C(L,T)$ depends on $L$ and $T$ but not on $n$.
 \end{thm}

For the proof of Theorem~\ref{thm.approx} we will need the following application
the Central Limit Theorem, which follows from \cite[Lemma~4.4 and Corollary~4.5]{Med13b}.

\begin{lem}\lbl{lem.lln}
Let $p\in (0,1)$, $T\in [0,\infty)$ and $f_{nij}\in L^\infty([0,T]).$ 
Suppose that RVs $\xi_{nij}, i,j\in [n],$ $n\in \N$, are such that for fixed $n\in \N$ and $i\in [n]$,
$\{\xi_{nij},\; j\in[n]\}$ are independent identically distributed binomial RVs with parameter $p\in (0,1)$.

Denote
\begin{eqnarray*}
& \sigma_{ni}^2(t) = n^{-1}\sum_{i=1}^n f_{nij}^2(t) p(1-p), i\in [n], &\sigma_n^2(t)= n^{-1}\sum_{i=1}^n \sigma_{ni}^2(t),\\
& z_{ni}^2(t) ={1\over n}\sum_{j=1}^n (\xi_{nij} - p)f_{nij}(t), & S_n(t) =\sum_{i=1}^n z_{ni}^2(t),
\end{eqnarray*}
and assume that for some $0<C_3\le C_4$
\be \lbl{2-positive-sigma}
C_3\le\liminf_{n\to\infty} \min_{t\in [0,T]}  \sigma_n^2(t)\le
\limsup_{n\to\infty} \max_{t\in [0,T]}  \sigma_n^2(t)\le C_4.
\ee

Then 
\be\lbl{2-clt}
{ S_n(t)-\sigma_n^2(t) \over  n^{-1/2}\sqrt{ 5\sigma_n^4(t)+O(n^{-1})}   }
\overset{d}{\longrightarrow} \mathcal{N}(0,1) \;\mbox{as}\; n\to\infty,
\ee
where the convergence in (\ref{2-clt}) is in distribution.
\end{lem}

\pf (Theorem~\ref{thm.approx}) Denote $w_n=u_n-v_n$. 
By subtracting Equation~$i$ in (\ref{kn}) from the corresponding
equation in (\ref{gnp}), we have
\be\lbl{subtract}
{d\over dt} w_{ni}= 
{1\over np} \sum_{j=1}^n a_{nij}  
\left[D(u_{nj}-u_{ni}) -  D(v_{nj}-v_{ni})\right] +z_{ni}, \; i\in [n],
\ee
where
\be\lbl{zni}
z_{ni}= (np)^{-1} \sum_{j=1}^n (a_{nij} -p)D(v_{nj}-v_{ni}).
\ee 

By multiplying both sides of (\ref{subtract}) by $n^{-1} w_{ni}$ and summing over $i$, we have
\be\lbl{next-w}
{1\over 2}{d\over dt} \|w_n\|_{2,n}^2= {1\over n^2} \sum_{i,j=1}^n
a_{ij} \left[ D(u_{nj}-u_{ni}) - D(v_{nj}-v_{ni})\right] w_{ni} + (z_n, w_n)_{n}.
\ee
We bound the first term on the right hand side of (\ref{next-w})  using the Lipschitz continuity of 
$D$, $|a_{ij}|\le 1$, the Cauchy-Schwarz inequality, and the triangle inequality 
$$
\left|
{1\over n^2} \sum_{i,j=1}^n
a_{ij} 
\left[ D(u_{nj}-u_{ni}) - D(v_{nj}-v_{ni})\right]w_{ni} 
\right|\le 
$$
\be\lbl{w-preGron}
{L\over n^2} \sum_{i,j=1}^n \left(|w_{nj}|+|w_{ni}|\right) |w_{ni}|
\le 2L \|w_{ni}\|^2_{2,n}.
\ee
We bound the  second term using the Cauchy-Schwarz inequality
\be\lbl{bound-second}
\left|(z_n, w_n)_{n} \right|\le \|z_n\|_{2,n} \|\eta_n\|_{2,n}\le {1\over 2} 
\left(\|z_n\|_{2,n}^2+ \|w_n\|_{2,n}^2\right),
\ee
where  $z_n=(z_{n1}, z_{n2},\dots, z_{nn}).$
The combination of (\ref{next-w}), (\ref{w-preGron}), and (\ref{bound-second}) yields
 \be\lbl{Gron-ready}
{d\over dt} \|w_n\|^2_{2,n} \le (4L+1) \|w_n\|^2_{2,n} + \|z_n\|^2_{2,n}.
\ee
By Gronwall's inequality,
\be\lbl{use-Gron}
\max_{t\in[0,T]} \|w\|^2_{2,n} \le {\max_{t\in[0,T]} \|z_n(t)\|^2_{2,n}\over 4L+1} \exp\{ (4L+1)T\}.
\ee
Thus,
\be\lbl{use-Gron-1}
\max_{t\in[0,T]} \|w \|_{2,n} \le {\max_{t\in[0,T]} \|z_n(t)\|_{2,n}\over \sqrt{4L+1}} 
\exp\{ (2L+1)T\}.
\ee

It remains to estimate $\|z_n(t)\|_{2,n}$ (see (\ref{zni})). 
To this end, we use 
Lemma~\ref{lem.lln} with
$$
f_{nij}(t):= D(v_{nj}(t)-v_{ni}(t)).
$$
Using the assumption (\ref{nondeg}), we verify that
(\ref{2-positive-sigma}) holds. Thus, we apply 
Lemma~\ref{lem.lln},  to show that
$$
\P\{ |n\|z_n(t)\|_{2,n}^2-\sigma_n^2(t)|>1\} = 
\P\left\{ \left|{ n\|z_n(t)\|_{2,n}^2-\sigma_n^2(t)\over
n^{-1/2} \sqrt{5\sigma_n^4(t)+O(n^{-1})}}\right|>
{n^{1/2}\over \sqrt{5\sigma_n^4(t)+O(n^{-1})}}\right\}
$$
\be\lbl{pre-bound-zn}
\le\P\left\{ \left|{ n\|z_n(t)\|_{2,n}^2-\sigma_n^2(t)\over
n^{-1/2} \sqrt{5\sigma_n^4(t)+O(n^{-1})}}\right|>
{n^{1/2}\over \sqrt{5 C_2^2+O(n^{-1})}}\right\}\to 0 \;\mbox{as}\; n\to\infty
\ee
uniformly in $t\in [0,T]$.

Using (\ref{nondeg}), from (\ref{pre-bound-zn}) we have
\be\lbl{bound-zn}
\P\{ \|z_n(t)\|_{2,n}^2\le (C_4+1)n^{-1}\} \le \P\{ |n\|z_n(t)\|_{2,n}^2-\sigma_n^2(t)|>1\} 
\to 0 \;\mbox{as}\; n\to\infty
\ee
uniformly in $t\in [0,T]$.
Thus,
\be\lbl{final-bound}
\lim_{n\to\infty} \P\{ \max_{t\in [0,T]} \|z_n(t)\|_{2,n}\le C_5 n^{-1/2}\}=0, \; C_5=\sqrt{C_4+1}.
\ee 
$\qed$

\begin{figure}
\begin{center}
{\bf a}\hspace{0.1 cm}\includegraphics[height=1.8in,width=2.0in]{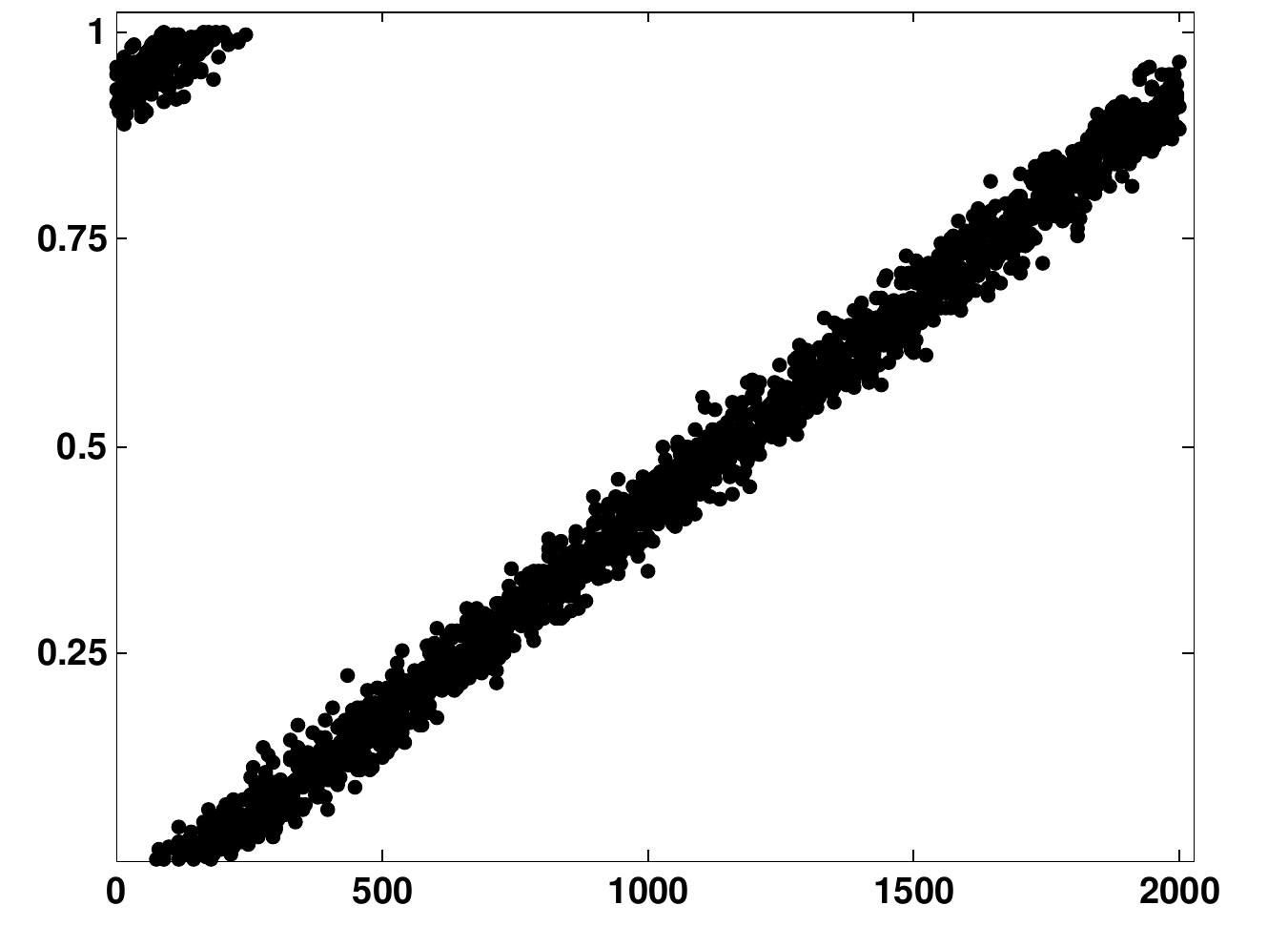}
{\bf b}\hspace{0.1 cm}\includegraphics[height=1.8in,width=2.0in]{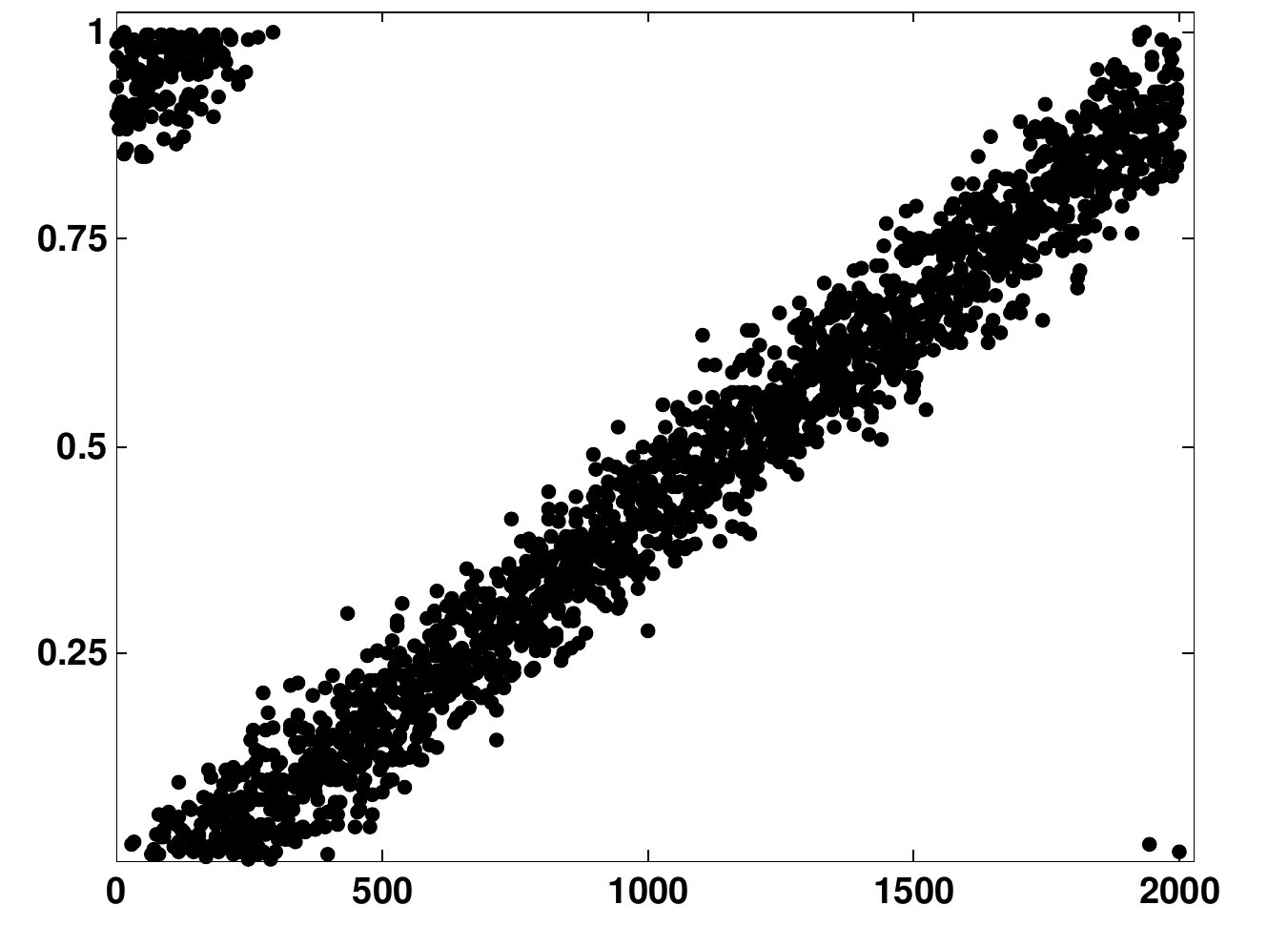}
{\bf c}\hspace{0.1 cm}\includegraphics[height=1.8in,width=2.0in]{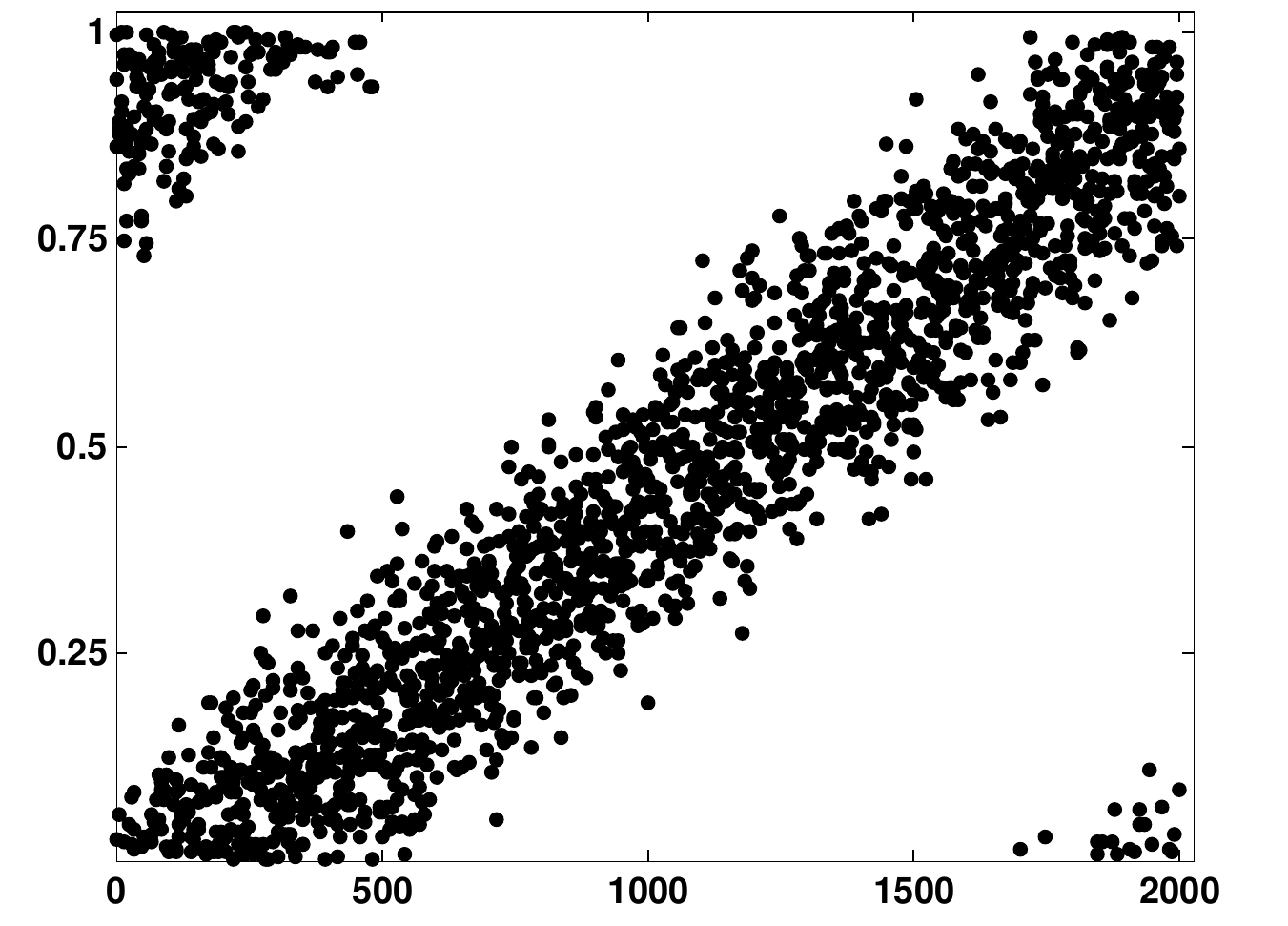}\\
{\bf d}\hspace{0.1 cm}\includegraphics[height=1.8in,width=2.0in]{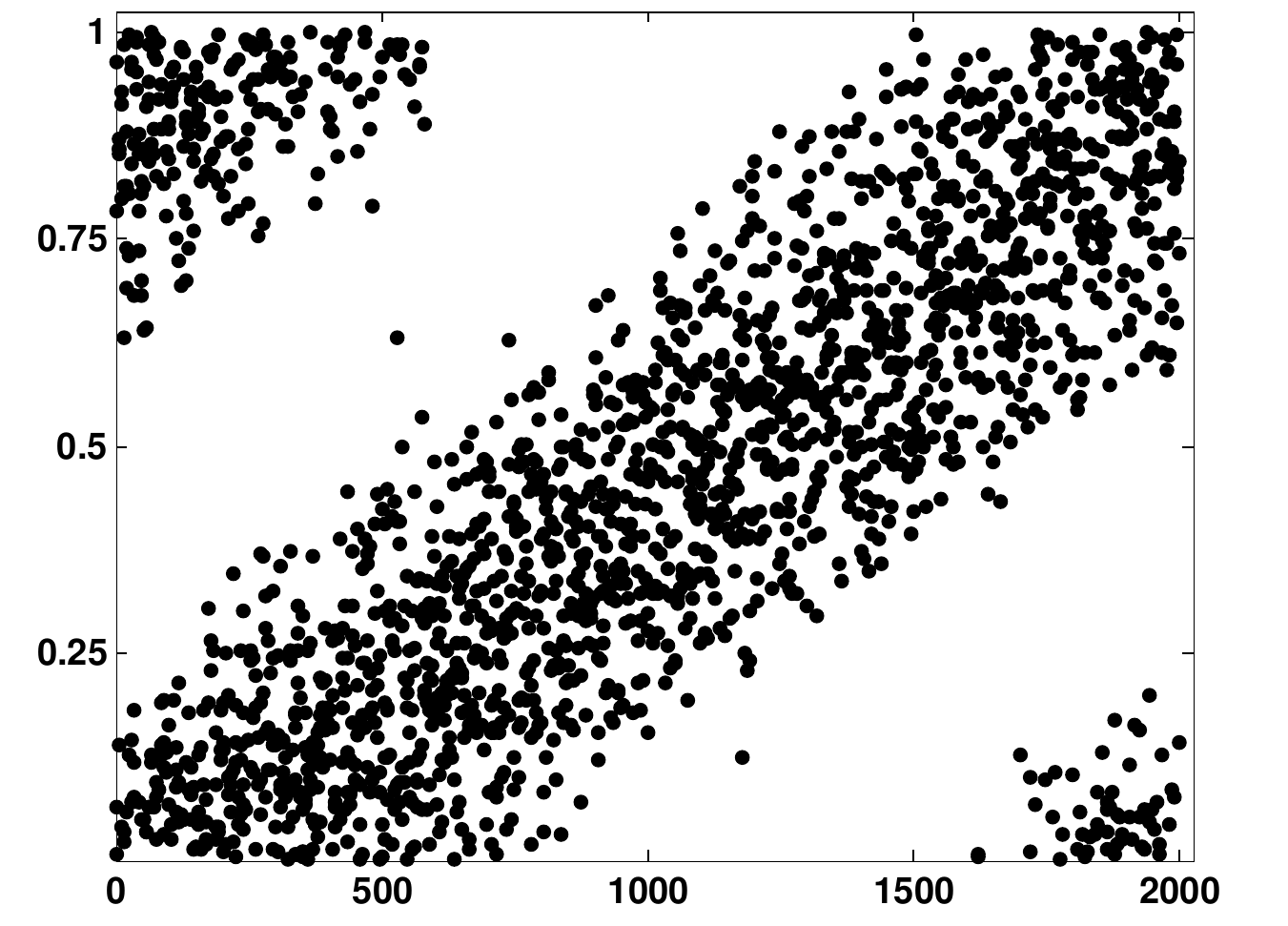}
{\bf e}\hspace{0.1 cm}\includegraphics[height=1.8in,width=2.0in]{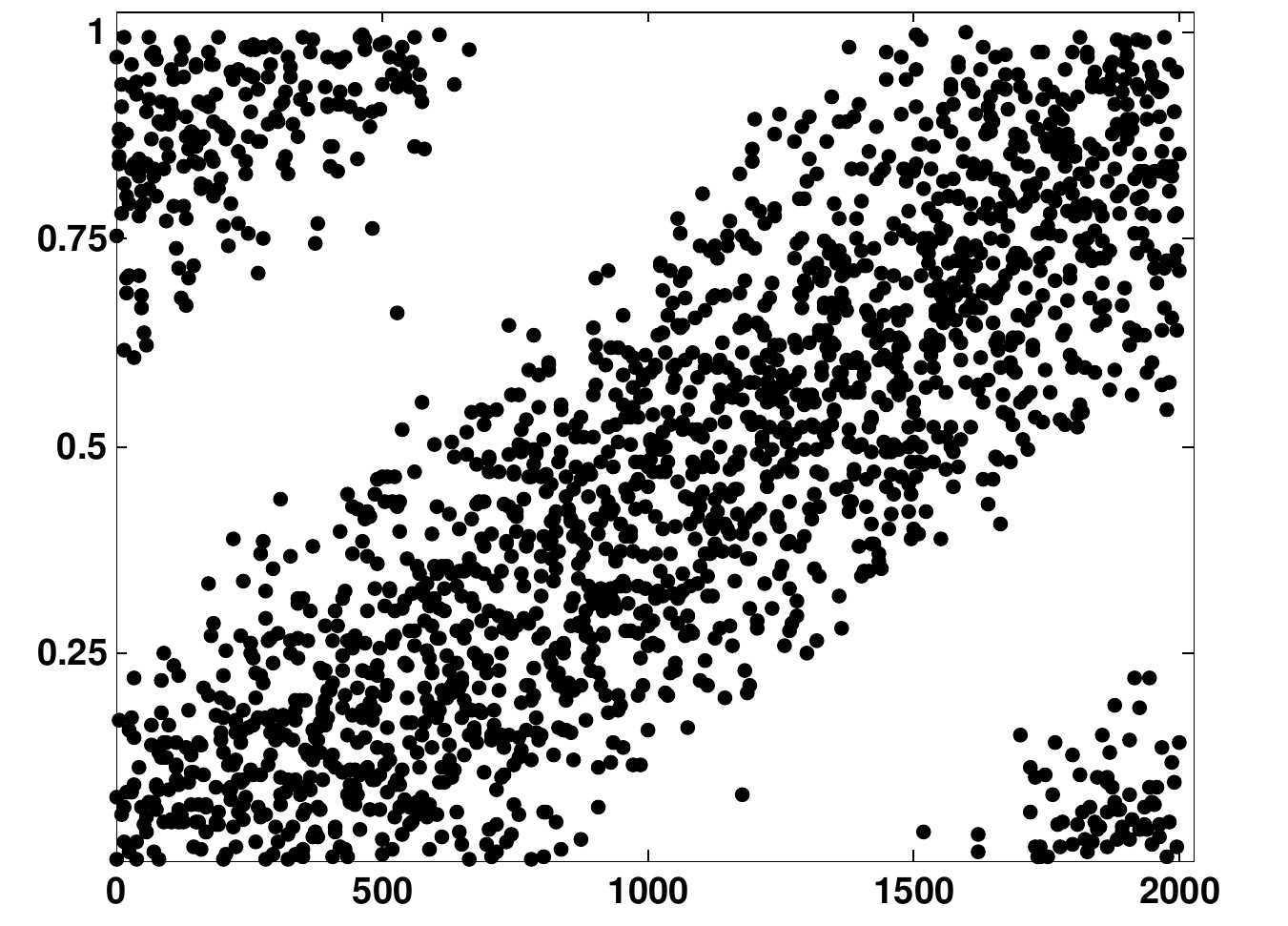}
{\bf f}\hspace{0.1 cm}\includegraphics[height=1.8in,width=2.0in]{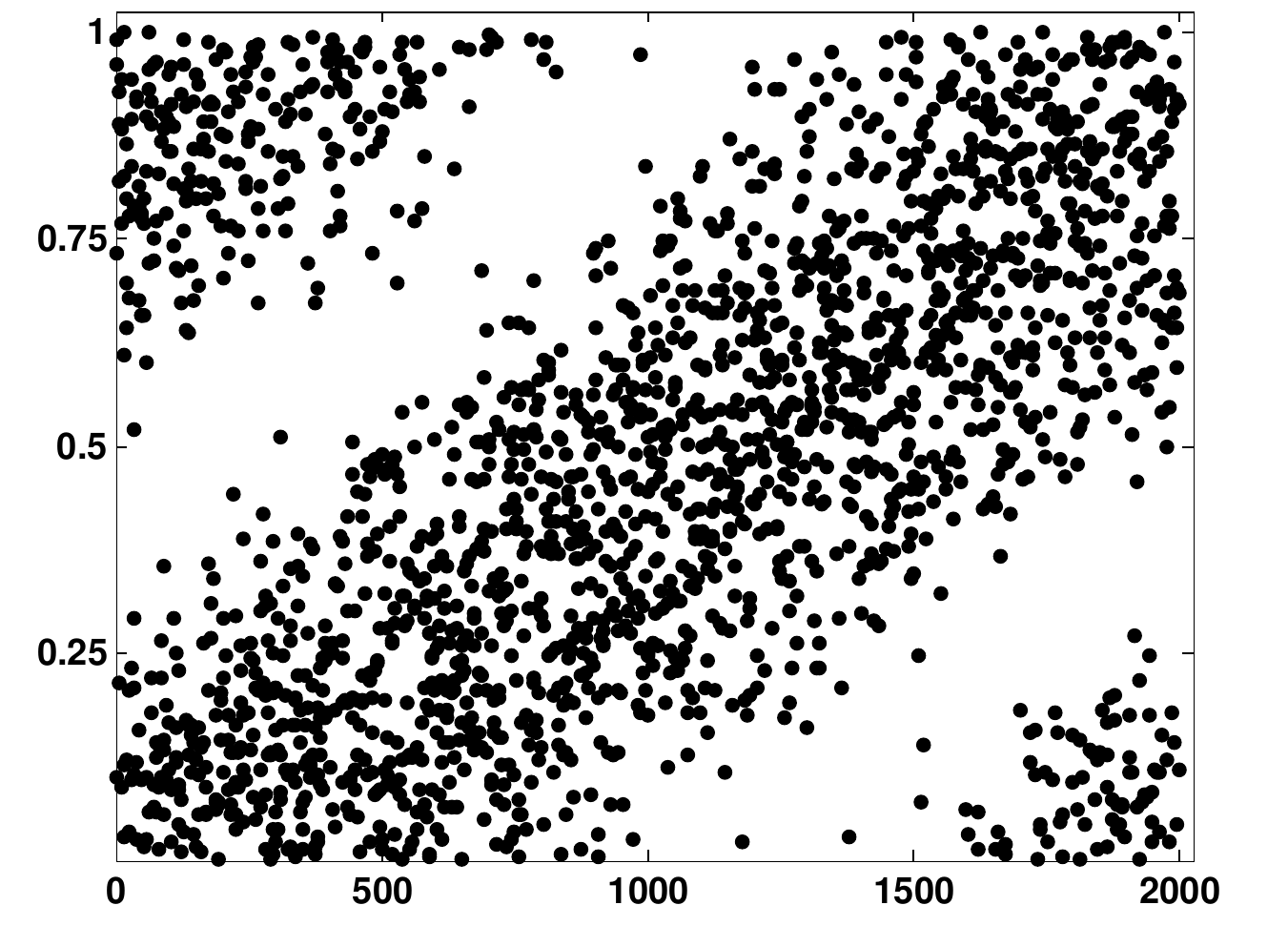}\\
{\bf g}\hspace{0.1 cm}\includegraphics[height=1.8in,width=2.0in]{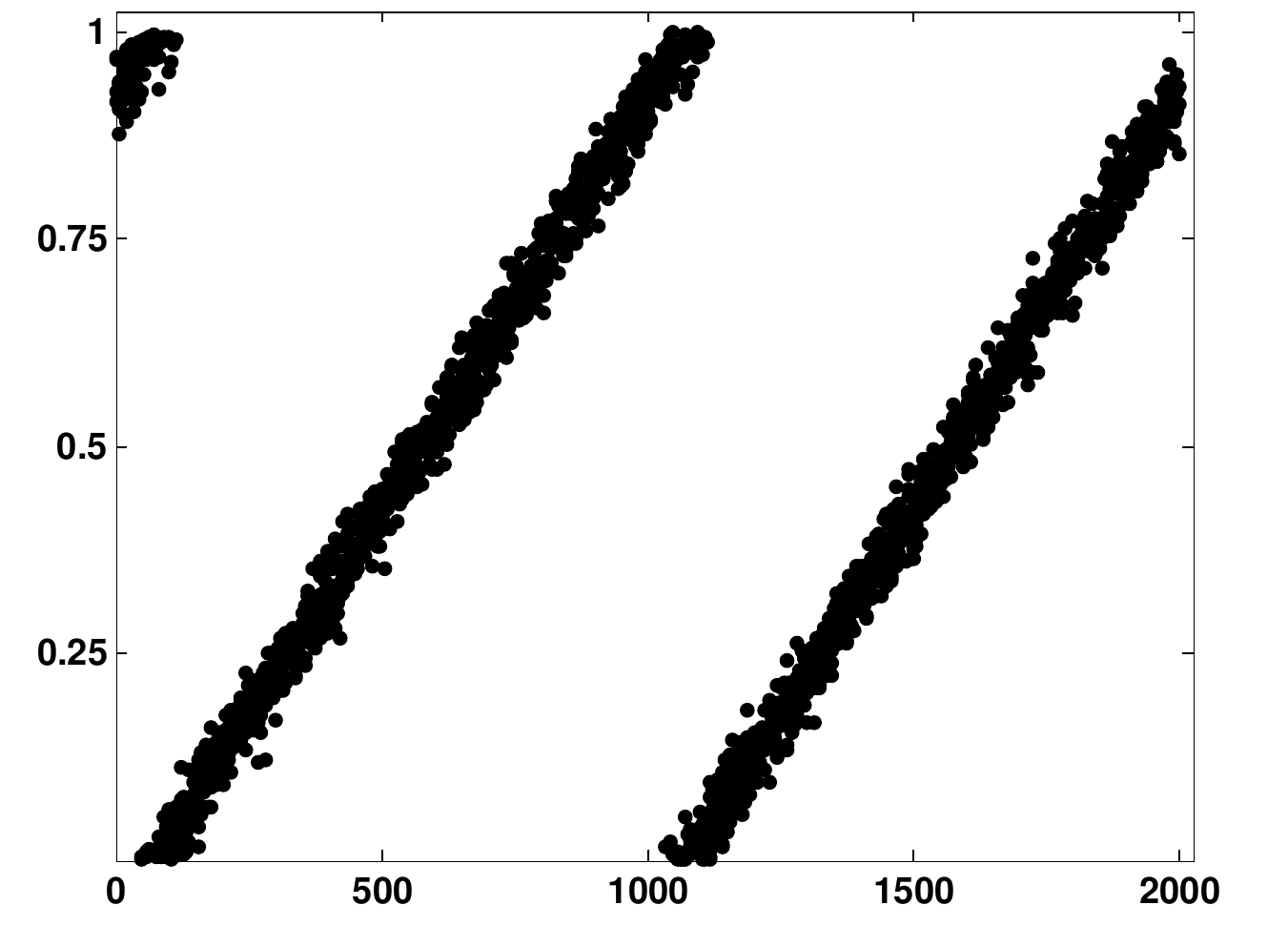}
{\bf h}\hspace{0.1 cm}\includegraphics[height=1.8in,width=2.0in]{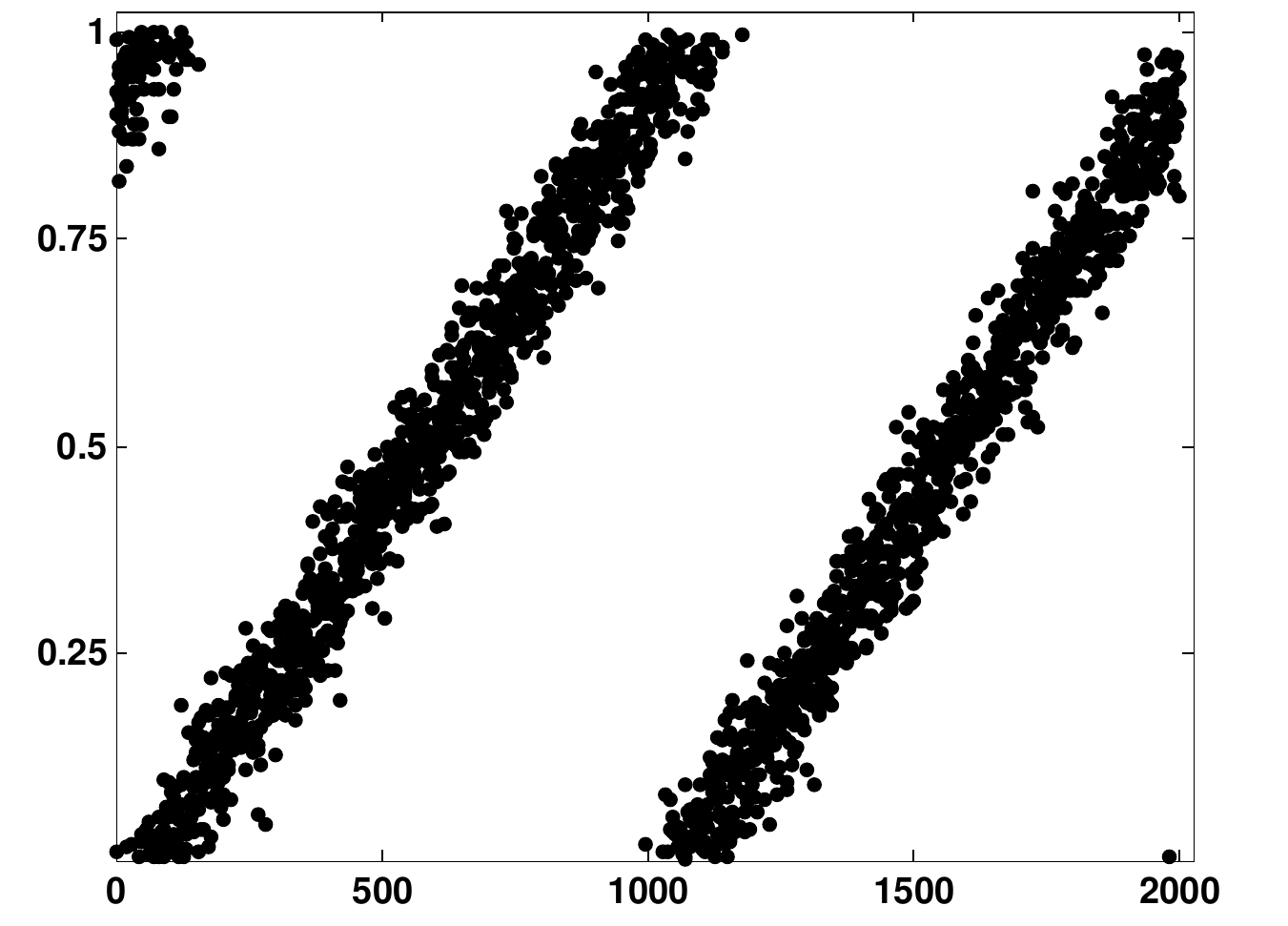}
{\bf i}\hspace{0.1 cm}\includegraphics[height=1.8in,width=2.0in]{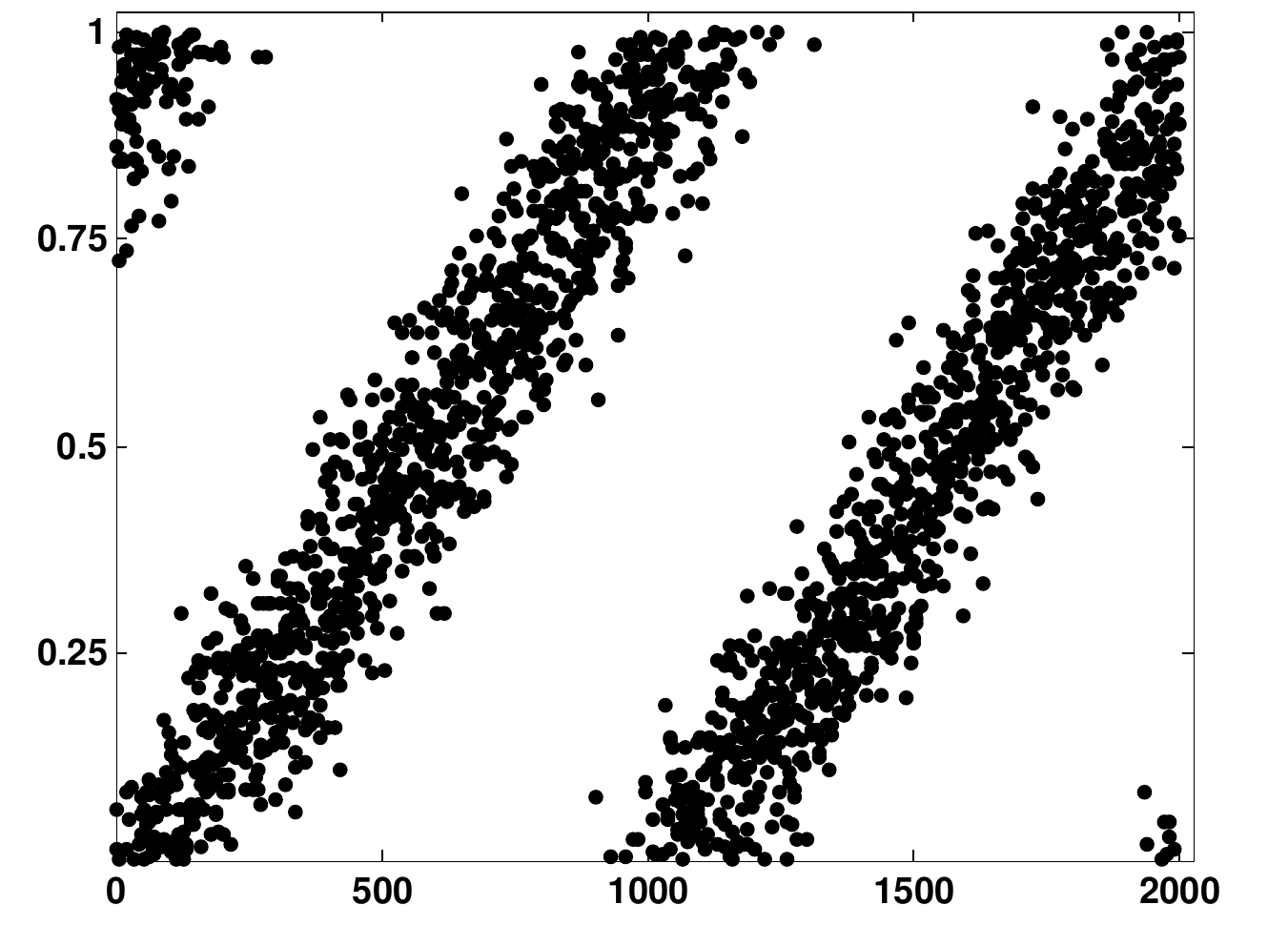}\\
{\bf j}\hspace{0.1 cm}\includegraphics[height=1.8in,width=2.0in]{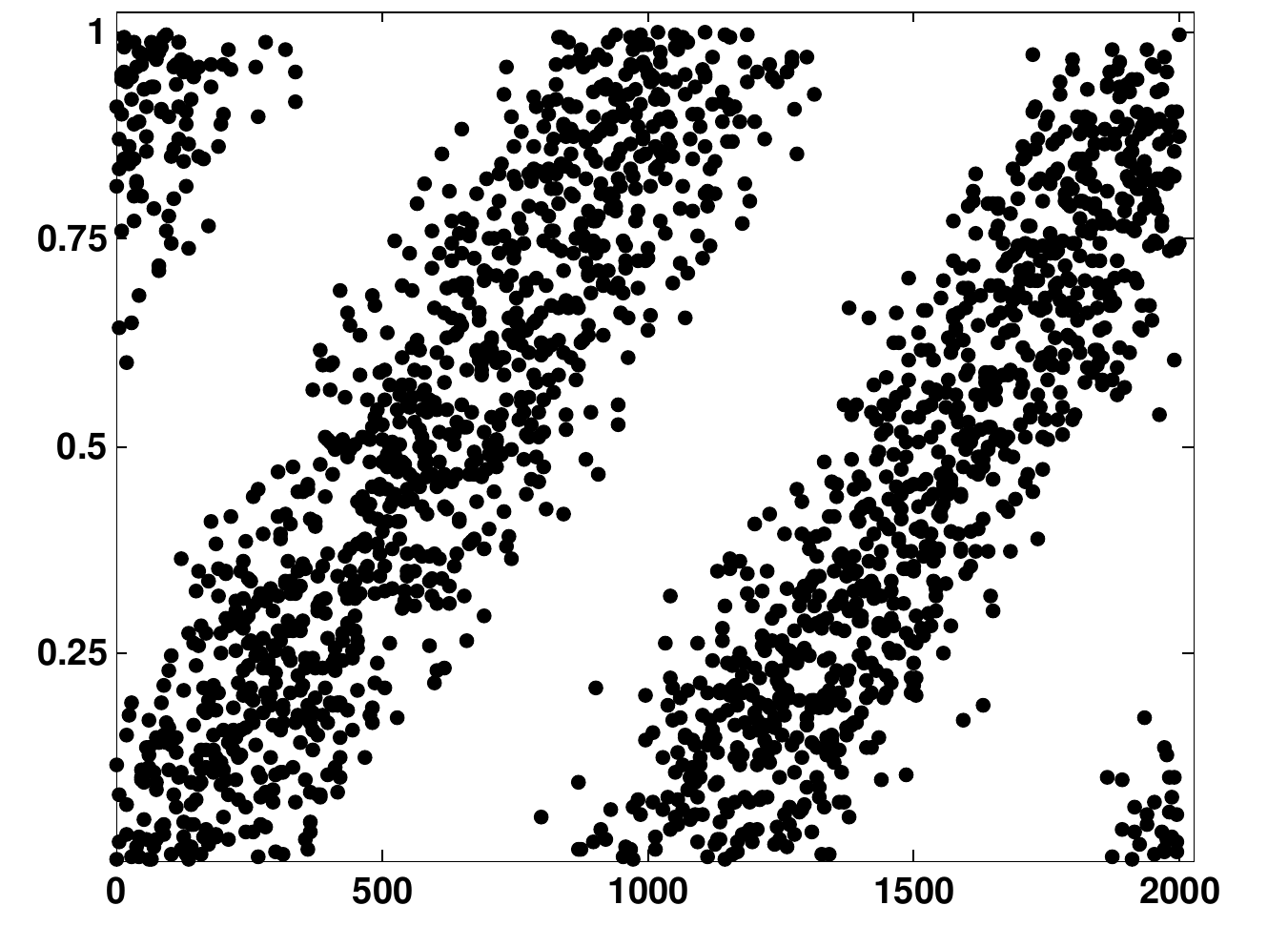}
{\bf k}\hspace{0.1 cm}\includegraphics[height=1.8in,width=2.0in]{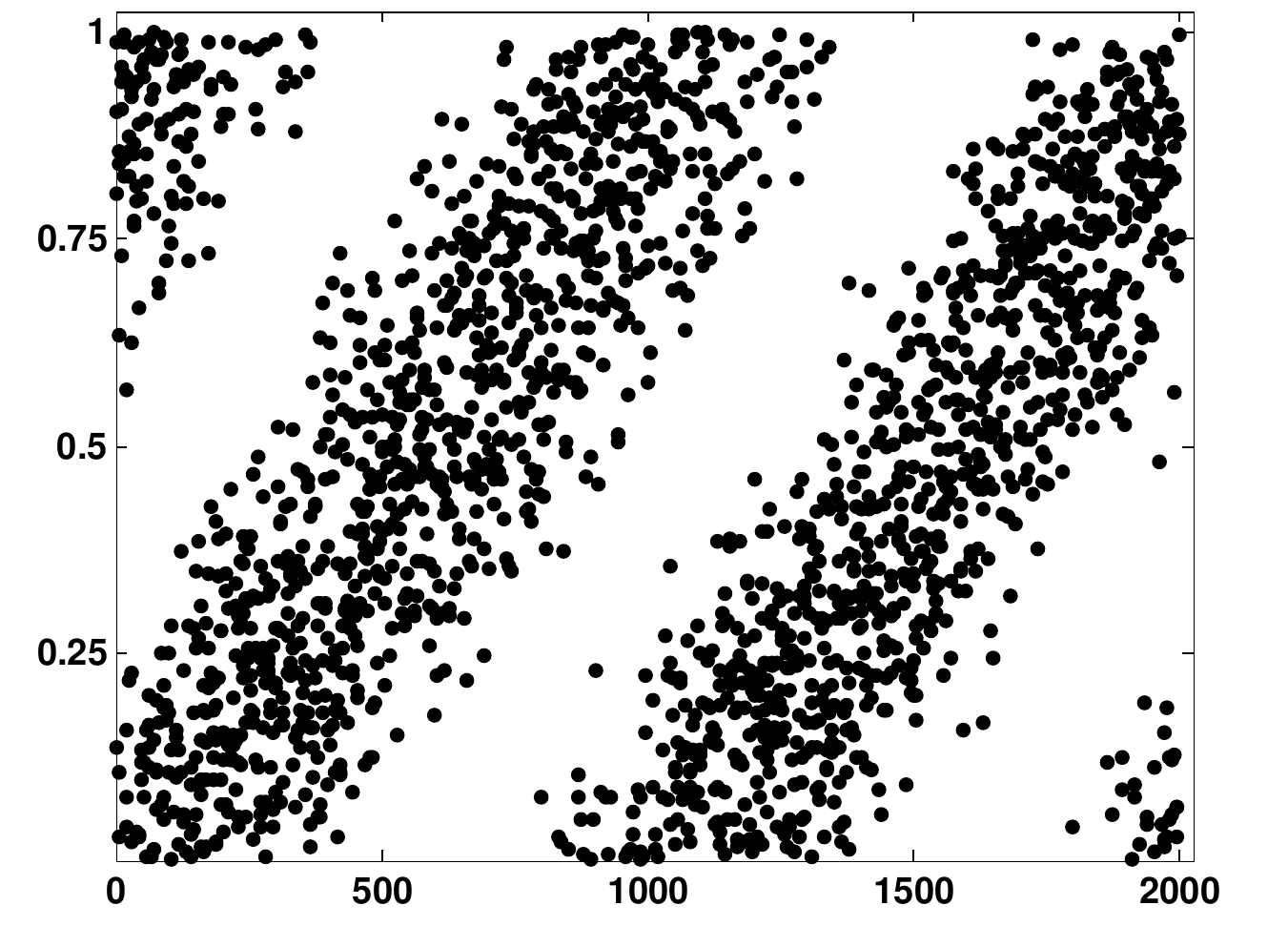}
{\bf l}\hspace{0.1cm}\includegraphics[height=1.8in,width=2.0in]{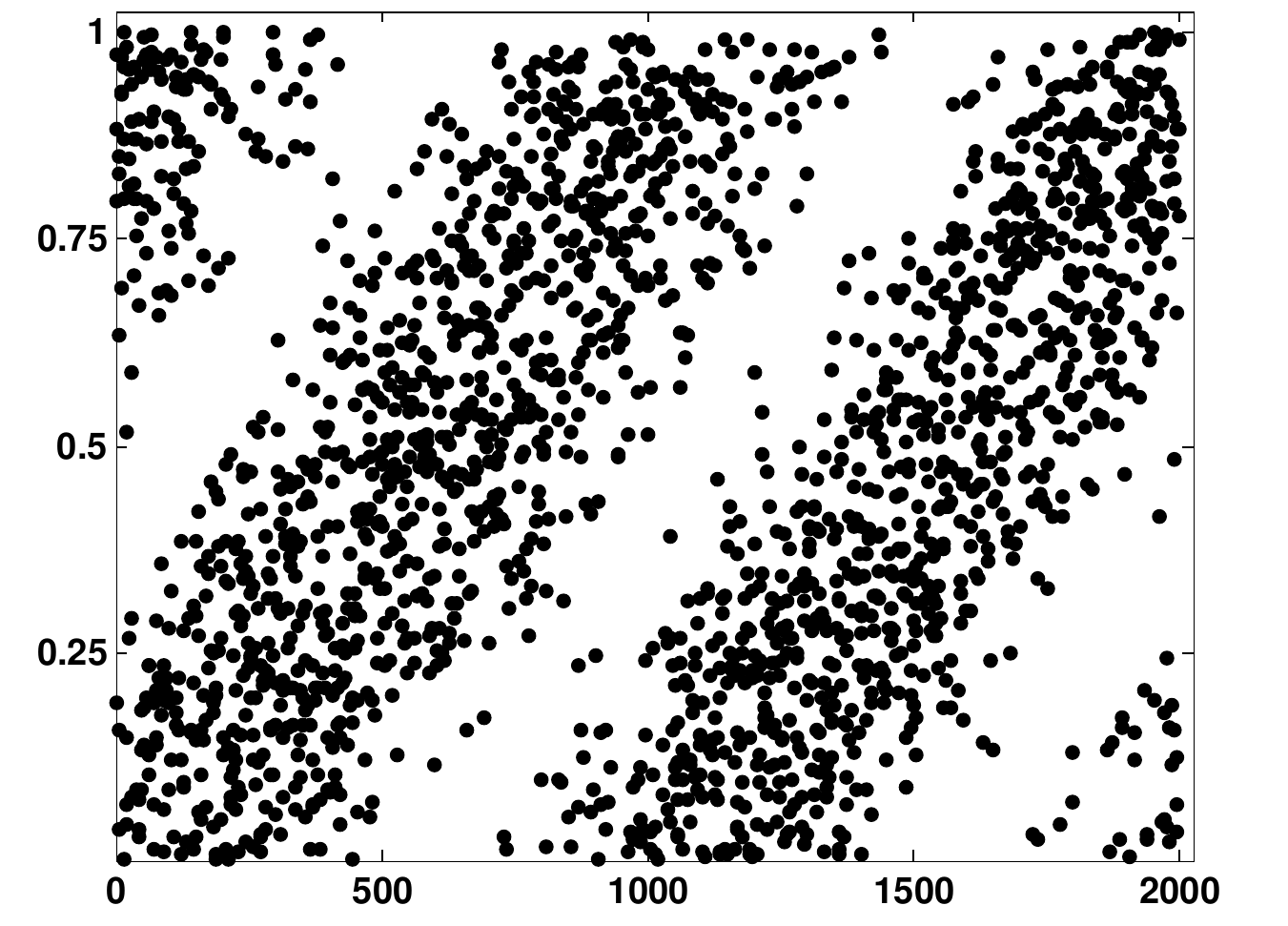}
\end{center}
\caption{Numerical solution of the IVP for the Kuramoto model with
repulsive coupling on $G(2000,0.5)$. Plots \textbf{a}-\textbf{f}
show the solution of the problem initialized with a $1$-twisted state
at times  $t=20, 40, 80, 160, 200,$ and $300$ respectively.
The plots in \textbf{g}-\textbf{l} show the results for the problem
initialized by a $2$-twisted state at the same times.
}
\lbl{f.2000}
\end{figure}

\subsection{Twisted states on the random network}

In this subsection, we prove two results on the existence of the
twisted states for the Kuramoto model on ER graphs.
First, we show that in the limit of large $n$, twisted states become steady state
solutions of the model on random graph and study the stability of twisted states.
Furthermore, for finite albeit large $n\in\N$, we show that the solution
of the IVP for the Kuramoto model on $G(n,p)$ with the initial condition sufficiently
close to a given twisted state remains near this twisted state for a long time.
This means that while twisted generically are not the equilibria of the Kuramoto model
on $G(n,p)$ for finite $n$, they can be interpreted as metastable states 
of the random network provided $n$ is sufficiently large. Both results
elucidate the relation between the Kuramoto models on $K_n$
and $G(n,p)$.

The existence of twisted states for the Kuramoto model on $G(n,p)$  in the limit as $n\to\infty$ 
follows from the argument that was used in the analysis of the Kuramoto model on small-world networks 
\cite{Med13c}.

\begin{thm} \lbl{thm.steady}
Let $\{G(n,p)\}$ be a sequence of the ER random graphs with adjacency matrices
$A_n=(a_{nij})$, and 
\be\lbl{rhs-dKur}
\mathcal{R}_{ni} (u_n)=\lim_{n\to\infty} {1\over np}  \sum_{j=1}^n 
a_{nij}\sin\left(2\pi( u_{nj}-u_{ni})\right), \; u_n\in\R^n, i\in [n].
\ee
Then for any $q\in \N\bigcup \{0\}$ and $i\in\N$
\be\lbl{conv-rhs}
\lim_{n\to\infty}\;\mathcal{R}_{ni} (u_n^{(q)})=0 
\ee
almost surely.
\end{thm}
\pf \;
Let $i\in\N$ be arbitrary but fixed and $n\ge i$. Recall that $a_{nij}, \; 1\le i<j\le n,$
are independent identically distributed binomial (with parameter $p$) RVs
and consider
\begin{eqnarray}
\lbl{def-eta}
\eta_{nij} &=&a_{nij} \sin (2\pi(u^{(q)}_{nj}-u^{(q)}_{ni}))=
a_{nij} \sin \left({2\pi q(j-i)\over n}\right),\\
\lbl{def-Sn}
S_{ni} & = &\sum_{j=1}^n  \eta_{nij} .
\end{eqnarray}

Using (\ref{rhs-dKur}) and the definitions of $\{\eta_{nij}\}$, we have
\be\lbl{rewrite-dKur}
\mathcal{R}_{ni}(u^{(q)}_n)= (np)^{-1}S_{ni}.
\ee

Using the binomial distribution of $a_{nij}$ and (\ref{def-Sn}), it is straightforward to compute
\begin{eqnarray}
\lbl{2nd}
\E {\eta_{nij}}^2 &=& \sin \left({2\pi q(j-i)\over n}\right)^2 p \le 1,\\
\lbl{fourth}
\E {\eta_{nij}}^4 &=& \sin \left({2\pi q(j-i)\over n}\right)^4 p \le 1,  j\in [n].
\end{eqnarray}
Further, using (\ref{def-Sn}), we have
\be\lbl{est-ESn}
\E S_{ni} = p \sum_{j=1}^n \sin \left({2\pi q(j-i)\over n}\right) =0, 
\ee
and
\begin{eqnarray}\nonumber
\E {S_{ni}}^4 &=& \sum_{j_1, j_2,j_3, j_4 \in [n]} 
\E \left[ \eta_{nij_1} \eta_{nij_2} \eta_{nij_3} \eta_{nij_4} \right] \\
\nonumber  
                       &=&  \sum_{j\in [n]} \E \left[{\eta_{nij}}^4\right]+
\dbinom{4}{2}
\sum_{j_1<j_2, j_1,j_2\in [n]} \E \left[{\eta_{nij_1}}^2{\eta_{nij_2}}^2\right] \\
\lbl{S4}
&\le& n+3n(n-1)< 3n^2,
\end{eqnarray}
where we used independence of $\{\eta_{nij}:\, j\in [n]\},$ 
(\ref{2nd}), (\ref{fourth}), and (\ref{est-ESn}).

By Markov inequality, using (\ref{est-ESn}) and (\ref{S4}), for any $\epsilon>0$
we have
\be\lbl{Markov}
\P\{ |S_{ni}|\ge n\epsilon \}\le 
3 \epsilon^{-4} n^{-2},
\ee
which in turn implies via the first Borel-Cantelli lemma \cite[Theorem~4.3]{Bill-Prob}
$$
\P \{ | n^{-1} S_{ni} | \ge \epsilon \; \mbox{holds for infinitely many}\; n\}=0, 
$$
By Theorem~5.2(i) in \cite{Bill-Prob}, the last statement is equivalent to convergence of 
$n^{-1}S_{ni}$ to $0$
almost surely.\\ 
$\qed$

Next, we show that in the Kuramoto model on $G(n,p)$ with finite $n$,
 twisted states are metastable.


\begin{thm}\lbl{thm.gnp-twist}
For any $\epsilon_1,\epsilon_2\in (0,1),\; T>0$ and $q\in\N\cup \{0\}$ there exists
$N=N(\epsilon_1,\epsilon_2)$ such that for every $n\ge N$
one can find $\delta=\delta(n,\epsilon_2)$ so that 
$$
\|u_n(0)-u_n^{(q)}\|_{n,2} <\delta \quad\implies\quad
\P\{\max_{t\in [0,T]} \| u_n(t)-u_n^{(q)}\|_{n,2}>\epsilon_1\}<\epsilon_2.
$$
Here, 
\be\lbl{recall-twist}
u_n^{(q)}=(u_n^{(q)}(0), u_n^{(q)}(1),\dots,u_n^{(q)}(n-1)),
\quad u_n^{(q)}(x)={2\pi qx\over n}+c \pmod{n}, \; x\in \Z_n, 
\ee
denotes a $q$-twisted state for some $c\in\R$.
\end{thm}
\begin{rem}
For $q=0$ the (in)stability follows from Theorem~\ref{thm.sync}: a synchronous
solution is stable if the coupling is attractive and is unstable otherwise.
\end{rem}
\pf
Let $\epsilon_1,\epsilon_2\in (0,1),\; T>0$ and $q\in\N/\{0\}$ be arbitrary
but fixed. 

Since $u_n^{(q)} (n>q)$ is an equilibrium of (\ref{kn}),  
by continuous dependence on initial data,
there exists 
$\delta=\delta(n,\epsilon_2)$ such that 
\be\lbl{v-stable}
\|v_n(0)-u_n^{(q)}\|_{2,n} <\delta \quad\implies \quad 
\max_{i\in [n]}\max_{t\in [0,T]}  |v_{ni}(t)-u^{(q)}_n(i-1)|<\epsilon_1/8.
\ee
Using  (\ref{v-stable}) and the definition of $u_n^{(q)}$ (\ref{recall-twist}), 
we have
$$
{2\pi q (i-j)\over n}-{\epsilon_1\over 4}\le v_{ni}(t)-v_{nj}(t) \le {2\pi q (i-j)\over n}+{\epsilon_1\over 4}, t\in [0,T],
$$
and
$$
\sin\left({2\pi q (i-j)\over n}\right)^2 -\epsilon_1/2 \le \sin\left(2\pi(v_{ni}-v_{nj})\right)^2 \le 
\sin\left({2\pi q (i-j)\over n}\right)^2 +\epsilon_1/2.
$$
Thus,
\begin{eqnarray}\nonumber
{1\over n}\sum_{j=1}^n \sin\left(2\pi(v_{ni}-v_{nj})\right)^2 &\ge& {1\over n}\sum_{j=1}^n 
\sin\left({2\pi q (i-j)\over n}\right)^2 -\epsilon_1/2\\
\lbl{verify-1}
&\ge& 
{1\over 2n} \sum_{j=1}^n \left(1-\cos\left({4\pi q (i-j)\over n}\right)\right)-
\epsilon_1/2=(1-\epsilon_1)/2.
\end{eqnarray}
Similarly, we show that 
\be\lbl{verify-2}
{1\over n}\sum_{j=1}^n \sin\left(2\pi(v_{ni}-v_{nj})\right)^2 \le (1+\epsilon_1)/2.
\ee

Let $u_n(t)$ and $v_n(t)$ denote the solutions of (\ref{gnp}) and (\ref{kn}) respectively, that start
from the same  initial condition as in (\ref{v-stable})
$
u_n(0)=v_n(0).
$
By Theorem~\ref{thm.approx},  there exists $N=N(\epsilon_1,\epsilon_2)$
such that 
\be\lbl{almost-thm}
\P\{\max_{t\in [0,T]} \| v_n(t)-u_n(t)\|_{n,2}>\epsilon_1/2\}<\epsilon_2 \;\mbox{for}\;
n>N(\epsilon_1,\epsilon_2).
\ee
Using (\ref{v-stable}), (\ref{almost-thm}), and the triangle inequality, we have
$$
\|u_n(t)-u_n^{(q)}\|_{n,2} \le \|u_n(t)- v_n(t)\|_{n,2}  +\epsilon_1/ 8.
$$
Thus, 
$$
\P\{\max_{t\in [0,T]} \| u_n(t)-u_n^{(q)}\|_{n,2}>\epsilon_1\}\le 
\P\{\max_{t\in [0,T]} \| v_n(t)-u_n^{(q)}\|_{n,2}>\epsilon_1/2\}<\epsilon_2.
$$
$\qed$

To illustrate Theorem~\ref{thm.gnp-twist} with numerical results, we integrated
the repulsively coupled Kuramoto model on a large random graph $G(2000,0.5)$. 
The results in Figure~\ref{f.2000} (\textbf{a}-\textbf{f}) show a very slow 
evolution starting from a $1$-twisted state. The initial  pattern gradually coarsens until the 
oscillators fill up the phase space randomly.
Plots \textbf{g}-\textbf{l} show that the evolution starting from a 
$2$-twisted state follows a similar pattern.
Theorem~\ref{thm.gnp-twist} applies to both models with repulsive and attractive 
coupling. In the latter case, the trajectory spends a long time  near a twisted state, 
its initial position, before converging to a synchronous solution.

\begin{figure}
\begin{center}
{\bf a}\hspace{0.1 cm}\includegraphics[height=1.8in,width=2.0in]{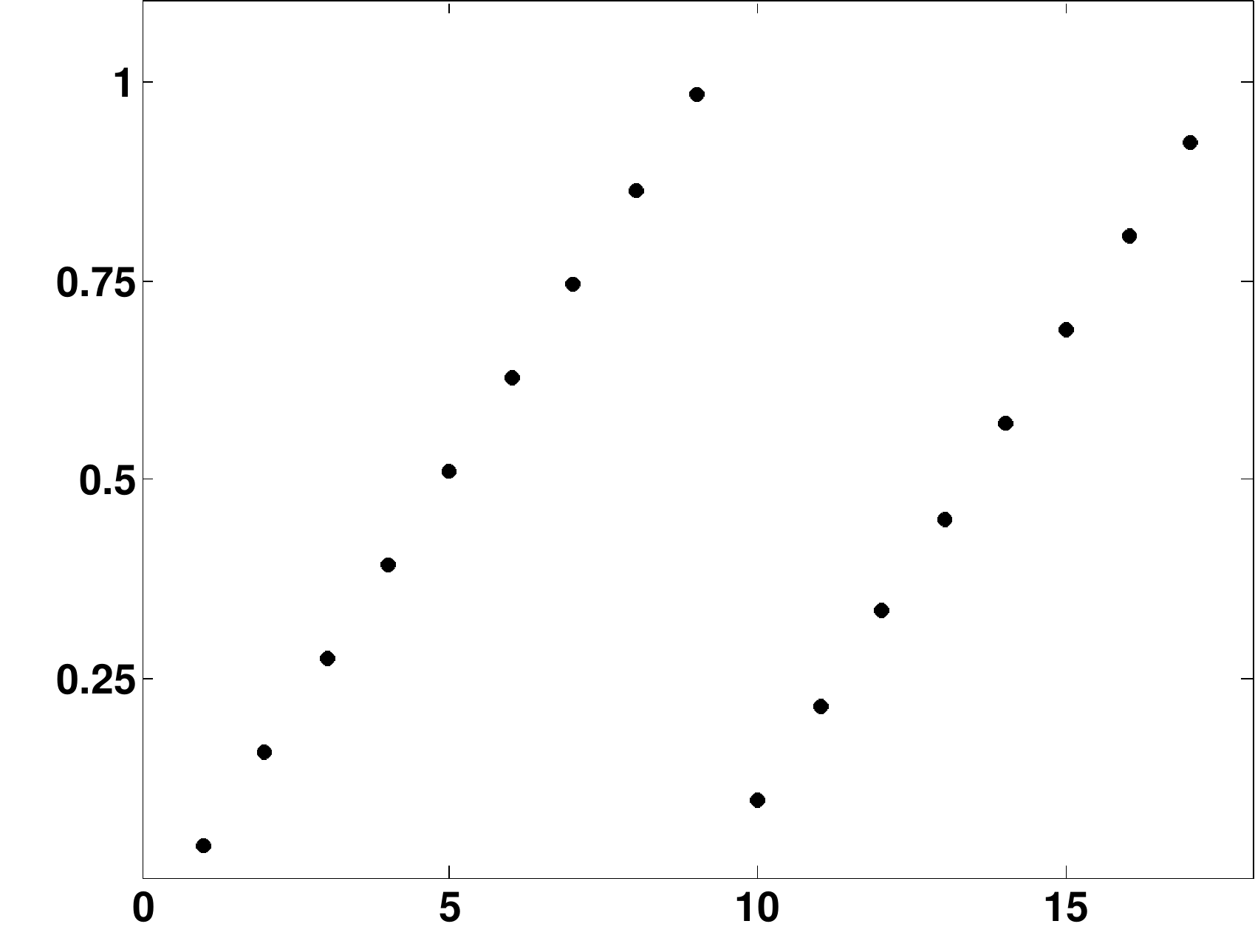}
{\bf b}\hspace{0.1 cm}\includegraphics[height=1.8in,width=2.0in]{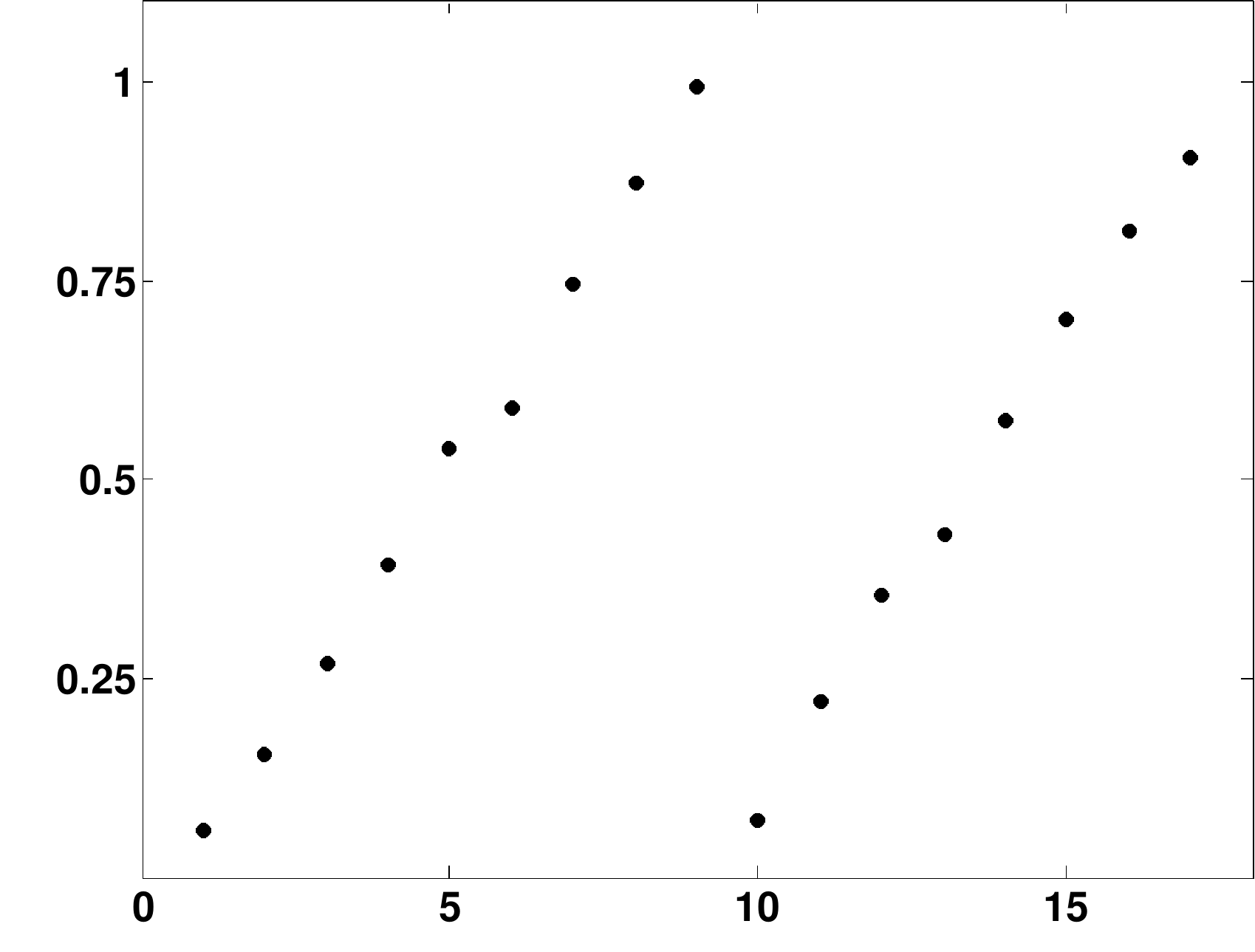}
{\bf c}\hspace{0.1 cm}\includegraphics[height=1.8in,width=2.0in]{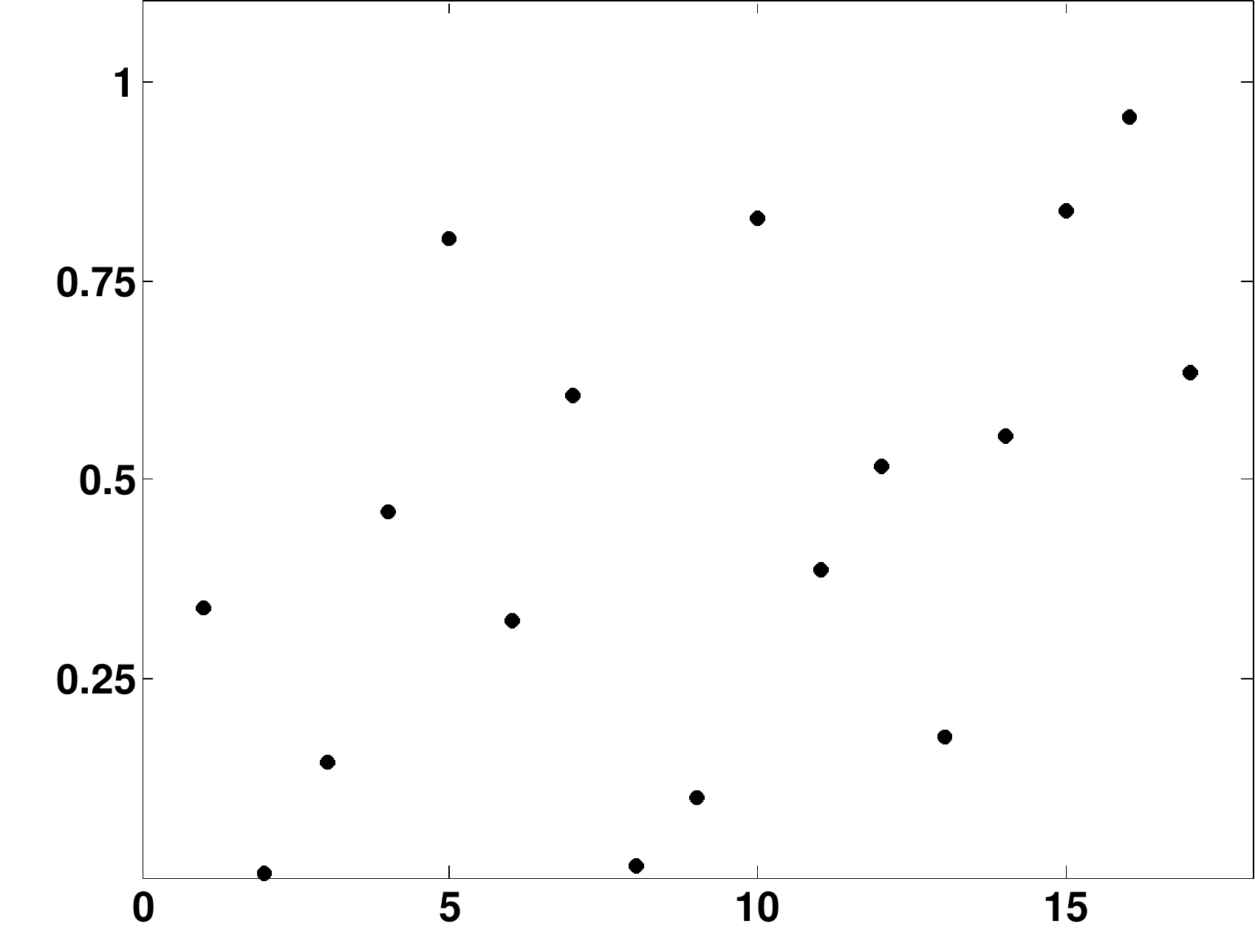}\\
{\bf d}\hspace{0.1 cm}\includegraphics[height=1.8in,width=2.0in]{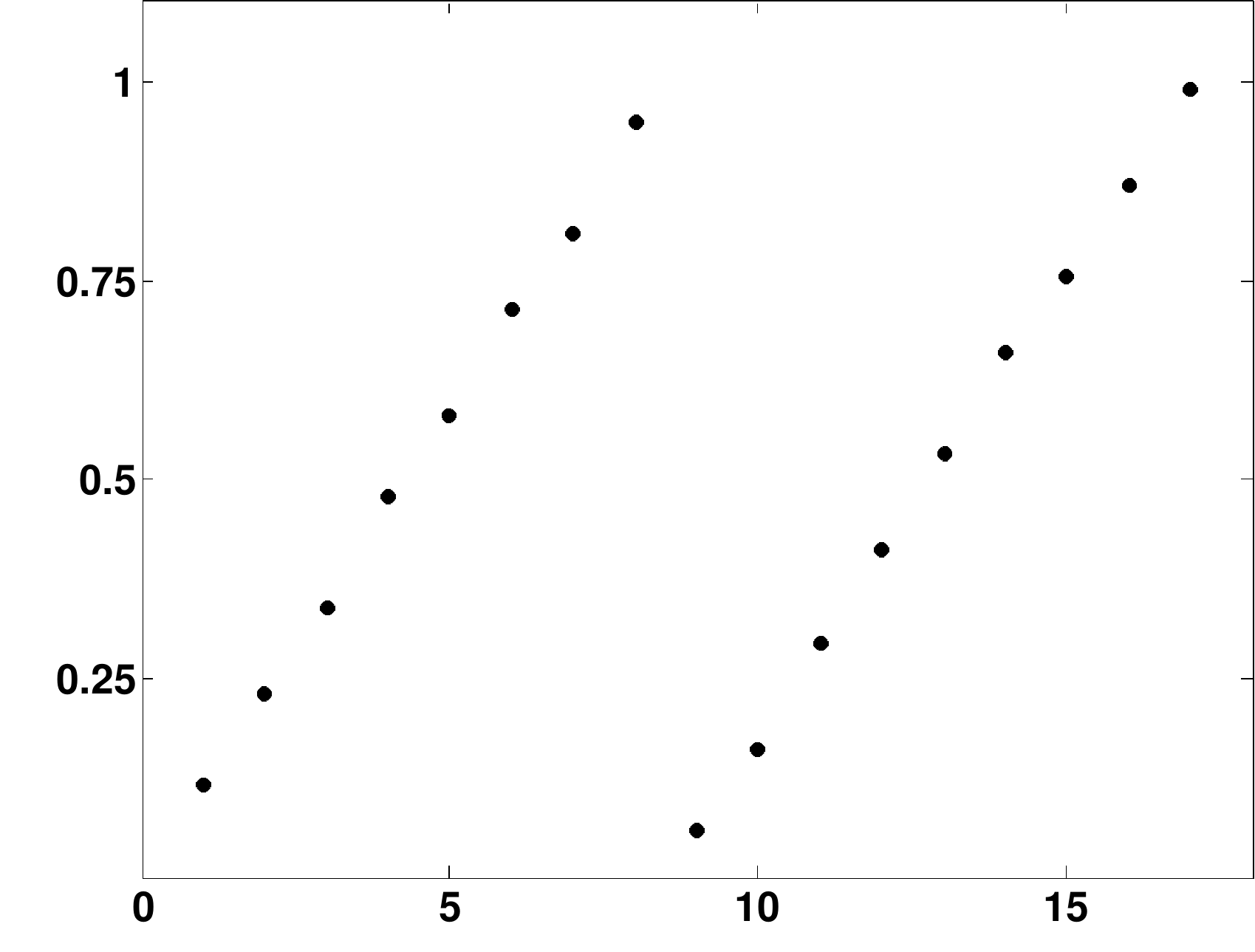}
{\bf e}\hspace{0.1 cm}\includegraphics[height=1.8in,width=2.0in]{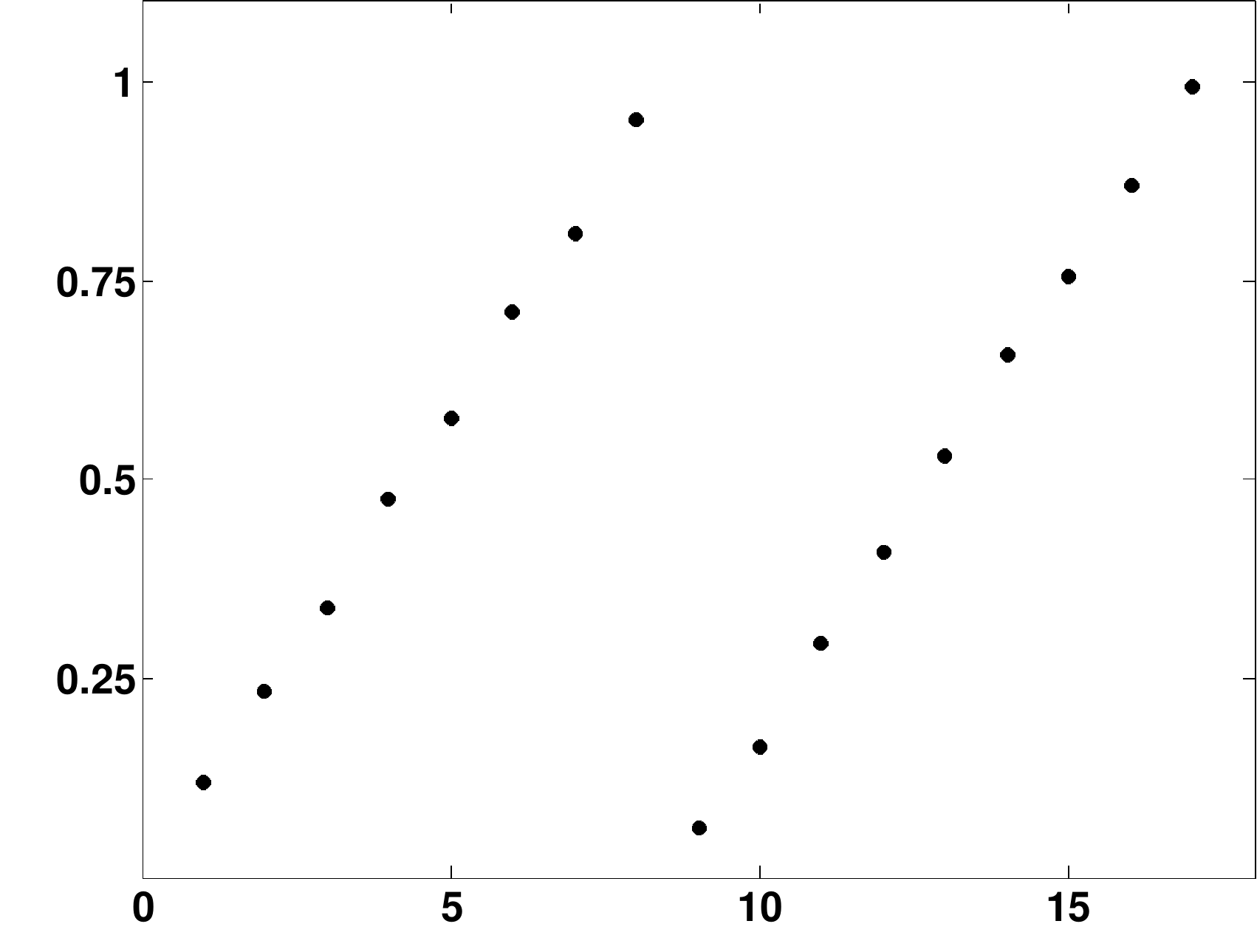}
{\bf f}\hspace{0.1 cm}\includegraphics[height=1.8in,width=2.0in]{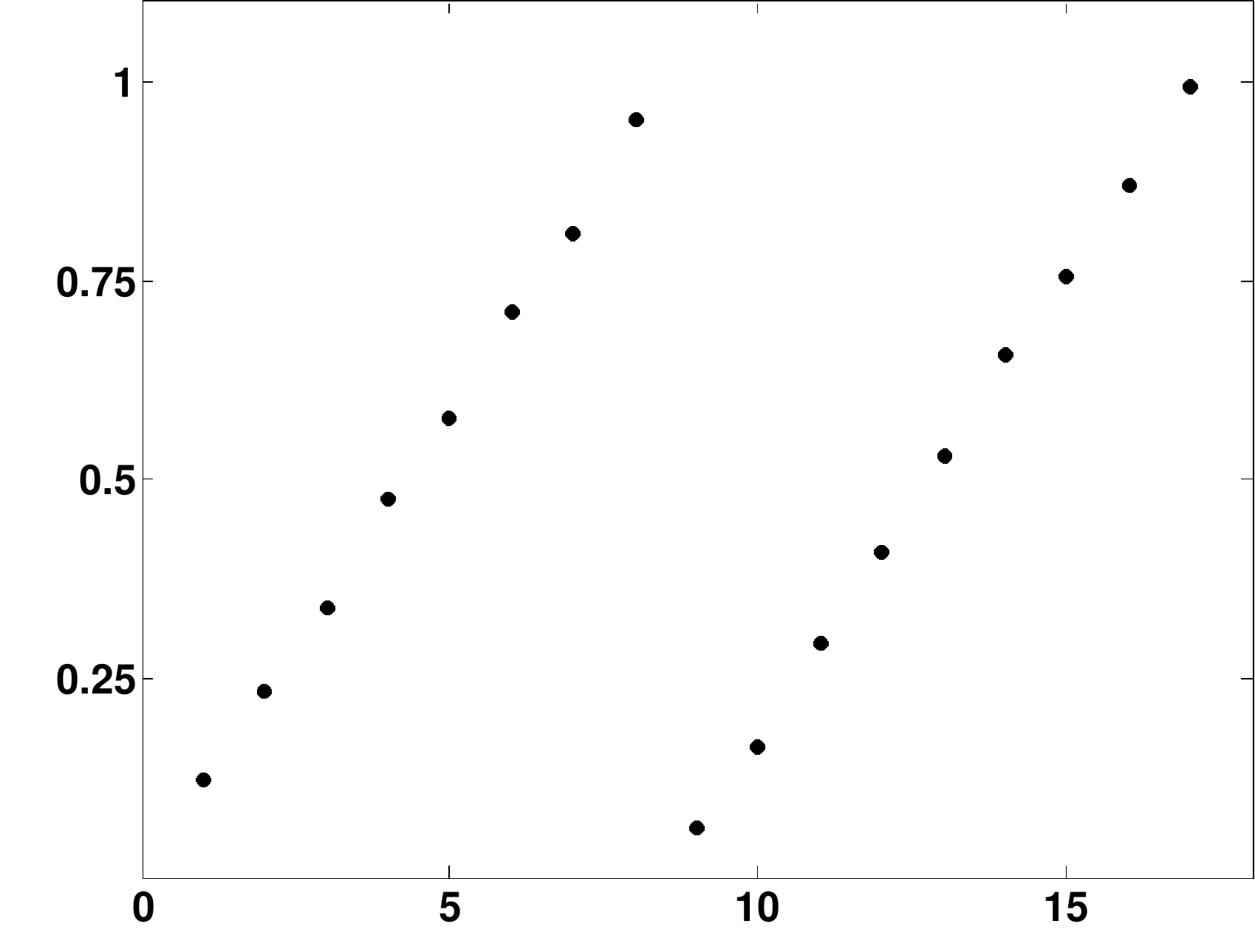}\\
\end{center}
\caption{Numerical solutions of the IVP for the Kuramoto model with repulsive coupling on
Paley ({\bf a}-{\bf c}) and complete ({\bf d}-{\bf f}) graphs.
Each model involves $17$ coupled oscillators. 
The initial conditions for each problem are taken near a $2$-twisted state ({\bf a}, {\bf d}).
These numerics illustrate that the $2$-twisted state is unstable for the Paley graph and stable for the
complete graph.
}
\lbl{f.trace}
\end{figure}

\section{Discussion}~\lbl{sec.discuss}
\setcounter{equation}{0}
In this paper, we studied stability of twisted states on certain Cayley and random graphs.
The motivation for this study was twofold. First, we wanted to extend the stability 
analysis in \cite{WilStr06} to the Kuramoto model on Cayley graphs generated by cyclic groups and  
to find out what can be said about twisted states on random graphs. Our second goal
was to compare dynamics of the Kuramoto model on the families of complete, Paley, and ER graphs,
which are  structurally very  similar. In particular, these graph sequences 
exhibit asymptotically equivalent edge distributions, graph spectra, and have the same
limits (cf. Section~\ref{sec.graphs}). Nonetheless, as our results show the relation between 
network structure and dynamics of coupled nonlinear systems can be quite subtle.
On one hand, we found that the dynamics on these graph sequences are similar in many 
respects. All three models have twisted states as steady state solutions, albeit for 
random graphs $G(n,p)$ this statement holds almost surely in the limit as $n\to\infty$.  
Further, the synchronization subspace is asymptotically stable for all three models with
attractive coupling. Moreover, the rates of convergence to the synchronization subspace,
which are determined by the first nonzero eigenvalues of the graph Laplacians (at least when 
coupling is sufficiently strong \cite{Med12}), are approximately the same for all three
graph sequences. On the other hand, the stability of the same twisted states on large 
complete and Paley graphs may be different despite the strong similarity between these graphs 
(see Figure~\ref{f.trace}).
In particular, we found that in the Kuramoto model with repulsive coupling on $K_n$, all nontrivial
twisted states are stable. In the same model on $P_n$, $q$-twisted states are unstable if
$q$ is not a QR modulo $n$. Thus, half of all twisted states are unstable in the Kuramoto model with
repulsive coupling on $P_n$\footnote{Note, however, that the positive eigenvalues in the spectrum
of the linearized problem in this case are all $o(1)$. Thus, the instability
is rather weak for large $n$.}. This example shows that the asymptotic behavior of solutions
of coupled models on graphs with close structural properties may still be very different.
This example also cautions about the validity of conclusions, one can draw from the analysis
of formal continuum limits of large networks. Recall that the sequences of complete, Paley,
and ER random graphs have the same graph limits (see \S\ref{sec.limit}). Thus, one
might expect that in the limit as the size of the network goes to infinity, the dynamics
of all three models are approximated by the same continuum model. In fact, for the Kuramoto models
on $K_n$ and $G(n,p)$ such limit was established in \cite{Med13a} and \cite{Med13b}
respectively. It was shown that the solutions of the IVPs for these models for large $n$
are approximated by the solutions of the IVP for the continuum equation
\be\lbl{cont}
{\p\over\p x} u(x,t)=(-1)^\alpha\int_I \sin (u(y,t)-u(x,t)) dy.
\ee
The justification of the continuum limit for $P_n$ is not covered by the analysis in these
papers. However, even for the Kuramoto models on $K_n$ and $G(n,p)$, the results in \cite{Med13a, Med13b}
establish the proximity of solutions of the IVPs for discrete and continuum models only 
on finite time intervals, which is not sufficient to guarantee that the solutions of the discrete
and continuum models have the same asymptotic behavior.  

\vskip 0.2cm
\noindent
{\bf Acknowledgements.}
This work was supported in part by the NSF grants  DMS 1109367 
and DMS 1412066  (to GSM).

\vfill\newpage
\bibliographystyle{amsplain}

\providecommand{\bysame}{\leavevmode\hbox to3em{\hrulefill}\thinspace}
\providecommand{\MR}{\relax\ifhmode\unskip\space\fi MR }
\providecommand{\MRhref}[2]{%
  \href{http://www.ams.org/mathscinet-getitem?mr=#1}{#2}
}
\providecommand{\href}[2]{#2}

\end{document}